\documentclass[notitlepage,openbib,12pt]{article}
\usepackage{amsfonts}
\usepackage{amsmath}
\usepackage{amssymb}
\usepackage{graphicx}
\usepackage{color}
\usepackage[colorlinks,citecolor=blue]{hyperref}
\usepackage{natbib}
\usepackage{footmisc}
\usepackage{geometry}

\newtheorem{theo}{Theorem}
\newtheorem{prop}{Proposition}

\newtheorem{define}{Definition}

\newtheorem{lemma}{Lemma}

\let\OLDthebibliography\thebibliography
\renewcommand\thebibliography[1]{
  \OLDthebibliography{#1}
  \setlength{\parskip}{0pt}
  \setlength{\itemsep}{0pt plus 0.3ex}
}
\makeindex
\input{tcilatex}

\begin{document}

\title{\textbf{Common Agency with Non-Delegation or Imperfect Commitment}%
\thanks{%
We are very grateful to Maxim Ivanov, Anirudh Iyer, Andrew Leal, and Mike
Peters for their insightful comments. We also thank seminar participants at
the 2023 Canadian Economic Theory Conference, Caltech, Monash, Southern
Methodist, UCLA and USC for helpful comments and feedback.}}
\author{Seungjin Han\thanks{
Department of Economics, McMaster University, Canada, hansj@mcmaster.ca}
\and Siyang Xiong\thanks{
Department of Economics, University of California, Riverside, United States,
siyang.xiong@ucr.edu}}
\maketitle

\begin{abstract}
In classical contract theory, we usually impose two assumptions: delegated
contracts and perfect commitment. While the second assumption is demanding,
the first one suffers no loss of generality. Following this tradition,
current common-agency models impose delegated contracts and perfect
commitment. We first show that non-delegated contracts expand the set of
equilibrium outcomes under common agency. Furthermore, the powerful menu
theorem for common agency (\cite{mp2001} and \cite{ms2002}) fails for either
non-delegated contracts or imperfect commitment. We identify canonical
contracts in such environments, and re-establish generalized menu theorems.
Given imperfect commitment, our results for common-agency models are
analogous to those in \cite{bs2001} and \cite{ds2021} for the classical
contract theory, which re-establish the revelation principle.
\end{abstract}


\baselineskip= 18.2pt

\section{Introduction}

\label{sec:intro}

Classical contract theory has achieved huge success in explaining economic
phenomena (e.g., adverse selection, moral hazard). It studies a contractual
relationship between a single principal and a single agent in isolation,
even though an agent may have contractual relationships with multiple
principals simultaneously in many real-life strategic situations (e.g., 
\emph{lobbying} (\cite{gh1994}) and \emph{financial contracting} (\cite%
{pr2001}).

Starting from \cite{bw1985, bw1996b, bw1996a}, the common-agency model has
been an extremely useful model to analyze strategic interaction between
several principals and a common agent in markets. Among many, its
applications are \emph{U.S. health care system} (\cite{fm2019}), \emph{%
capital tax competition} (\cite{kk2013}, \cite{cw2018}), \emph{lobbying} (%
\cite{dgh1997}, \cite{ms2008}, \cite{egr2012}), \emph{oligopolistic
competition} (\cite{df2010}), and \emph{financial contracting} (\cite%
{kmp2007}).

While common agency models are rich enough to be applied to many different
economic problems where an agent non-exclusively communicates with all
principals, it conceptually poses several modeling challenges. Notably, the
presence of multiple principals implies that the agent not only has private
information about her (payoff) type but also is informed about contracts
offered by all principals, messages she sends to them, actions or terms of
trades chosen by principals, etc.\footnote{%
Throughout the paper, we use "he" and "she" to denote a principal and the
agent, respectively.} The latter are called market information. This is why
the revelation principle - the powerful tool for equilibrium analysis in the
classical contract theory - fails under common agency. Given common agency,
there is therefore loss of generality to focus on direct mechanisms that ask
the agent to report her type only.

A principal may adopt a general mechanism that asks the agent to report both
her type and market information, but the celebrated menu theorem (\cite%
{mp2001} and \cite{ms2002}) shows that we do not need to. Instead of
offering a general mechanism, a principal can offer a menu contract (which
is the image set of the mechanism) without loss of generality and let the
agent choose an action directly from the menu. The idea is that the agent's
choice itself reveals all the relevant information including both her type
and market information. The cardinality of the equilibrium menu can be
bigger than that of the agent's type set,\footnote{%
The equilibrium menu may contain some actions (i.e., latent actions) which
are intended to be chosen only when a competing principal deviates from the
equilibrium path.} but it is substantially smaller than the cardinality of
general mechanisms, which simplifies our analysis in common-agency models.
Therefore, the menu theorem for common agency is analogous to the revelation
principle for the classical contract theory.

\subsection{Delegated contracts and perfect commitment}

It is important to note that the classical contract theory implicitly
imposes two assumptions: delegated contracts and perfect commitment. In a
delegated contract, the agent's message fully determines the principal's
action, whereas, in a non-delegated contract, the agent's message determines
only a subset of actions, and the principal may choose his action freely in
the designated subset afterwards. Despite no legal definition of what
constitutes a legitimate contract, the classical contract theory focuses on
delegated contracts only. Given a single principal in the model, this
suffers no loss of generality, because every non-delegated contract can be
replicated by a delegated contract without hurting incentive compatibility.
However, the assumption of perfect commitment suffers loss of generality. 
\cite{bs2001} and \cite{ds2021} show how the revelation principle should be
adapted under imperfect commitment.

Current common-agency models inherit the assumptions of delegated contracts
and perfect commitment from the classical contract theory, and the menu
theorem (\cite{mp2001} and \cite{ms2002}) are established under these two
assumptions. However, there is an interesting phenomenon in two prominent
applications of common agency. First, lobbying activities or campaign
contributions must be disclosed in the U.S. and Canada. Second, in U.S.
health care system, secret pharmaceutical contracts are often blamed for
causing enormously high prices of drugs and there has been much support for
transparent contracts in the pharmaceutical industry (\cite{rf2019}).
Theoretically, if the two assumptions of delegated contracts and perfect
commitment hold, it is easy to show that transparency of contracts does not
have any impact on equilibria.\footnote{%
One may argue that transparency of contract may have impact on payoffs
induced by future contracts.---This is the type of imperfect commitment
studied in \cite{ds2021}. The definition of perfect commitment is that all
payoff relevant factors are contractible. Therefore, transparency does not
have any impact on equilibria, given perfect commitment.} This immediately
begs the question: does it suffer loss of generality to impose the two
assumptions under common agency? We provide a positive answer in this paper.
In fact, non-delegated contracts and imperfect commitment fundamentally
change equilibrium allocations under common agency.

For simplicity, we focus on non-delegated contracts first. With
non-delegated contracts, a principal needs to take an action (from the
designated subset) after all of the contracts have been offered by
principals and all of the messages have been transmitted from the agent to
principals. Thus, it is crucial whether a principal observes his opponents'
contracts and messages, before taking his final action.

To study different contracting environments determined by differing
regulations and laws, we adopt different models based on their announcement
and communication features.\footnote{%
For example, it would be proper to analyze lobbying in the U.S. and Canada
with the public-announcement-public-communication model.} There are two
(extreme) communication protocols: public communication (i.e., principals'
messages from the agent are revealed publicly) and private communication
(i.e., each principal's message is sent privately from the agent).
Furthermore, there are two (extreme) announcement protocols: public
announcement (i.e., principals offer contracts publicly to the agent) and
private announcement (i.e., each principal offers a contract privately to
the agent). Thus, we have four common-agency models with different
announcement and communication protocols.

\subsection{Our results}

In this paper, we aim to answer the following three questions.%
\begin{gather*}
\text{(1) }\left( 
\begin{array}{c}
\text{Can non-delegated contracts sustain normatively good equilibria, } \\ 
\text{which cannot be sustained by delegated contracts?}%
\end{array}%
\right) \\
\text{(2) Does the classical menu theorem (\cite{mp2001}) still hold with
non-delegated contracts?} \\
\text{(3) }\left( 
\begin{array}{c}
\text{If not, what canonical contracts should we focus on?} \\ 
\text{Or equivalently, how should the menu theorem be adapted?}%
\end{array}%
\right)
\end{gather*}%
The answer for the first question is: it depends on the announcement
protocol and the communication protocol. With private announcement and
private communication, every equilibrium allocation induced by non-delegated
contracts is an equilibrium allocation with delegated contracts (Theorem \ref%
{thm:non-delegated_vs_delegated}). The intuition is that no information
updating exists under private announcement and private communication, and as
a result, every non-delegated contract can be translated to a delegated
contract without hurting incentive compatibility. However, for the other 3
(of the 4) models with different announcement or communication protocols, we
show that non-delegated contracts expand equilibrium allocations. In section %
\ref{sec:menu:fail}, we provide two examples with public announcement in
which a Pareto efficient allocation cannot be induced by any equilibrium
with delegated contracts, but it can be supported by non-delegated
contracts. The intuition is that principals may offer menu-of-menu contracts
(i.e., a contract contains a collection of menus), in which the agent
chooses one menu for each principal from the collection. Thus, principals
get a chance to coordinate their choices in the designated menus later to
sustain a Pareto efficient allocation. However, this is impossible in an
equilibrium with delegated contracts, when principals and the agent have
conflict of interest.

Clearly, the answer is no for the second question in the three models which
possess a positive answer for the first question.\footnote{%
A menu contract is a delegated contract. As a result, if the answer is yes
for the first question, the answer must be no for the second question.} With
private announcement and private communication, the answer for the second
question is still no, in spite of the negative answer for the first
question. The reason is that there are many more non-delegated contracts
than delegated contracts, and it suffers loss of generality to focus on the
latter only on off-equilibrium paths.

Failure of the menu theorem in all of the four models leads to the question:
what are the canonical contracts that we should focus on in common agency
with non-delegated contracts? One conjecture is that it suffers no loss of
generality to focus on menu-of-menu contracts as described above. This
conjecture is false, because menu-of-menu contracts do not have messages
rich enough to replicate all of the equilibria. For instance, different
types of agents may send different messages $m_{j}$ and $m_{j}^{\prime }$ to
choose the same menu $E_{j}$ from principal $j$'s general non-delegated
contract in an equilibrium, which leads to principal $j$'s different action
choices from $E_{j}$. This kind of action choices is not possible if
principal $j$ offers the corresponding menu-of-menu contract.

Besides the classical menu contracts (denoted by $C^{P}$), we identify three
additional classes of canonical contracts:%
\begin{gather*}
\text{menu-of-menu-with-recommendation contracts, denoted by }C^{R}\text{
(Definition \ref{def:menu_of_menu_with_recommendation}),} \\
\text{menu-of-menu-with-full-recommendation contracts, denoted by }C^{F}%
\text{ (Definition \ref{def:menu_of_menu_with_full_recommendation}),} \\
\text{generalized menu-of-menu-with-full-recommendation contracts, denoted
by }C^{F^{\ast }}\text{ (See (\ref{tttf3})),} \\
\text{with }C^{F}\subsetneqq C^{R}\text{ and }C^{F}\subsetneqq C^{F^{\ast }}%
\text{.}
\end{gather*}%
For contracts in $C^{R}$, principals can manipulate the recommendations for
each menu, while for contracts in $C^{F}$, principals have less freedom to
do so. Compared to $C^{F}$, $C^{F^{\ast }}$ includes some additional
degenerate contracts.\footnote{%
We use an example in Section \ref{sec:private-public} to show that such
degenerate contracts are necessary, in order to establish generalized menu
theorem for private announcement and public communication.}

With these additional canonical contracts, we are able to establish
generalized menu theorems for all of the four models. Similar to the
classical menu theorem, we prove that it suffers no loss of generality to
focus on the following contracts:%
\begin{equation*}
\left[ 
\begin{array}{c}
\text{private announcement} \\ 
\text{private communication}%
\end{array}%
\right] :\left( 
\begin{array}{c}
\text{focus on }C^{P}\text{on the equilibrium path,} \\ 
\text{ focus on }C^{F}\text{on off-equilibrium paths}%
\end{array}%
\right) \text{ }
\end{equation*}%
\begin{equation*}
\left[ 
\begin{array}{c}
\text{public announcement} \\ 
\text{private communication}%
\end{array}%
\right] :\left( 
\begin{array}{c}
\text{focus on }C^{R}\text{on the equilibrium path,} \\ 
\text{ focus on }C^{F}\text{on off-equilibrium paths}%
\end{array}%
\right) \text{,}
\end{equation*}%
\begin{equation*}
\left[ 
\begin{array}{c}
\text{public announcement} \\ 
\text{public communication}%
\end{array}%
\right] :\left( 
\begin{array}{c}
\text{focus on }C^{R}\text{on the equilibrium path,} \\ 
\text{ focus on }C^{F}\text{on off-equilibrium paths}%
\end{array}%
\right) \text{,}
\end{equation*}%
\begin{equation*}
\left[ 
\begin{array}{c}
\text{private announcement} \\ 
\text{public communication}%
\end{array}%
\right] :\left( 
\begin{array}{c}
\text{focus on }C^{R}\text{on the equilibrium path,} \\ 
\text{ focus on }C^{F^{\ast }}\text{on off-equilibrium paths}%
\end{array}%
\right) \text{.}
\end{equation*}%
Our full characterization is very subtle. For instance, the generalized menu
theorem for public announcement and public (or private) communication does
not extend to private announcement and public communication, which is proved
by an example in Section \ref{sec:private-public}. Furthermore, our analysis
uncovers a feature of our full characterization, which is not shared by the
classical menu theorem: asymmetric contract spaces between on and off the
equilibrium paths. Specifically, to prove our full characterization, we need
to replicate an equilibrium on the general contract space with an
equilibrium on a simple contract space: on the equilibrium path, we just
mimic \emph{one contract profile} in the former space with one contract
profile in the latter space, and we show $C^{R}$ suffices for the latter
space (but $C^{F}$ (or $C^{F^{\ast }}$) does not); on off-equilibrium paths,
we mimic \emph{all possible contract profiles} (due to all possible
deviations) in the former space with those in the latter space, and we show $%
C^{F}$ (or $C^{F^{\ast }}$) suffices for the latter space (but $C^{R}$ does
not). We embed this subtle strategic difference into the different
requirements in the definitions of $C^{R}$, $C^{F}$ and $C^{F^{\ast }}$
discussed above.

Finally, we show common agency with non-delegated contracts is a special
case of common agency with imperfect commitment \emph{\`{a} la} \cite%
{bs2000, bs2001, bs2007}. A rigorous relationship between the two models is
provided in Section \ref{sec:commitment}.\ All of our analysis and full
characterization in the former model can be easily extended to the latter.
The revelation principle in mechanism design with imperfect commitment has
been scrutinized by recent papers (e.g., \cite{bs2001}, \cite{ds2021}),
which provide various economic applications. To the best of our knowledge,
this paper is the first one to study the menu theorem (i.e., the counterpart
of revelation principle) for common agency with imperfect commitment.

\subsection{Related Literature}

In his work on competing auctions, \cite{pm1993} discusses the difficulty in
defining the large strategy space, \textquotedblleft since mechanisms must
map the set of mechanisms into outcomes\ldots \textquotedblright . Unlike
single-principal models, an agent's private information is partly endogenous
in models with multiple principals because, in addition to any exogenous
private information on their types, agents also observe the contracts
offered by all principals. While it seems natural to allow a principal to
write contracts that ask agents to report the entirety of their private
information, such a contract invites the agent to report intricate contracts
(received by the agent) and becomes even more complicated itself; the
complexity of such contracts escalates rapidly -- this is what McAfee refers
to as the \textquotedblleft infinite-regress problem\textquotedblright .

\cite{ep1999} propose a class of universal mechanisms that allow agents to
describe competing principals' mechanisms using a universal language, so
that there is no loss of generality to focus on such mechanisms on and off
the equilibrium paths. However, this language itself is quite complex,
reflecting the infinite-regress problem embedded in the structure of a
principal's mechanism. Given a set of mechanisms allowed in a competing
mechanism game, \cite{ty2010} shows that any incentive compatible profile of
direct mechanisms can be supported as an equilibrium allocation if and only
if each principal's payoff associated with it is no less than his
min-max-min payoff value derived with respect to his opponents' mechanisms,
his own mechanisms, and the agents' continuation equilibria. This
folk-theorem-type result is derived using a recommendation mechanism as an
equilibrium mechanism that delegates the choice of his incentive compatible
direct mechanism to agents. The mechanism chosen by most agents is
implemented. Recommendation mechanisms can be thought of as an extension of
menus proposed for common agency (Peters (2001); Martimort and Stole (2002)).

\cite{ep1999}, \cite{ty2010}, \cite{mp2001}, and \cite{ms2002} aim to find
canonical mechanisms and/or full characterization of equilibrium allocations
in a contracting environment where principals offer delegated contracts. 
\cite{bs2009} noticed that some equilibrium allocations with delegated
contracts disappear if non-delegated contracts are allowed in a game of
complete information with multiple principles and agents. His example
highlights this, showing that the lower bound of a principal's equilibrium
payoff with non-delegated contracts is the minmax value over deterministic
actions but the lower bound with delegated contracts is the maxmin value. As 
\cite{mp2014} points out, when principals are permitted to act randomly,
Szentes argument no longer holds since the two lower bounds are identical
under random actions. Our examples show that in common agency models with
non-delegated contracts, there are new equilibrium allocations that cannot
be supported in a common agency model with delegated contracts and the logic
is independent of whether random actions are permitted.


\section{Examples\label{sec:examples}}

\label{sec:menu:fail}

We aim to show that equilibria differ substantially for the four models with
different announcement and communication protocols. Given private
announcement and private communication, Theorem \ref%
{thm:non-delegated_vs_delegated} will show that non-delegated contracts do
not expand the set of equilibria. Thus, our examples focus on the other
three models. In particular, we provide two examples with public
announcement in this section, in which some Pareto efficient allocations
cannot be induced by any equilibrium with delegated contracts,\footnote{%
These new equilibrium allocations are based on deterministic contracts and
pure-strategy equilibria. However, even if we allow for random contracts and
consider mixed strategies, these new equilibrium allocations cannot be
supported by any equilibrium in the common-agency model with delegated
contracts.} but they can be supported by equilibria with non-delegated
contracts. In the first example, a Pareto efficient allocation can be
sustained under both private communication and public communication. In the
second example, the Pareto efficient allocation can be sustained only under
public communication, but not under private communication. We will provide a
third example for private announcement and public communication in Section %
\ref{sec:private-public:counterexample}.

\subsection{Example 1}

\label{sec:menu:fail:public-private}

There are two principals, $\mathcal{J}=\left\{ 1,2\right\} $ and each
principal can take an action of either $0$ or $1$ (i.e., $%
Y_{1}=Y_{2}=\left\{ 0,1\right\} $). We consider a case of complete
information. The state is $\theta $ and it is publicly known to every
player: $\Theta =\left\{ \theta \right\} .$ Principals' preferences are
described as follows.%
\begin{equation*}
\begin{tabular}{|c|c|c|}
\hline
$\left( v_{1}\left[ \left( y_{1},y_{2}\right) ,\theta \right] ,\text{ }v_{2}%
\left[ \left( y_{1},y_{2}\right) ,\theta \right] \right) :$ & $y_{2}=0$ & $%
y_{2}=1$ \\ \hline
$y_{1}=0$ & $0,0$ & $8,-8$ \\ \hline
$y_{1}=1$ & $0,0$ & $1,1$ \\ \hline
\end{tabular}%
\end{equation*}%
The agent's preference is perfectly aligned with principal 1's: $u\left[
\left( y_{1},y_{2}\right) ,\theta \right] =v_{1}\left[ \left(
y_{1},y_{2}\right) ,\theta \right] $ for any $(y_{1},y_{2})\in Y_{1}\times
Y_{2}$.

A Pareto efficient allocation $(y_{1}=1,y_{2}=1)$ cannot be induced by an
equilibrium with delegated contracts. To see this, suppose otherwise. By the
menu theorem, we can use menu contracts to sustain an equilibrium which
induces $(y_{1}=1,y_{2}=1)$. Thus, principal $2$'s equilibrium menu contract
must contain the action of $1$. Then, principal $1$ finds it profitable to
deviate to the degenerate menu $\left\{ 0\right\} $, which would induce the
continuation equilibrium $\left( y_{1}=0,y_{2}=1\right) $, i.e., both
principal $1$ and the agent achieve the highest utility, $8$. Therefore, we
reach a contradiction.

However, $(y_{1}=1,y_{2}=1)$ can be induced by an equilibrium with
non-delegated contract with public announcement, \emph{regardless of
communication protocols}. Fix any $\left( m_{1}^{\ast },m_{2}^{\ast }\right)
\in M_{1}\times M_{2}$, and consider the following equilibrium.

\begin{enumerate}
\item On the equilibrium path, each principal $j\in \mathcal{J}$ offers a
degenerate contract $c_{j}^{\ast }:M_{j}\rightarrow 2^{Y_{j}}$ such that $%
c_{1}^{\ast }\left( m_{1}\right) =\left\{ 1\right\} $ and $c_{2}^{\ast
}\left( m_{2}\right) =\left\{ 0,1\right\} $ for any $\left(
m_{1},m_{2}\right) \in M_{1}\times M_{2}$,

\item the agent sends $m_{j}^{\ast }$ to every principal $j\in \mathcal{J}$.

\item Principal $1$ chooses $1$ from the subset $\left\{ 1\right\} $ and
principal $2$ chooses $1$ from the subset $\left\{ 0,1\right\} $.
\end{enumerate}

To sustain this as an equilibrium, the agent follows the table below to send
messages at Stage 2, and principals also follow this table to play
continuation equilibrium at Stage 3.%
\begin{equation*}
\begin{tabular}{|c|c|c|c|c|}
\hline
Ranking: & 1 & 2 & 3 & 4 \\ \hline
$c_{1}\left( m_{1}\right) \times c_{2}\left( m_{2}\right) =$ & $\left\{
0\right\} \times \left\{ 1\right\} $ & $\left\{ 0,1\right\} \times \left\{
1\right\} $ & $\left\{ 1\right\} \times \left\{ 1\right\} $ & $\left\{
1\right\} \times \left\{ 0,1\right\} $ \\ \hline
continuation equilibrium at Stage 3 & $\left( 0,1\right) $ & $\left(
0,1\right) $ & $\left( 1,1\right) $ & $\left( 1,1\right) $ \\ \hline
$v_{j_{1}}=u=$ & $8$ & $8$ & $1$ & $1$ \\ \hline
\end{tabular}%
\end{equation*}%
\begin{equation*}
\begin{tabular}{|c|c|c|c|c|}
\hline
5 & 6 & 7 & 8 & 9 \\ \hline
$\left\{ 0,1\right\} \times \left\{ 0,1\right\} $ & $\left\{ 0\right\}
\times \left\{ 0,1\right\} $ & $\left\{ 0\right\} \times \left\{ 0\right\} $
& $\left\{ 1\right\} \times \left\{ 0\right\} $ & $\left\{ 0,1\right\}
\times \left\{ 0\right\} $ \\ \hline
$\left( 0,0\right) $ & $\left( 0,0\right) $ & $\left( 0,0\right) $ & $\left(
1,0\right) $ & $\left( 1,0\right) $ \\ \hline
$0$ & $0$ & $0$ & $0$ & $0$ \\ \hline
\end{tabular}%
\end{equation*}%
That is, there are 9 non-empty subsets of $Y_{1}\times Y_{2}$, and at Stage
3, for each subset (listed at Row 2) that is pinned down by the agent's
messages, we let principals play the corresponding continuation (Nash)
equilibrium (listed at Row 3). Row 4 lists the payoffs of these continuation
equilibria for the agent and principal $1$. Furthermore, row 1 ranks these
continuation equilibria (and the corresponding subsets of $Y_{1}\times Y_{2}$%
), according to the agent's payoffs (at Row 4). Clearly, principals'
incentive compatibility hold at Stage 3.

We say a contract profile $\left( c_{1},c_{2}\right) $ is ranked $n$-th if
and only if $n$ is the smallest integer such that there exists $\left(
m_{1},m_{2}\right) $ and $c_{1}\left( m_{1}\right) \times c_{2}\left(
m_{2}\right) $ is ranked $n$-th in the table above. First, we consider
public communication. When principals offer a contract profile that is
ranked $n$-th at Stage 1, we let the agent send messages that induce the $n$%
-th ranked subsets of $Y_{1}\times Y_{2}$ at Stage 2, and principals play
the corresponding continuation equilibrium at Stage 3 (as described in row 3
in the table). As a result, the agent's incentive compatibility holds at
Stage 2. Furthermore, Principal 2's incentive compatibility also holds
because he achieves the maximal utility on the equilibrium path. Finally,
suppose that principal $1$ unilaterally deviates from the equilibrium. Since
Principal $2$'s equilibrium contract leads only to $\left\{ 0,1\right\} $,
by the table above, the contract profile induced by principal $1$'s
deviation can be ranked only lower, i.e., not a profitable deviation for
principal $1$.

Second, we consider private communication. Since Principal 2 achieves the
maximal utility on the equilibrium path, he does not want to deviate.
Suppose Principal $1$ deviates to $c_{1}:M_{1}\rightarrow 2^{Y_{1}}$ with $%
0\in c_{1}\left( m_{1}\right) $ for some $m_{1}\in M_{1}$. Let Principal $2$
play $0$ at Stage 3, and hold the belief that the agent sends $m_{1}$ to
Principal $1$ and Principal $1$ plays $0$. Giving Principal $2$ playing $0$
at Stage 3, both the agent and Principal $1$ always get $0$ for any actions,
and hence, it is incentive compatible for them to send $m_{1}$ and play $0$.
As a result, such a deviation is not profitable for Principal $1$.

\subsection{Example 2\label{sec:example_public_public}}

\label{sec:menu:fail:public-public}

There are two principals, $\mathcal{J}=\left\{ 1,2\right\} $ and each
principal can take one of four possible actions (i.e., $Y_{1}=Y_{2}=\left\{
1,2,3,4\right\} $). There are two possible states $\Theta =\left\{
1,4\right\} $, which is privately observed by the agent.

Both principals have the same preferences. If both principals' actions match
the true state, both principals get utility $8$. Otherwise, they both get $1$
if the sum of their actions is even, and they get $-1$ if the sum is odd,
i.e.,

\begin{equation*}
v_{1}\left[ \left( y_{1},y_{2}\right) ,\theta \right] =v_{2}\left[ \left(
y_{1},y_{2}\right) ,\theta \right] =\left\{ 
\begin{tabular}{ll}
$8$, & if $\left( y_{j},y_{j}\right) =\left( \theta ,\theta \right) $, \\ 
$1$, & if $\left( y_{1},y_{2}\right) \neq \left( \theta ,\theta \right) $
and $\left( y_{1}+y_{2}\right) $ is even, \\ 
$-1$, & if $\left( y_{1}+y_{2}\right) $ is odd,%
\end{tabular}%
\right. \text{ }\forall \theta \in \Theta \text{.}
\end{equation*}%
The agent gets $1$ if $\left( y_{1},y_{2}\right) =\left( 1,4\right) $, and
gets $0$ otherwise, i.e., 
\begin{equation}
u\left[ \left( y_{1},y_{2}\right) ,\theta \right] =\left\{ 
\begin{tabular}{ll}
$1$, & if $\left( y_{1},y_{2}\right) =\left( 1,4\right) $, \\ 
$0$, & otherwise,%
\end{tabular}%
\right. \text{ }\forall \theta \in \Theta \text{.}  \label{htt4}
\end{equation}

Principals' best allocation is $z^{\ast \ast }\equiv \left[ \left(
y_{1},y_{2}\right) =\left( \theta ,\theta \right) \text{ at each }\theta \in
\Theta \right] $. We show $z^{\ast \ast }$ cannot be induced by any
equilibrium with delegated contracts. By the menu theorem, each principal's
equilibrium menu must include both $1$ and $4$, in order to sustain $z^{\ast
\ast }$. Given this, it is uniquely optimal for the agent to choose $1$ from
principal 1's menu and $4$ from principal 2's menu regardless of the state.
Therefore, $z^{\ast \ast }$ cannot be sustained by delegated contracts.

\subsubsection{Private communication}

Given private communication, $z^{\ast \ast }$ cannot be induced by any
equilibrium with non-delegated contracts. To see this, suppose otherwise,
i.e., there exists an equilibrium which induces $z^{\ast \ast }$. On the
equilibrium path, suppose principals offer $(c_{1},c_{2})$, and the agent
sends $\left( m_{1}^{\theta },m_{2}^{\theta }\right) $ at state $\theta \in
\Theta $, and given private communication, principal $j\in \mathcal{J}$
takes the action $\theta \in c_{j}\left( m_{j}^{\theta }\right) $, upon
observing $\left[ (c_{1},c_{2}),m_{j}^{\theta }\right] $. Regardless of
states, the agent thus finds it profitable to deviate to send $%
m_{j=1}^{\theta =1}$ and $m_{j=2}^{\theta =4}$ to principals $1$ and $2$,
respectively, which induces $\left( y_{1},y_{2}\right) =\left( 1,4\right) $
and achieves the maximal utility. Therefore, we reach a contradiction.

\subsubsection{Public communication}

Now, we show that $z^{\ast \ast }$ is induced by an equilibrium with
non-delegated contracts under public communication. Let $M_{j}$ denote the
set of messages that the agent can send to principal $j$, and $M\equiv
M_{1}\times M_{2}$. For each $j\in \mathcal{J}$, fix any two distinct
messages, $m_{j}^{1}$ and $m_{j}^{4}$ in $M_{j}$, and define a contract, $%
c^{\ast \ast }\equiv \left[ c_{j}^{\ast \ast }:M_{j}\rightarrow 2^{Y_{j}}%
\right] _{j\in \mathcal{J}}$ as follows.%
\begin{equation*}
c_{j}^{\ast \ast }\left( m_{j}\right) =\left\{ 
\begin{tabular}{ll}
$\left\{ 1,2\right\} $, & if $m_{j}=m_{j}^{1}$, \\ 
$\left\{ 3,4\right\} $, & if $m_{j}=m_{j}^{4}$, \\ 
$\left\{ 3,4\right\} $, & otherwise%
\end{tabular}%
\right. \text{.}
\end{equation*}

Consider the following equilibrium:

\begin{enumerate}
\item On the equilibrium path, every principal $j\in \mathcal{J}$ offers $%
c_{j}^{\ast \ast }$,

\item the agent sends $m_{j}^{\theta }$ to every principal $j\in \mathcal{J}$
at every state $\theta \in \Theta $,

\item upon receiving $\left( m_{1}^{\theta },m_{2}^{\theta }\right) ,$
principals choose $(\theta ,\theta )$.
\end{enumerate}

On the equilibrium path, the induced outcome is $\left( 1,1\right) $ at
state $1$ and $\left( 4,4\right) $ at state $4$. Therefore, $z^{\ast \ast }$
is implemented and principals achieve their maximal utility at both states,
and hence their incentive compatibility holds at Stage 1.

Let $C_{j}$ be the set of all possible non-delegated contracts for principal 
$j$ and $C\equiv C_{1}\times C_{2}.$ To sustain this as an equilibrium, the
agent and principals take the following (behavior) strategies at Stages 2
and 3. The agent's communication strategy at Stage 2 is denoted by $s\equiv
\lbrack s_{k}:C\times \Theta \rightarrow M_{k}]_{k\in \mathcal{J}}$ and
principal $j$'s action choice strategy at Stage 3 is denoted by $%
t_{j}:C\times M\rightarrow Y_{j}$.\footnote{%
Note that the domain of $t_{j}$ is $C\times M$ because all contracts and
messages are publicly observable.} 
\begin{gather}
\text{At Stage 2, the agent takes }s=[s_{k}:C\times \Theta \rightarrow
M_{k}]_{k\in \mathcal{J}}\text{ such that}  \notag \\
s\left( c,\theta \right) =\left\{ 
\begin{tabular}{ll}
$\left( m_{1}^{\theta },m_{2}^{\theta }\right) ,$ & if $c=c^{\ast \ast }$ \\ 
$m$, & if $c\neq c^{\ast \ast }$ and $\exists m\in M\text{ such that }%
c\left( m\right) =\left\{ 1\right\} \times \left\{ 4\right\} \text{,}$ \\ 
$\left( m_{1}^{1},m_{2}^{4}\right) $, & otherwise,%
\end{tabular}%
\right. \text{ }\forall \theta \in \Theta \text{;}  \label{htt3}
\end{gather}%
\begin{gather}
\text{At Stage 3, principal }1\text{ takes }t_{1}:C\times M\rightarrow Y_{1}%
\text{ and principal }2\text{ takes }t_{2}:C\times M\rightarrow Y_{2}\text{
such that}  \notag \\
\text{if }c=c^{\ast \ast }\text{ and }m=\left( m_{1}^{\theta },m_{2}^{\theta
}\right) \text{ for some }\theta \in \Theta \text{, }t_{1}\left( c,m\right)
=t_{2}\left( c,m\right) =\theta \text{,}  \notag \\
\text{otherwise, (i) }c\left( m\right) =\left\{ 1\right\} \times \left\{
4\right\} \Rightarrow \left[ t_{1}\left( c,m\right) ,t_{2}\left( c,m\right) %
\right] =\left( 1,4\right) \text{,}  \label{htt1} \\
\text{and}  \notag \\
\text{(ii) }c\left( m\right) \neq \left\{ 1\right\} \times \left\{ 4\right\}
\Rightarrow \left[ t_{1}\left( c,m\right) ,t_{2}\left( c,m\right) \right]
\in \arg \max_{\left( y_{1},y_{2}\right) \in c\left( m\right) \diagdown
\left\{ \left( 1,4\right) \right\} }v_{1}\left[ \left( y_{1},y_{2}\right)
,\theta =1\right] \text{.}  \label{htt2}
\end{gather}%
Since principals have the same utility, they can coordinate their actions
under public announcement and public communication. At Stage 3, if the
agent's message pins down principals' action profile as $\left( 1,4\right) $%
, (\ref{htt1}) says that principals would take $\left( 1,4\right) $, and
otherwise, (\ref{htt2}) says that principals would believe the state is $1$
and take an optimal action profile in $c\left( m\right) \diagdown \left\{
\left( 1,4\right) \right\} $. As a result, incentive compatibility of
principals at Stage 3 holds.

At Stage 2, (\ref{htt3}) says that the agent would send a message to pin
down $\left\{ 1\right\} \times \left\{ 4\right\} $ for principals if such a
message exists, and thus achieve her maximal utility. If there is no message
that can pin down $\left\{ 1\right\} \times \left\{ 4\right\} $ for
principals, the agent is indifferent among all of the messages, because in
this case, by (\ref{htt2}), principals would choose an action profile
different from $\left( 1,4\right) $, and by (\ref{htt4}), the agent is
indifferent between any action profiles in $Y\diagdown \left\{ \left(
1,4\right) \right\} $. Therefore, incentive compatibility of the agent at
Stage 2 holds.

\section{Preliminaries\label{sec:preliminaries}}

\label{sec:model}

\subsection{Primitives}

\label{sec:model:primitive}

A single agent privately observes her type $\theta \in \Theta $, which is
drawn from a common prior $p\in \Delta \left( \Theta \right) $ with full
support. Let $\mathcal{J}\equiv \left\{ 1,\ldots ,J\right\} $ be the set of
principals, with $J\geq 2$. Each principal $j$ takes an action $y_{j}\in
Y_{j}$. Let $Y\equiv \times _{j\in \mathcal{J}}Y_{j}$. Principal $j$'s
utility function is $v_{j}:Y\times \Theta \rightarrow 
\mathbb{R}
$. The agent's utility function is $u:Y\times \Theta \rightarrow 
\mathbb{R}
$. We assume $\left\vert Y\times \Theta \right\vert <\infty $.

In order to incorporate the structure of delegation into a contract, we use $%
M_{j}^{\mathcal{A}}$ as the message set in principal $j$'s contract if
necessary, instead of generic notation $M_{j}$. For each $j\in \mathcal{J}$,
principal $j$'s contract is a function $c_{j}:M_{j}^{\mathcal{A}}\rightarrow 
\mathcal{A}_{j}$, where $M_{j}^{\mathcal{A}}$ is an infinite set of messages
that the agent can send to principal $j$ and 
\begin{gather*}
\mathcal{A}_{j}^{delegated}\equiv \left\{ \left\{ y_{j}\right\} :y_{j}\in
Y_{j}\right\} \text{, }\mathcal{A}_{j}^{non-delegated}\equiv
2^{Y_{j}}\diagdown \left\{ \varnothing \right\} \text{,} \\
\mathcal{A}_{j}\in \left\{ \mathcal{A}_{j}^{delegated}\text{, }\mathcal{A}%
_{j}^{non-delegated}\right\} \text{.}
\end{gather*}%
$M_{j}^{\mathcal{A}}$ can be very general and we do not impose any other
restriction on it. Let $M^{\mathcal{A}}\equiv \times _{k\in \mathcal{J}%
}M_{k}^{\mathcal{A}}$ and $M_{-j}^{\mathcal{A}}\equiv \times _{k\in \mathcal{%
J\diagdown }\left\{ j\right\} }M_{k}^{\mathcal{A}}$.\footnote{%
We use $M^{\mathcal{A}}$ as a generic element in $\left\{ M^{\mathcal{A}%
^{delegated}},\text{ }M^{\mathcal{A}^{non-delegated}}\right\} $. The
superscript $\mathcal{A}$ is used simply to distinguish it from $M^{R}$ and $%
M^{F}$, which will be defined later.} If $\mathcal{A}_{j}=\mathcal{A}%
_{j}^{delegated}$, principal $j$ delegates her\textbf{\ }action to the
agent, i.e., the agent's message fully determines $j$'s action. If $\mathcal{%
A}_{j}=\mathcal{A}_{j}^{non-delegated}$, principal $j$ has to strategically
choose an action in $c_{j}(m_{j})\in 2^{Y_{j}}\diagdown \left\{ \varnothing
\right\} $ after receiving a message $m_{j}$. Define $\mathcal{A}\equiv
\times _{k\in \mathcal{J}}\mathcal{A}_{k}$ and $\mathcal{A}_{-j}\equiv
\times _{k\in \mathcal{J\diagdown }\left\{ j\right\} }\mathcal{A}_{k}$. Let $%
C_{j}^{\mathcal{A}}\equiv \left( \mathcal{A}_{j}\right) ^{M_{j}^{\mathcal{A}%
}}$ be the set of all possible contracts available to principal $j$, $C^{%
\mathcal{A}}\equiv \times _{k\in \mathcal{J}}C_{k}^{\mathcal{A}}$ and $%
C_{-j}^{\mathcal{A}}\equiv \times _{k\in \mathcal{J\diagdown }\left\{
j\right\} }C_{k}^{\mathcal{A}}$.

\subsection{A generic game}

\label{sec:model:game}

We will consider different contract spaces (and message spaces). Principals
and the agent will play a generic game under different contract spaces,
which is defined as follows.

For each $j\in \mathcal{J}$, let $M_{j}$ be a generic set of messages that
the agent may send to principal $j$, and a contract of principal $j$ is $%
c_{j}:M_{j}\rightarrow \mathcal{A}_{j}$. Thus, $C_{j}\equiv \left( \mathcal{A%
}_{j}\right) ^{M_{j}}$ is a generic contract space for principal $j$, which
may represent different contract spaces (e.g., $C_{j}^{P}$, $C_{j}^{R}$, $%
C_{j}^{F}$ defined later), besides $C_{j}^{\mathcal{A}}$ in Section \ref%
{sec:model:primitive}. Denote $C\equiv \times _{k\in \mathcal{J}}C_{k}$ and $%
M\equiv \times _{k\in \mathcal{J}}M_{k}$.

Throughout this subsection, we fix a generic profile $\left( M,C\right) $ to
define a generic game.

\subsubsection{Models and timeline}

\label{sec:model:game:timeline}

When principal $j$ chooses his action in the set $c_{j}\left( m_{j}\right) $%
, his decision depends not only on the message he\textbf{\ }receives (i.e., $%
m_{j}$), but also on what he\textbf{\ }observes regarding the contracts
offered by the other principals and the messages that the agent sends to all
principals.

Let $\Gamma \equiv \left[ \Gamma _{k}:C\rightarrow 2^{C}\right] _{k\in 
\mathcal{J}}$ denote a potential announcement structure. We focus on two
structures: (1) public announcement (denoted by $\Gamma ^{public}=\left(
\Gamma _{k}^{public}\right) _{k\in \mathcal{J}}$) with%
\begin{equation}
\Gamma _{k}^{public}\left( c_{k},c_{-k}\right) =\left\{ \left(
c_{k},c_{-k}\right) \right\} \text{, }\forall k\in \mathcal{J}\text{, }%
\forall \left( c_{k},c_{-k}\right) \in C\text{,}  \label{pub_ann}
\end{equation}%
and (2) private announcement (denoted by $\Gamma ^{private}=\left( \Gamma
_{k}^{private}\right) _{k\in \mathcal{J}}$) with%
\begin{equation}
\Gamma _{k}^{private}\left( c_{k},c_{-k}\right) =\left\{ c_{k}\right\}
\times C_{-k}\text{, }\forall k\in \mathcal{J}\text{, }\forall \left(
c_{k},c_{-k}\right) \in C\text{.}  \label{pri_ann}
\end{equation}%
I.e., principals observe all of the contracts offered under public
announcement, whereas, under private announcement, each principal observes
only his own contract.

Similarly, let $\Psi \equiv \left[ \Psi _{k}:M\rightarrow 2^{M}\right]
_{k\in \mathcal{J}}$ denote a communication structure. We focus on two
structures: (1) public communication (denoted by $\Psi ^{public}=\left( \Psi
_{k}^{public}\right) _{k\in \mathcal{J}}$) with%
\begin{equation}
\Psi _{k}^{public}\left( m_{k},m_{-k}\right) =\left\{ \left(
m_{k},m_{-k}\right) \right\} \text{, }\forall k\in \mathcal{J}\text{, }%
\forall \left( m_{k},m_{-k}\right) \in M\text{,}  \label{pub_com}
\end{equation}%
and (2) private communication (denoted by $\Psi ^{private}=\left( \Psi
_{k}^{private}\right) _{k\in \mathcal{J}}$) with%
\begin{equation}
\Psi _{k}^{private}\left( m_{k},m_{-k}\right) =\left\{ m_{k}\right\} \times
M_{-k}\text{, }\forall k\in \mathcal{J}\text{, }\forall \left(
m_{k},m_{-k}\right) \in M\text{.}  \label{pri_com}
\end{equation}%
I.e., under public communication, all principals observe all messages,
whereas, under private communication, each principal observes only the
message he receives.

Thus, a model is fully characterized by a tuple $\left\langle \mathcal{A}%
,\Gamma ,\Psi \right\rangle $, where 
\begin{equation*}
\mathcal{A}\in \left\{ \mathcal{A}^{delegated}\text{, }\mathcal{A}%
^{non-delegated}\right\} ,\text{ }\Gamma \in \left\{ \Gamma ^{private},\text{
}\Gamma ^{public}\right\} ,\text{ }\Psi \in \left\{ \Psi ^{private},\text{ }%
\Psi ^{public}\right\} \text{.}
\end{equation*}

Given a model $\left\langle \mathcal{A},\text{ }\Gamma ,\text{ }\Psi
\right\rangle $, the game proceeds according to the following timeline.

\begin{enumerate}
\item Before the game starts, Nature draws the agent's type according to the
common prior $p\in \Delta (\Theta )$ and the realized type is the agent's
private information;

\item At Stage 1, each principal $j\in \mathcal{J}$ simultaneously offers a
contract $c_{j}\in C_{j}$ to the agent. The agent observes $c\equiv
(c_{1},\ldots ,c_{J})$, whereas each principal $j\in \mathcal{J}$ knows that
only a contract profile in $\Gamma _{j}\left( c\right) $ is possibly chosen
by the principals.

\item At Stage 2, the agent sends messages $m\equiv (m_{1},\ldots ,m_{J})\in
M$, one for each principal. Each principal $j\in \mathcal{J}$ knows that
only a message profile in $\Psi _{j}\left( m\right) $ is possibly sent by
the agent to the principals;

\item At Stage 3, each principal $j\in \mathcal{J}$ simultaneously chooses
an action in $c_{j}(m_{j})$;

\item Finally, payoffs are realized.
\end{enumerate}

In the model of \cite{mp2001}, the agent also chooses an effort. For
simplicity, we choose not to include the agent's effort in our model.
Another reason for this modeling choice is that the menu theorem already
fails in such a simple model with non-delegated contracts. In Appendix \ref%
{sec:agent_effort}, we show how our results can be extended to a model with
the agent's effort.

\subsubsection{Strategies}

At Stage 1, each principal $j$ chooses $c_{j}\in C_{j}$. At Stage 2, the
agent chooses a function $s\equiv \lbrack s_{k}:C\times \Theta \rightarrow
M_{k}]_{k\in \mathcal{J}}$. Let $S_{k}$ be the set of all possible $%
s_{k}:C\times \Theta \rightarrow M_{k}$. Denote $s\equiv \left( s_{1},\ldots
,s_{J}\right) \in S\equiv \times _{k\in \mathcal{J}}S_{k}$ and $s_{-j}\in
S_{-j}\equiv \times _{k\in \mathcal{J\diagdown }\left\{ j\right\} }S_{k}$.
Denote $s\left( c,\theta \right) =\left( s_{1}\left( c,\theta \right)
,\ldots ,s_{J}\left( c,\theta \right) \right) $ for each $\left( c,\theta
\right) \in C\times \Theta $. At Stage 3, each principal $j$'s chooses a
function $t_{j}:\Gamma _{j}\left( C\right) \times \Psi _{j}\left( M\right)
\rightarrow Y_{j}$ such that%
\begin{equation*}
t_{j}\left[ \Gamma _{j}\left( c_{j},c_{-j}\right) ,\Psi _{j}\left(
m_{j},m_{-j}\right) \right] \in c_{j}(m_{j})\text{, }\forall \left[ \left(
c_{j},c_{-j}\right) ,\left( m_{j},m_{-j}\right) \right] \in C\times M\text{,}
\end{equation*}%
and let $T_{j}$ be the set of all such functions. Denote $T\equiv \left(
T_{k}\right) _{k\in \mathcal{J}}$ and $T_{-j}\equiv \left( T_{k}\right)
_{_{k\in \mathcal{J\diagdown }\left\{ j\right\} }}$.

Given any $(c,s,t)\in C\times S\times T$, the utility for the agent of type $%
\theta $ is 
\begin{equation*}
U\left( c,s,t,\theta \right) \equiv u\left( t_{1}\left[ \Gamma _{1}\left(
c\right) ,\text{ }\Psi _{1}\left( s\left( c,\theta \right) \right) \right]
,\ldots ,t_{J}\left[ \Gamma _{J}\left( c\right) ,\text{ }\Psi _{J}\left(
s\left( c,\theta \right) \right) \right] ,\text{ }\theta \right) \text{,}
\end{equation*}%
and principal $j$'s expected utility is 
\begin{equation*}
V_{j}(c,s,t)\equiv \sum_{\theta \in \Theta }p\left( \theta \right) \times
v_{j}\left( t_{1}\left[ \Gamma _{1}\left( c\right) ,\text{ }\Psi _{1}\left(
s\left( c,\theta \right) \right) \right] ,\ldots ,t_{J}\left[ \Gamma
_{J}\left( c\right) ,\text{ }\Psi _{J}\left( s\left( c,\theta \right)
\right) \right] ,\text{ }\theta \right) \text{.}
\end{equation*}

\subsubsection{Legitimate beliefs}

At Stage 3, principal $j$ must form a belief on $\left( C\times M\times
\Theta \right) $ conditional on $\Gamma _{j}\left( c\right) $ she\textbf{\ }%
observes at Stage 1 and $\Psi _{j}\left( m\right) $ she\textbf{\ }observes
at Stage 2. It is described by a function $b_{j}:\Gamma _{j}\left( C\right)
\times \Psi _{j}\left( M\right) \rightarrow \Delta \left( C\times M\times
\Theta \right) $. Given belief $b_{j}$, principal $j$'s expected utility
conditional on $\left( \alpha _{j},\beta _{j}\right) \in \Gamma _{j}\left(
C\right) \times \Psi _{j}\left( M\right) $ is%
\begin{equation*}
V_{j}\left( t_{j},t_{-j}|\alpha _{j},\beta _{j},b_{j}\right) \equiv
\dint\limits_{b_{j}\left( \alpha _{j},\beta _{j}\right) }v_{j}\left( t_{1} 
\left[ \Gamma _{1}\left( c\right) ,\text{ }\Psi _{1}\left( m\right) \right]
,\ldots ,t_{J}\left[ \Gamma _{J}\left( c\right) ,\text{ }\Psi _{J}\left(
m\right) \right] ,\text{ }\theta \right) d\left( c,m,\theta \right) \text{.}
\end{equation*}

For each principal $j$'s belief, we apply Bayes' rule if and only if she%
\textbf{\ }cannot confirm that the other players have deviated from an
equilibrium. Given $(c,s)$ being played in an equilibrium, define $\mathcal{B%
}_{j}^{\left( c,\text{ }s\right) }$ as the set of principal $j$' valid
beliefs. For every $j\in \mathcal{J}$,%
\begin{equation}
\mathcal{B}_{j}^{\left( c,\text{ }s\right) }\equiv \left\{ 
\begin{array}{c}
b_{j}:\Gamma _{j}\left( C\right) \times \Psi _{j}\left( M\right) \rightarrow
\Delta \left( C\times M\times \Theta \right) \text{ such that} \\ 
b_{j}\left[ \Gamma _{j}\left( c^{\prime }\right) ,\Psi _{j}\left( m\right) %
\right] \left( \Gamma _{j}\left( c^{\prime }\right) \times \Psi _{j}\left(
m\right) \times \Theta \right) =1\text{, }\forall \left( c^{\prime
},m\right) \in C\times M\text{,} \\ 
\\ 
\text{and }\left( 
\begin{array}{c}
\forall c_{j}^{\prime }\in C_{j},\text{ }\forall \theta \in \Theta \text{, }
\\ 
\text{set }\beta _{j}=\Psi _{j}\left( s\left( c_{j}^{\prime },c_{-j},\theta
\right) \right) \text{, and we have} \\ 
b_{j}\left[ \Gamma _{j}\left( c_{j}^{\prime },c_{-j}\right) ,\text{ }\beta
_{j}\right] \left( c_{j}^{\prime },c_{-j},\text{ }s\left( c_{j}^{\prime
},c_{-j},\theta \right) ,\text{ }\theta \right) =\frac{p\left( \theta
\right) }{\dsum\limits_{\theta ^{\prime \prime }\in \left\{ \theta ^{\prime
}\in \Theta :\text{ }\beta _{j}=\Psi _{j}\left( s\left( c_{j}^{\prime
},c_{-j},\theta ^{\prime }\right) \right) \right\} }p\left( \theta ^{\prime
\prime }\right) }%
\end{array}%
\right)%
\end{array}%
\right\} \text{.}  \label{belief}
\end{equation}%
First, 
\begin{equation}
b_{j}\left[ \Gamma _{j}\left( c^{\prime }\right) ,\Psi _{j}\left( m\right) %
\right] \left( \Gamma _{j}\left( c^{\prime }\right) \times \Psi _{j}\left(
m\right) \times \Theta \right) =1\text{, }\forall \left( c^{\prime
},m\right) \in C\times M  \label{perfect-recall}
\end{equation}%
in (\ref{belief}) is the classic perfect recall condition. For instance,
given public announcement, if principal $j$ observes $c^{\prime }\in C$ at
Stage 1, $j$ must believe in $c^{\prime }$ with probability 1 at Stage 3
(even on off-equilibrium paths).\footnote{%
If we do not impose the perfect recall condition, our results and proofs
remain true.}

Second, we will adopt the solution concept of (weak) Perfect Bayesian
equilibrium, and hence, players will use Bayes' rule to update their beliefs
whenever possible. As usual, when one principal deviates unilaterally, she%
\textbf{\ }assumes that the other players follows the equilibrium strategy
profile. This is rigorously described in the set $\mathcal{B}_{j}^{\left( c,%
\text{ }s\right) }$ in (\ref{belief}). Specifically, $\mathcal{B}%
_{j}^{\left( c,\text{ }s\right) }$ contains any belief function $b_{j}$
which satisfies the following condition. Given $(c,s)$ being played in an
equilibrium, suppose principal $j$ unilaterally deviates to $c_{j}^{\prime
}\in C_{j}$. If $j$ observes $\Gamma _{j}\left( c_{j}^{\prime
},c_{-j}\right) $ (i.e., $j$ cannot confirm that principals $-j$ have
deviated from $(c,s)$) and observes $\beta _{j}=\Psi _{j}\left( s\left(
c_{j}^{\prime },c_{-j},\theta \right) \right) $ for some $\theta \in \Theta $
(i.e., $j$ cannot confirm that the agent has deviated from $(c,s)$),
principal $j$ believes that principals $-j$ have offered $c_{-j}$, and the
agent has followed $s$. As a result, the following set contains all possible
states,%
\begin{equation*}
\left\{ \theta ^{\prime }\in \Theta :\text{ }\Psi _{j}\left( s\left(
c_{j}^{\prime },c_{-j},\theta \right) \right) =\beta _{j}=\Psi _{j}\left(
s\left( c_{j}^{\prime },c_{-j},\theta ^{\prime }\right) \right) \right\} 
\text{,}
\end{equation*}%
and by Bayes' rule, $j$'s updated belief is%
\begin{equation*}
b_{j}\left[ \Gamma _{j}\left( c_{j}^{\prime },c_{-j}\right) ,\text{ }\beta
_{j}\right] \left( c_{j}^{\prime },c_{-j},\text{ }s\left( c_{j}^{\prime
},c_{-j},\theta \right) ,\text{ }\theta \right) =\frac{p\left( \theta
\right) }{\dsum\limits_{\theta ^{\prime \prime }\in \left\{ \theta ^{\prime
}\in \Theta :\text{ }\beta _{j}=\Psi _{j}\left( s\left( c_{j}^{\prime
},c_{-j},\theta ^{\prime }\right) \right) \right\} }p\left( \theta ^{\prime
\prime }\right) }\text{.}
\end{equation*}

If principal $j$ can confirm that either principals $-j$ or the agent have
deviated from $(c,s)$, we impose no requirement on $b_{j}\in \mathcal{B}%
_{j}^{\left( c,\text{ }s\right) }$ (besides the perfect recall condition),
because this happens with probability $0$ in an equilibrium, and Bayes rule
does not apply.

\subsubsection{Perfect Bayesian equilibrium}

For simplicity, we adopt the solution concept of pure-strategy perfect
Bayesian equilibrium and just call it an equilibrium. Our results can be
extended to mixed-strategy equilibria (See Appendix \ref%
{sec:mixed_strategy_eq}).

\begin{define}
\label{def:pbe}Given $\left( C,M\right) $ in a model $\left\langle \mathcal{A%
},\text{ }\Gamma ,\text{ }\Psi \right\rangle $, $\left( c,s,t\right) \in
C\times S\times T$ is a $C$-equilibrium if there exists $\left( b_{k}\right)
_{k\in \mathcal{J}}\in \times _{k\in \mathcal{J}}\mathcal{B}_{k}^{\left( c,%
\text{ }s\right) }$ such that (i) for every $j\in \mathcal{J}$,%
\begin{equation*}
V_{j}\left( \left( c_{j},c_{-j}\right) ,s,t\right) \geq V_{j}\left( \left(
c_{j}^{\prime },c_{-j}\right) ,s,t\right) \text{, }\forall c_{j}^{\prime
}\in C_{j}\text{,}
\end{equation*}%
and (ii) for every $\theta \in \Theta $,%
\begin{equation*}
U\left( c^{\prime },s,t,\theta \right) \geq U\left( c^{\prime },s^{\prime
},t,\theta \right) \text{, }\forall j\in \mathcal{J}\text{, }\forall \left(
c^{\prime },s^{\prime }\right) \in C\times S\text{,}
\end{equation*}%
and (iii) for every $j\in \mathcal{J}$,%
\begin{equation*}
V_{j}\left( t_{j},t_{-j}|\alpha _{j},\beta _{j},b_{j}\right) \geq
V_{j}\left( t_{j}^{\prime },t_{-j}|\alpha _{j},\beta _{j},b_{j}\right) \text{%
, }\forall t_{j}^{\prime }\in T_{j}\text{, }\forall \left( \alpha _{j},\beta
_{j}\right) \in \Gamma _{j}\left( C\right) \times \Psi _{j}\left( M\right) 
\text{.}
\end{equation*}%
Let $\mathcal{E}^{\left\langle \mathcal{A},\text{ }\Gamma ,\text{ }\Psi
\right\rangle \text{-}C}$ denote the set of $C$-equilibria in the model $%
\left\langle \mathcal{A},\text{ }\Gamma ,\text{ }\Psi \right\rangle $.
\end{define}

\subsubsection{Allocation}

Let $Z\equiv \times _{k\in \mathcal{J}}\left[ Z_{k}:\Theta \rightarrow Y_{k}%
\right] $ denote the set of allocations. Given any $\left( c,s,t\right) \in
C\times S\times T$, define $z^{\left( c,s,t\right) }:\Theta \rightarrow Y$ as%
\begin{equation*}
z^{\left( c,s,t\right) }\left( \theta \right) =\left[ z_{k}^{\left(
c,s,t\right) }\left( \theta \right) \right] _{k\in \mathcal{J}}=\left( t_{k}%
\left[ \Gamma _{k}\left( c\right) ,\text{ }\Psi _{k}\left( s\left( c,\theta
\right) \right) \right] \right) _{k\in \mathcal{J}}\text{, }\forall \theta
\in \Theta \text{,}
\end{equation*}%
i.e., $z^{\left( c,s,t\right) }$ is the allocation induced by $\left(
c,s,t\right) $. Define%
\begin{equation}
Z^{\mathcal{E}^{\left\langle \mathcal{A},\text{ }\Gamma ,\text{ }\Psi
\right\rangle \text{-}C}}\equiv \left\{ z^{\left( c,s,t\right) }\in Z:\left(
c,s,t\right) \in \mathcal{E}^{\left\langle \mathcal{A},\text{ }\Gamma ,\text{
}\Psi \right\rangle \text{-}C}\right\} \text{,}  \label{eq_all}
\end{equation}%
i.e., $Z^{\mathcal{E}^{\left\langle \mathcal{A},\text{ }\Gamma ,\text{ }\Psi
\right\rangle \text{-}C}}$ is the set of all $C$-equilibrium allocations in
the model $\left\langle \mathcal{A},\text{ }\Gamma ,\text{ }\Psi
\right\rangle $.

\subsection{Goals\label{sec:goals}}

The primitive contract space is $C^{\mathcal{A}}$ in Section \ref%
{sec:model:primitive}, and our goal is a simple full characterization of $Z^{%
\mathcal{E}^{\left\langle \mathcal{A},\text{ }\Gamma ,\text{ }\Psi
\right\rangle \text{-}C^{\mathcal{A}}}}$. Given $\mathcal{A}=\mathcal{A}%
^{delegated}$, the menu theorem in \cite{mp2001} has achieved this goal. A
menu contract introduced in \cite{mp2001} is a function, $%
c_{j}:E_{j}\rightarrow Y_{j}$, such that $E_{j}\in 2^{Y_{j}}\diagdown
\left\{ \varnothing \right\} $ and $c_{j}\left( y_{j}\right) =y_{j}$ for all 
$y_{j}\in E_{j}$. Let $C_{j}^{P}$ denote the set of all menu contracts for
principal $j$, $C^{P}\equiv \times _{k\in \mathcal{J}}C_{k}^{P}$, and $%
C_{-j}^{P}\equiv \times _{k\in \mathcal{J\diagdown }\left\{ j\right\}
}C_{k}^{P}$. Let $M_{j}^{P}\equiv Y_{j}$ denote the set of messages used in
all possible menus, $M^{P}\equiv \times _{k\in \mathcal{J}}M_{k}^{P}$, and $%
M_{-j}^{P}\equiv \times _{k\in \mathcal{J\diagdown }\left\{ j\right\}
}M_{k}^{P}$.

\begin{theo}[The menu Theorem, \protect\cite{mp2001}]
\label{thm:peters}We have%
\begin{equation}
Z^{\mathcal{E}^{\left\langle \mathcal{A}^{delegated},\text{ }\Gamma ,\text{ }%
\Psi \right\rangle \text{-}C^{\mathcal{A}^{delegated}}}}=Z^{\mathcal{E}%
^{\left\langle \mathcal{A}^{delegated},\text{ }\Gamma ,\text{ }\Psi
\right\rangle \text{-}C^{P}}}\text{, }\forall \left\langle \Gamma ,\text{ }%
\Psi \right\rangle \in \left\{ \Gamma ^{private},\Gamma ^{public}\right\}
\times \left\{ \Psi ^{private},\Psi ^{public}\right\} \text{.}
\label{menu_equality}
\end{equation}
\end{theo}

With delegated contracts (i.e., $\mathcal{A}=\mathcal{A}^{delegated}$), the
announcement and communication structures do not have impact on equilibria.%
\footnote{%
Different announcement and communication structures lead to different
information for principals \emph{only after they offer their contracts}.
With delegated contracts, principals do not make any strategic decision
after offering their contracts, and hence, the information (induced by
different announcement and communication structures) is irrelevant.} The
implication of Theorem \ref{thm:peters} is that, given delegated contracts,
it suffers no loss of generality for principals to offer menus both on and
off the equilibrium paths. Since $C^{P}$ is a much simpler set than $C^{%
\mathcal{A}}$, this result substantially simplifies the characterization of
equilibrium allocations.

Section \ref{sec:examples} provides two examples where the common-agency
models $\left\langle \mathcal{A}^{non-delegated},\Gamma ^{public},\text{ }%
\Psi ^{public}\right\rangle $ and $\left\langle \mathcal{A}%
^{non-delegated},\Gamma ^{public},\text{ }\Psi ^{private}\right\rangle $
generate equilibrium allocations that cannot be generated by any delegated
contracts (including all menu contracts). The example in Section \ref%
{sec:private-public:counterexample} will show this point in the model $%
\left\langle \mathcal{A}^{non-delegated},\Gamma ^{private},\text{ }\Psi
^{public}\right\rangle $ as well.\footnote{%
The example in Section \ref{sec:private-public:counterexample} shows a
stronger point.} Thus, given $\mathcal{A}=\mathcal{A}^{non-delegated}$, we
aim to identify canonical contracts in each model $\left\langle \mathcal{A},%
\text{ }\Gamma ,\text{ }\Psi \right\rangle $ that lead to a simple full
characterization of $Z^{\mathcal{E}^{\left\langle \mathcal{A},\text{ }\Gamma
,\text{ }\Psi \right\rangle \text{-}C^{\mathcal{A}}}}$.

\section{Simpler contract spaces than $C^{\mathcal{A}}$}

\label{sec:contract}

For a model $\left\langle \mathcal{A}^{non-delegated},\text{ }\Gamma ,\text{ 
}\Psi \right\rangle $, one may consider a menu-of-menu contract as a
canonical contract that principals may use. This is a contract that includes
subsets of actions and let the agent choose one of them, and after a subset
of actions is chosen, the principal subsequently chooses his own action from
it.

However, a menu-of-menu contract is not general enough to generate all
equilibrium allocations. For instance, in an equilibrium under general
non-delegated contracts, suppose types $\theta $ and $\theta ^{\prime }$ of
the agent send two distinct messages $m_{j}$ and $m_{j}^{\prime }$ to
principal $j$ at Stage 2, which pin down the same subset (menu) $E_{j}\in
2^{Y_{j}}\diagdown \left\{ \varnothing \right\} $, while principal $j$ takes
distinct actions $y_{j}\in E_{j}$ and $y_{j}^{\prime }\in E_{j}$ at Stage 3,
upon observing distinct messages $m_{j}$ and $m_{j}^{\prime }$,
respectively. If we replicate this equilibrium by a menu-of-menu contract,
types $\theta $ and $\theta ^{\prime }$ must choose the same $E_{j}\in
2^{Y_{j}}\diagdown \left\{ \varnothing \right\} $ at Stage 2, and as a
result, principal $j$ must take the same action at Stage 3, upon observing
the same message $E_{j}$ from the agent.

We will define two classes of contracts which suffer no loss of generality
in $\left\langle \mathcal{A}^{non-delegated},\text{ }\Gamma ,\text{ }\Psi
\right\rangle $. Before introducing these two classes in Sections \ref%
{sec:contract:menu-of-menu} and \ref{sec:contract:menu-of-menu:full}, we
rigorously define the goal of our generalized menu theorem in \ref%
{sec:contract:goal}.

\subsection{$\left[ C^{I},C^{II}\right] $-equilibrium and the goal}

\label{sec:contract:goal}

Given $\mathcal{A=A}^{non-delegated}$, we follow the same strategy of \cite%
{mp2001} to characterize equilibrium allocations. That is, we will identify
two simple contract spaces, $C^{I}$ and $C^{II}$, and prove that it suffers
no loss of generality for principals to focus on $C^{I}$ and $C^{II}$, one
for the equilibrium path and the other for off-equilibrium paths. Thus, we
first define a new notion of $\left[ C^{I},C^{II}\right] $-equilibrium as
follows.

\begin{define}
\label{def:pbe:on-off}Given two generic contract spaces, $C^{I}$ and $C^{II}$%
, in a model $\left\langle \mathcal{A},\text{ }\Gamma ,\text{ }\Psi
\right\rangle $, $\left( c,s,t\right) \in C\times S\times T$ is a $\left[
C^{I},C^{II}\right] $-equilibrium if $c\in C^{I}$ and $\left( c,s,t\right) $
is a $\widehat{C}$-equilibrium, where%
\begin{equation*}
\widehat{C}\equiv \times _{k\in \mathcal{J}}\widehat{C}_{k}\equiv \times
_{k\in \mathcal{J}}\left( \left\{ c_{k}\right\} \cup C_{k}^{II}\right) \text{%
.}
\end{equation*}
\end{define}

Given $\left\langle \mathcal{A},\text{ }\Gamma ,\text{ }\Psi \right\rangle $%
, let $\mathcal{E}^{\left\langle \mathcal{A},\text{ }\Gamma ,\text{ }\Psi
\right\rangle \text{-}\left[ C^{I},C^{II}\right] }$ denote the set of all $%
\left[ C^{I},C^{II}\right] $-equilibria, and%
\begin{equation*}
Z^{\mathcal{E}^{\left\langle \mathcal{A},\text{ }\Gamma ,\text{ }\Psi
\right\rangle \text{-}\left[ C^{I},C^{II}\right] }}\equiv \left\{ z^{\left(
c,s,t\right) }\in Z:\left( c,s,t\right) \in \mathcal{E}^{\left\langle 
\mathcal{A},\text{ }\Gamma ,\text{ }\Psi \right\rangle \text{-}\left[
C^{I},C^{II}\right] }\right\} \text{.}
\end{equation*}%
Clearly, $\mathcal{E}^{\left\langle \mathcal{A},\text{ }\Gamma ,\text{ }\Psi
\right\rangle \text{-}C^{\mathcal{A}}}\equiv \mathcal{E}^{\left\langle 
\mathcal{A},\text{ }\Gamma ,\text{ }\Psi \right\rangle \text{-}\left[ C^{%
\mathcal{A}},C^{\mathcal{A}}\right] }$. Thus, the classical menu theorem
(i.e., Theorem \ref{thm:peters}) is that, for any $\left\langle \Gamma ,%
\text{ }\Psi \right\rangle \in \left\{ \Gamma ^{private},\Gamma
^{public}\right\} \times \left\{ \Psi ^{private},\Psi ^{public}\right\} $,
we have 
\begin{equation*}
Z^{\mathcal{E}^{\left\langle \mathcal{A}^{delegated},\text{ }\Gamma ,\text{ }%
\Psi \right\rangle \text{-}C^{\mathcal{A}^{delegated}}}}=Z^{\mathcal{E}%
^{\left\langle \mathcal{A}^{delegated},\text{ }\Gamma ,\text{ }\Psi
\right\rangle \text{-}\left[ C^{\mathcal{A}^{delegated}},C^{\mathcal{A}%
^{delegated}}\right] }}=Z^{\mathcal{E}^{\left\langle \mathcal{A}^{delegated},%
\text{ }\Gamma ,\text{ }\Psi \right\rangle \text{-}\left[ C^{P},C^{P}\right]
}}\text{.}
\end{equation*}%
For each model, we aim to find the simple contract spaces, $C^{I}$ and $%
C^{II}$, such that 
\begin{equation*}
Z^{\mathcal{E}^{\left\langle \mathcal{A}^{non-delegated},\text{ }\Gamma ,%
\text{ }\Psi \right\rangle \text{-}\left[ C^{\mathcal{A}^{non-delegated}},C^{%
\mathcal{A}^{non-delegated}}\right] }}=Z^{\mathcal{E}^{\left\langle \mathcal{%
A}^{non-delegated},\text{ }\Gamma ,\text{ }\Psi \right\rangle \text{-}\left[
C^{I},C^{II}\right] }}\text{.}
\end{equation*}%
We propose such simple contract spaces in Sections \ref%
{sec:contract:menu-of-menu} and \ref{sec:contract:menu-of-menu:full}.

\subsection{Menu-of-menu-with-recommendation contracts}

\label{sec:contract:menu-of-menu}

Given $j\in \mathcal{J}$, pick any $E_{j}\in 2^{Y_{j}}\diagdown \left\{
\varnothing \right\} $, and we say that $\left[ E_{j},y_{j}\right] $ is a
menu with a recommendation if and only if $y_{j}\in E_{j}$. Define%
\begin{equation*}
M_{j}^{R}\equiv \left\{ \left[ E_{j},y_{j}\right] :E_{j}\in
2^{Y_{j}}\diagdown \left\{ \varnothing \right\} \text{ and }y_{j}\in
E_{j}\right\} \text{,}
\end{equation*}%
i.e., $M_{j}^{R}$ is the set of all menus with a recommendation. Let $%
M^{R}\equiv \times _{k\in \mathcal{J}}M_{k}^{R}$.

\begin{define}
\label{def:menu_of_menu_with_recommendation}A
menu-of-menu-with-recommendation contract for principal $j$ is a function, $%
c_{j}:K_{j}\rightarrow 2^{Y_{j}}\diagdown \left\{ \varnothing \right\} $
such that $K_{j}\in 2^{M_{j}^{R}}\diagdown \left\{ \varnothing \right\} $
and $c_{j}\left( \left[ E_{j},y_{j}\right] \right) =E_{j}$, $\forall \left[
E_{j},y_{j}\right] \in K_{j}$.
\end{define}

When the agent sends a message $\left[ E_{j},y_{j}\right] \in K_{j}$, the
interpretation is that she chooses the menu of actions $E_{j}$ along with
recommending $y_{j}$ to principal $j$. Nonetheless, this recommendation is
not binding, and principal $j$ can still choose any action in $E_{j}$. For
example, suppose that $K_{j}=\left\{ \left[ \left\{ a,b\right\} ,a\right] ,%
\left[ \left\{ c,d\right\} ,c\right] ,\left[ \left\{ e,f,g\right\} ,e\right] 
\text{, }\left[ \left\{ e,f,g\right\} ,f\right] \right\} .$ Then, a
menu-of-menu-with-recommendation contract $c_{j}:K_{j}\rightarrow
2^{Y_{j}}\diagdown \left\{ \varnothing \right\} $ has the following property:%
\begin{gather*}
c_{j}\left( \left[ \left\{ a,b\right\} ,a\right] \right) =\left\{
a,b\right\} \text{, }c_{j}\left( \left[ \left\{ c,d\right\} ,c\right]
\right) =\left\{ c,d\right\} \text{,} \\
c_{j}\left( \left[ \left\{ e,f,g\right\} ,e\right] \right) =c_{j}\left[
\left\{ e,f,g\right\} ,f\right] =\left\{ e,f,g\right\} \text{.}
\end{gather*}%
Let $C_{j}^{R}$ be the set of all possible menu-of-menu-with-recommendation
contracts for principal $j$, $C^{R}\equiv \times _{k\in \mathcal{J}%
}C_{k}^{R} $ and $C_{-j}^{R}\equiv \times _{k\neq j}C_{k}^{R}$. Given
non-delegated contracts, we will show that it suffers no loss of generality
to focus on contracts in $C_{j}^{R}$ on the equilibrium path.

\subsection{Menu-of-menu-with-full-recommendation contracts}

\label{sec:contract:menu-of-menu:full}

We now define another class of contracts. For any $E_{j}\in
2^{Y_{j}}\diagdown \left\{ \varnothing \right\} $, a menu of $E_{j}$ with 
\emph{full recommendation} is $\left\{ \left[ E_{j},y_{j}\right] :y_{j}\in
E_{j}\right\} $. For example, with $E_{j}=\left\{ a,b,c\right\} $, a menu of 
$E_{j}$ with full recommendation is $\left\{ \left[ \left\{ a,b,c\right\} ,a%
\right] \text{, }\left[ \left\{ a,b,c\right\} ,b\right] \text{, }\left[
\left\{ a,b,c\right\} ,c\right] \right\} $.

\begin{define}
\label{def:menu_of_menu_with_full_recommendation}A
menu-of-menu-with-full-recommendation contract for principal $j$ is a
function, $c_{j}:L_{j}\cup H_{j}\rightarrow 2^{Y_{j}}\diagdown \left\{
\varnothing \right\} $ with%
\begin{gather*}
H_{j}=\left\{ \left[ E_{j},y_{j}\right] :y_{j}\in E_{j}\right\} \text{ for
some }E_{j}\in 2^{Y_{j}}\diagdown \left\{ \varnothing \right\} \text{,} \\
L_{j}\subset 2^{Y_{j}}\diagdown \left\{ E_{j}\right\} \text{,}
\end{gather*}%
such that%
\begin{gather*}
c_{j}\left( E_{j}^{\prime }\right) =E_{j}^{\prime }\text{, }\forall
E_{j}^{\prime }\in L_{j}\text{,} \\
c_{j}\left( \left[ E_{j},y_{j}\right] \right) =E_{j}\text{, }\forall \left[
E_{j},y_{j}\right] \in H_{j}\text{.}
\end{gather*}
\end{define}

Let us illustrate the domain of a menu-of-menu-with-full-recommendation
contract, $c_{j}:L_{j}\cup H_{j}\rightarrow 2^{Y_{j}}\diagdown \left\{
\varnothing \right\} $. The set $L_{j}$ is a (possibly empty) subset of $%
2^{Y_{j}}$, and $H_{j}=\left\{ \left[ E_{j},y_{j}\right] :y_{j}\in
E_{j}\right\} $ is a menu of $E_{j}$ with full recommendation such that $%
E_{j}$ is non-empty and $E_{j}\notin L_{j}$. For example, $L_{j}\cup
H_{j}=\left\{ \left\{ a,b\right\} \text{, }\left\{ c,d,e\right\} \text{, }%
\left[ \left\{ f,g,h\right\} ,f\right] \text{, }\left[ \left\{ f,g,h\right\}
,g\right] \text{, }\left[ \left\{ f,g,h\right\} ,h\right] \right\} $, i.e., 
\begin{eqnarray*}
c_{j}\left( \left\{ a,b\right\} \right) &=&\left\{ a,b\right\} \text{, }%
c_{j}\left( \left\{ c,d,e\right\} \right) =\left\{ c,d,e\right\} \text{,} \\
c_{j}\left( \left[ \left\{ f,g,h\right\} ,f\right] \right) &=&c_{j}\left( %
\left[ \left\{ f,g,h\right\} ,g\right] \right) =c_{j}\left( \left[ \left\{
f,g,h\right\} ,h\right] \right) =\left\{ f,g,h\right\} \text{.}
\end{eqnarray*}%
The interpretation of $c_{j}$ is: principal $j$ asks the agent to choose a
subset in the menu of menus $\left\{ \left\{ a,b\right\} \text{, }\left\{
c,d,e\right\} \text{, }\left\{ f,g,h\right\} \right\} $; the agent may
choose $\left\{ a,b\right\} $, or $\left\{ c,d,e\right\} $, or $\left\{
f,g,h\right\} $; if and only if the agent chooses $\left\{ f,g,h\right\} $,
the agent must, in addition, recommend an action in $\left\{ f,g,h\right\} $%
, i.e., $f$ or $g$ or $h$. Nevertheless, the recommendation is not binding.

Let $C_{j}^{F}$ denote the set of all menu-of-menu-with-full-recommendation
contracts for principal $j$, $C^{F}\equiv \times _{k\in \mathcal{J}%
}C_{k}^{F} $, and $C_{-j}^{F}\equiv \times _{k\in \mathcal{J\diagdown }%
\left\{ j\right\} }C_{k}^{F}$. Let $M_{j}^{F}$ denote the set of all
messages that could possibly be included in the domain of a
menu-of-menu-with-full-recommendation contracts for principal $j$, i.e., $%
M_{j}^{F}=\left( 2^{Y_{j}}\diagdown \left\{ \varnothing \right\} \right)
\cup M_{j}^{R}$. Let $M^{F}\equiv \times _{k\in \mathcal{J}}M_{k}^{F}$, and $%
M_{-j}^{F}\equiv \times _{k\in \mathcal{J\diagdown }\left\{ j\right\}
}M_{k}^{F}$. We will exploit contracts in $C^{F}$ to see under what
announcement and communication structures those contracts are general enough
when principals consider unilateral deviations.

Any menu contract (as defined in Section \ref{sec:goals}), e.g., $\{a,b,c\}$%
, can be viewed as a menu-of-menu-with-full-recommendation contract because
we can set $L_{j}=\{\left\{ a\right\} ,\left\{ b\right\} \}$ and $%
H_{j}=\left\{ \left[ \{c\},c\right] \right\} $. Therefore, $C^{P}$ can be
viewed as a strict subset of $C^{F}$. Furthermore, given any
menu-of-menu-with-full-recommendation contract, $c_{j}:L_{j}\cup
H_{j}\rightarrow 2^{Y_{j}}\diagdown \left\{ \varnothing \right\} $, since a
recommendation is non-binding, for each $D_{j}\in L_{j}$, we can add an
arbitrary recommendation $y_{j}\in D_{j}$. Thus, any
menu-of-menu-with-full-recommendation contract can be viewed as a
menu-of-menu-with-recommendation contract, i.e., 
\begin{equation}
C^{P}\subsetneqq C^{F}\subsetneqq C^{R}\text{.}  \label{ktt1}
\end{equation}

\section{Full equilibrium characterization I}

\label{sec:resulst}

Given $\mathcal{A}=\mathcal{A}^{non-delegated}$, we focus on three models in
this section: $\left\langle \Gamma ^{private},\text{ }\Psi
^{private}\right\rangle ,$ $\left\langle \text{ }\Gamma ^{public},\text{ }%
\Psi ^{private}\right\rangle $ and $\left\langle \text{ }\Gamma ^{public},%
\text{ }\Psi ^{public}\right\rangle $. We consider the model of $%
\left\langle \Gamma ^{private},\text{ }\Psi ^{public}\right\rangle $ in the
next section because it requires a different full characterization.

In the three models, there is no loss of generality to focus on $\left[
C^{R},C^{F}\right] $-equilibria.

\begin{theo}
\label{thm:three}Suppose $\mathcal{A}=\mathcal{A}^{non-delegated}$. We have%
\begin{gather}
Z^{\mathcal{E}^{\left\langle \mathcal{A},\text{ }\Gamma ,\text{ }\Psi
\right\rangle \text{-}\left[ C^{\mathcal{A}},C^{\mathcal{A}}\right] }}=Z^{%
\mathcal{E}^{\left\langle \mathcal{A},\text{ }\Gamma ,\text{ }\Psi
\right\rangle \text{-}\left[ C^{R},C^{F}\right] }}\text{,}  \label{tth2} \\
\forall \left\langle \Gamma ,\text{ }\Psi \right\rangle \in \left\{
\left\langle \Gamma ^{private},\Psi ^{private}\right\rangle ,\text{ }%
\left\langle \Gamma ^{public},\Psi ^{private}\right\rangle ,\text{ }%
\left\langle \Gamma ^{public},\Psi ^{public}\right\rangle \right\} \text{.} 
\notag
\end{gather}
\end{theo}

We need the following two technical propositions to prove Theorem \ref%
{thm:three}, and their proofs are relegated to Section \ref%
{sec:resulst:illustrate:extension} and Appendix \ref{sec:prop:deviation}.

\begin{prop}
\label{prop:other:on}Suppose $\mathcal{A}=\mathcal{A}^{non-delegated}$. We
have 
\begin{gather*}
Z^{\mathcal{E}^{\left\langle \mathcal{A},\text{ }\Gamma ,\text{ }\Psi
\right\rangle \text{-}\left[ C^{\mathcal{A}},C^{\mathcal{A}}\right] }}=Z^{%
\mathcal{E}^{\left\langle \mathcal{A},\text{ }\Gamma ,\text{ }\Psi
\right\rangle \text{-}\left[ C^{R},C^{\mathcal{A}}\right] }}\text{,} \\
\forall \left\langle \Gamma ,\text{ }\Psi \right\rangle \in \left\{
\left\langle \Gamma ^{private},\Psi ^{private}\right\rangle ,\text{ }%
\left\langle \Gamma ^{public},\Psi ^{private}\right\rangle ,\text{ }%
\left\langle \Gamma ^{public},\Psi ^{public}\right\rangle \right\} \text{.}
\end{gather*}
\end{prop}

\begin{prop}
\label{prop:deviation}Suppose $\mathcal{A}=\mathcal{A}^{non-delegated}$. Fix
any $I\in \left\{ \mathcal{A},P,F,R\right\} $. We have%
\begin{gather*}
Z^{\mathcal{E}^{\left\langle \mathcal{A},\text{ }\Gamma ,\text{ }\Psi
\right\rangle \text{-}\left[ C^{I},C^{\mathcal{A}}\right] }}=Z^{\mathcal{E}%
^{\left\langle \mathcal{A},\text{ }\Gamma ,\text{ }\Psi \right\rangle \text{-%
}\left[ C^{I},C^{F}\right] }}\text{,} \\
\forall \left\langle \Gamma ,\text{ }\Psi \right\rangle \in \left\{
\left\langle \Gamma ^{private},\Psi ^{private}\right\rangle ,\text{ }%
\left\langle \Gamma ^{public},\Psi ^{private}\right\rangle ,\text{ }%
\left\langle \Gamma ^{public},\Psi ^{public}\right\rangle \right\} \text{.}
\end{gather*}
\end{prop}

\noindent \textbf{Proof of Theorem \ref{thm:three}.} Proposition \ref%
{prop:other:on} and Proposition \ref{prop:deviation} with $I=R$ imply
Theorem \ref{thm:three}. $\blacksquare $

\subsection{Illustration of Proposition \protect\ref{prop:other:on}:
reduction}

\label{sec:resulst:illustrate:reduction}

\begin{lemma}
\label{lem:on-path:public}Consider $\mathcal{A}=\mathcal{A}^{non-delegated}$%
. For any $\left\langle \Gamma ,\text{ }\Psi \right\rangle \in \left\{
\Gamma ^{private},\Gamma ^{public}\right\} \times \left\{ \Psi
^{private},\Psi ^{public}\right\} $, we have $Z^{\mathcal{E}^{\left\langle 
\mathcal{A},\text{ }\Gamma ,\text{ }\Psi \right\rangle \text{-}\left[ C^{%
\mathcal{A}},C^{\mathcal{A}}\right] }}\subset Z^{\mathcal{E}^{\left\langle 
\mathcal{A},\text{ }\Gamma ,\text{ }\Psi \right\rangle \text{-}\left[
C^{R},C^{\mathcal{A}}\right] }}$.
\end{lemma}

\begin{lemma}
\label{lem:on-path:private-private}Given $\left\langle \mathcal{A},\text{ }%
\Gamma ,\text{ }\Psi \right\rangle =\left\langle \mathcal{A}^{non-delegated},%
\text{ }\Gamma ^{private},\text{ }\Psi ^{private}\right\rangle $, we have%
\begin{equation*}
Z^{\mathcal{E}^{\left\langle \mathcal{A},\text{ }\Gamma ,\text{ }\Psi
\right\rangle \text{-}\left[ C^{\mathcal{A}},C^{\mathcal{A}}\right]
}}\subset Z^{\mathcal{E}^{\left\langle \mathcal{A},\text{ }\Gamma ,\text{ }%
\Psi \right\rangle \text{-}\left[ C^{P},C^{\mathcal{A}}\right] }}\text{.}
\end{equation*}
\end{lemma}

Because $C^{P}\subset C^{R}$ (in (\ref{ktt1})), Lemma \ref%
{lem:on-path:private-private} implies that Lemma \ref{lem:on-path:public}
also holds for $\left\langle \Gamma ,\text{ }\Psi \right\rangle
=\left\langle \Gamma ^{private},\text{ }\Psi ^{private}\right\rangle .$
However, we present Lemma \ref{lem:on-path:private-private} separately from
Lemma \ref{lem:on-path:public} to provide a sharper characterization in the
model $\left\langle \mathcal{A}^{non-delegated},\text{ }\Gamma ^{private},%
\text{ }\Psi ^{private}\right\rangle $ through Theorem \ref%
{thm:full-private-private} in Section \ref{sec:private_private}. The proofs
of Lemmas \ref{lem:on-path:public} and \ref{lem:on-path:private-private} are
relegated to Appendix \ref{sec:lem:on-path:public} and \ref%
{sec:lem:on-path:private-private}.

To see the insight behind Lemma \ref{lem:on-path:public}, consider any
equilibrium $(c,s,t)\in \mathcal{E}^{\left\langle \mathcal{A},\text{ }\Gamma
,\text{ }\Psi \right\rangle \text{-}\left[ C^{\mathcal{A}},C^{\mathcal{A}}%
\right] }$. Given principal $j$'s equilibrium contract $c_{j},$ we can
construct the corresponding menu-of-menu-with-recommendation contract $%
c_{j}^{\prime }:K_{j}\rightarrow 2^{Y_{j}}\backslash \{\varnothing \},$
using $(s_{j},t_{j}),$ such that 
\begin{gather*}
K_{j}\equiv \left\{ \left( c_{j}\left[ s_{j}\left( c_{j},\widetilde{c}%
_{-j},\theta \right) \right] ,\text{ }t_{j}\left[ \Gamma _{j}\left( c_{j},%
\widetilde{c}_{-j}\right) ,\Psi _{j}\left( s\left( c_{j},\widetilde{c}%
_{-j},\theta \right) \right) \right] \right) :\left( \widetilde{c}%
_{-j},\theta \right) \in C_{-j}^{\mathcal{A}}\times \Theta \right\} \text{,}
\\
c_{j}^{\prime }\left( \left[ E_{j},y_{j}\right] \right) =E_{j},\text{ }%
\forall \left[ E_{j},y_{j}\right] \in K_{j}\text{.}
\end{gather*}%
Note that $(s,t)$ is an equilibrium strategy profile at Stages 2 and 3 in
the $\left[ C^{\mathcal{A}},C^{\mathcal{A}}\right] $ game. Therefore, when
principal $j$ offers $c_{j}$ given any $\widetilde{c}_{-j}$, it is optimal
for the agent of type $\theta $ to choose a menu $E_{j}=c_{j}\left[
s_{j}\left( c_{j},\widetilde{c}_{-j},\theta \right) \right] $ along with a
recommendation $y_{j}=t_{j}\left[ \Gamma _{j}\left( c_{j},\widetilde{c}%
_{-j}\right) ,\Psi _{j}\left( s\left( c_{j},\widetilde{c}_{-j},\theta
\right) \right) \right] $, expecting $j$ to follow her recommendation at
Stage 3, which is indeed optimal for $j.$

For the intuition behind Lemma \ref{lem:on-path:private-private}, note that
there is no information update for principal $j$ at Stage 3 given $%
\left\langle \Gamma ,\text{ }\Psi \right\rangle =\left\langle \Gamma
^{private},\Psi ^{private}\right\rangle $, because he cannot observe the
other principals' contracts nor the messages that the agent sent to them.
Therefore, his action choice in $c_{j}\left[ s_{j}\left( c_{j},\widetilde{c}%
_{-j},\theta \right) \right] $ depends only on the message the agent sent to
him, and hence principal $j$ can delegate the action choice to the agent by
offering the corresponding menu contract $\overline{c}_{j}:E_{j}\rightarrow
Y_{j}$ such that 
\begin{gather*}
E_{j}\equiv \left\{ t_{j}\left[ \Gamma _{j}\left( c_{j},\widetilde{c}%
_{-j}\right) ,\Psi _{j}\left( s\left( c_{j},\widetilde{c}_{-j},\theta
\right) \right) \right] :\left( \widetilde{c}_{-j},\theta \right) \in
C_{-j}^{\mathcal{A}}\times \Theta \right\} , \\
\overline{c}_{j}(y_{j})=y_{j},\text{ }\forall y_{j}\in E_{j}.
\end{gather*}

\subsection{Illustration of Proposition \protect\ref{prop:other:on}:
extended contracts}

\label{sec:resulst:illustrate:extension}

To complete the proof of Proposition \ref{prop:other:on}, we introduce the
notion of \textquotedblleft extended contracts.\textquotedblright\ For each $%
j\in \mathcal{J}$, we say a contract $c_{j}^{\prime }:M_{j}^{\prime
}\longrightarrow 2^{Y_{j}}\diagdown \left\{ \varnothing \right\} $ is an
extension of another $c_{j}^{\prime \prime }:M_{j}^{\prime \prime
}\longrightarrow 2^{Y_{j}}\diagdown \left\{ \varnothing \right\} $ (denoted
by $c_{j}^{\prime }\geq c_{j}^{\prime \prime }$) if and only if there exists
a \emph{surjective} function $\iota _{j}:M_{j}^{\prime }\longrightarrow
M_{j}^{\prime \prime }$ such that $c_{j}^{\prime }\left( m_{j}\right)
=c_{j}^{\prime \prime }\left( \iota _{j}\left( m_{j}\right) \right) $ for
all $m_{j}\in M_{j}^{\prime }.$ Therefore, heuristically, $c_{j}^{\prime
}:M_{j}^{\prime }\longrightarrow 2^{Y_{j}}\diagdown \left\{ \varnothing
\right\} $ has redundant messages, compared to $c_{j}^{\prime \prime
}:M_{j}^{\prime \prime }\longrightarrow 2^{Y_{j}}\diagdown \left\{
\varnothing \right\} $, but $c_{j}^{\prime }\left( M_{j}^{\prime }\right)
=c_{j}^{\prime \prime }\left( M_{j}^{\prime \prime }\right) .$ Extended
contracts play a critical role in our full characterization.

Based on \textquotedblleft $\geq $\textquotedblright\ defined above, we
define a binary relation as follows.%
\begin{equation*}
C^{I}\sqsupset ^{\ast }C^{II}\Longleftrightarrow \left( 
\begin{array}{c}
\forall j\in \mathcal{J}\text{, }\forall c_{j}^{\prime \prime }\in C_{j}^{II}%
\text{,} \\ 
\exists c_{j}^{\prime }\in C_{j}^{I}\text{, }c_{j}^{\prime }\geq
c_{j}^{\prime \prime }%
\end{array}%
\right) \text{, }\forall I,II\in \left\{ \mathcal{A}^{non-delegated},P,F,R%
\right\} \text{.}
\end{equation*}%
Clearly, $C^{\mathcal{A}^{non-delegated}}\sqsupset ^{\ast }C^{R}\sqsupset
^{\ast }C^{F}\sqsupset ^{\ast }C^{P}$.

Lemma \ref{lem:extension:on} below shows that if $C^{I}\sqsupset ^{\ast
}C^{III}$, a principal's equilibrium contract in the $\left[ C^{III},C^{II}%
\right] $ game can be replaced by its extension in the $\left[ C^{I},C^{II}%
\right] $ game. While the proof is technical, it is intuitive because an
extension of a contract has redundant messages, compared to the original
contract but with the same image set.\textbf{\ }Players can treat a
redundant message in an extension of a contract as if it is a message in the
original contract with the same image. Because of this, the agent does not
find it profitable to deviate to redundant messages. The proof of Lemma \ref%
{lem:extension:on} is relegated to Appendix \ref{sec:lem:extension:on}.

\begin{lemma}
\label{lem:extension:on}Let $\mathcal{A=A}^{non-delegated}$. For any $%
I,II,III\in \left\{ \mathcal{A},P,F,F^{\ast },R\right\} $, we have%
\begin{gather*}
C^{I}\sqsupset ^{\ast }C^{III}\Longrightarrow Z^{\mathcal{E}^{\left\langle 
\mathcal{A},\text{ }\Gamma ,\text{ }\Psi \right\rangle \text{-}\left[
C^{I},C^{II}\right] }}\supset Z^{\mathcal{E}^{\left\langle \mathcal{A},\text{
}\Gamma ,\text{ }\Psi \right\rangle \text{-}\left[ C^{III},C^{II}\right] }}%
\text{,} \\
\forall \left\langle \Gamma ,\text{ }\Psi \right\rangle \in \left\{ \Gamma
^{private},\Gamma ^{public}\right\} \times \left\{ \Psi ^{private},\Psi
^{public}\right\} \text{,}
\end{gather*}%
where $C^{F^{\ast }}$ is defined in (\ref{tttf3}).
\end{lemma}

\noindent \textbf{Proof of Proposition \ref{prop:other:on}.} Given $\mathcal{%
A}=\mathcal{A}^{non-delegated}$, Lemma \ref{lem:extension:on} and $C^{%
\mathcal{A}}\sqsupset ^{\ast }C^{R}\sqsupset ^{\ast }C^{P}$ imply 
\begin{gather}
Z^{\mathcal{E}^{\left\langle \mathcal{A},\text{ }\Gamma ,\text{ }\Psi
\right\rangle \text{-}\left[ C^{\mathcal{A}},C^{\mathcal{A}}\right]
}}\supset Z^{\mathcal{E}^{\left\langle \mathcal{A},\text{ }\Gamma ,\text{ }%
\Psi \right\rangle \text{-}\left[ C^{R},C^{\mathcal{A}}\right] }}\supset Z^{%
\mathcal{E}^{\left\langle \mathcal{A},\text{ }\Gamma ,\text{ }\Psi
\right\rangle \text{-}\left[ C^{P},C^{\mathcal{A}}\right] }}\text{,}
\label{poo1} \\
\forall \left\langle \Gamma ,\text{ }\Psi \right\rangle \in \left\{
\left\langle \Gamma ^{private},\Psi ^{private}\right\rangle ,\text{ }%
\left\langle \Gamma ^{public},\Psi ^{private}\right\rangle ,\text{ }%
\left\langle \Gamma ^{public},\Psi ^{public}\right\rangle \right\} \text{,} 
\notag
\end{gather}%
which, together with Lemmas \ref{lem:on-path:public} and \ref%
{lem:on-path:private-private}, implies Proposition \ref{prop:other:on}.$%
\blacksquare $

\subsection{Illustration of Proposition \protect\ref{prop:deviation}}

\label{sec:resulst:illustrate}

For any $I,II\in \left\{ \mathcal{A}^{non-delegated},P,F,R\right\} $, we
define another binary relation as follows.%
\begin{equation*}
C^{I}\sqsupset ^{\ast \ast }C^{II}\Longleftrightarrow \left( 
\begin{array}{c}
\forall j\in \mathcal{J}\text{, }\forall c_{j}^{\prime }\in C_{j}^{I}\text{,}
\\ 
\exists c_{j}^{\prime \prime }\in C_{j}^{II}\text{, }c_{j}^{\prime }\geq
c_{j}^{\prime \prime }%
\end{array}%
\right) \text{.}
\end{equation*}%
Clearly, $C^{\mathcal{A}^{non-delegated}}\sqsupset ^{\ast \ast }C^{F}$ and $%
C^{P}\sqsupset ^{\ast \ast }C^{F}$, but "$C^{I}\sqsupset ^{\ast \ast }C^{P}$%
" fails for each $I\in \left\{ \mathcal{A}^{non-delegated},F,R\right\} $.

\begin{lemma}
\label{lem:extension:off}Let $\mathcal{A=A}^{non-delegated}$. For any $%
I,II,IV\in \left\{ \mathcal{A},P,F,R\right\} $, we have%
\begin{gather*}
C^{II}\sqsupset ^{\ast \ast }C^{IV}\Longrightarrow Z^{\mathcal{E}%
^{\left\langle \mathcal{A},\text{ }\Gamma ,\text{ }\Psi \right\rangle \text{-%
}\left[ C^{I},C^{II}\right] }}\supset Z^{\mathcal{E}^{\left\langle \mathcal{A%
},\text{ }\Gamma ,\text{ }\Psi \right\rangle \text{-}\left[ C^{I},C^{IV}%
\right] }}\text{,} \\
\forall \left\langle \Gamma ,\text{ }\Psi \right\rangle \in \left\{
\left\langle \Gamma ^{private},\Psi ^{private}\right\rangle ,\text{ }%
\left\langle \Gamma ^{public},\Psi ^{private}\right\rangle ,\text{ }%
\left\langle \Gamma ^{public},\Psi ^{public}\right\rangle \right\} \text{.}
\end{gather*}
\end{lemma}

To see the intuition behind Lemma \ref{lem:extension:off}, let us fix any $%
\left( c,s,t\right) \in \mathcal{E}^{\left\langle \mathcal{A},\text{ }\Gamma
,\text{ }\Psi \right\rangle \text{-}\left[ C^{I},C^{IV}\right] }$. Given $%
C^{II}\sqsupset ^{\ast \ast }C^{IV}$, for any potential deviation to $%
c_{j}^{\prime }\in C_{j}^{II}$, we can find $c_{j}^{\prime \prime }\in
C_{j}^{IV}$ with $c_{j}^{\prime }\geq c_{j}^{\prime \prime }$. I.e., $%
c_{j}^{\prime }$ is an extension of $c_{j}^{\prime \prime }$ so that $%
c_{j}^{\prime }$ includes redundant messages, compared to $c_{j}^{\prime
\prime }$. We thus can construct continuation equilibrium strategies at
Stages 2 and 3 upon $j$'s deviation to $c_{j}^{\prime }$ that preserve the
continuation equilibrium allocation upon $j$'s deviation to $c_{j}^{\prime
\prime }$. Since $\left( c,s,t\right) $ is a $\left[ C^{I},C^{IV}\right] $%
-equilibrium, $c_{j}^{\prime \prime }$ is not a profitable deviation, and as
a result, $c_{j}^{\prime }\in C_{j}^{II}$ is not a profitable deviation
either. The proof of Lemma \ref{lem:extension:off} is relegated to Appendix %
\ref{sec:lem:extension:off}.

To see the logic behind Proposition \ref{prop:deviation}.\textbf{\ }Fix $%
\mathcal{A}=\mathcal{A}^{non-delegated}$. Fix any $\left\langle \Gamma ,%
\text{ }\Psi \right\rangle \in \left\{ \left\langle \Gamma ^{private},\Psi
^{private}\right\rangle ,\text{ }\left\langle \Gamma ^{public},\Psi
^{private}\right\rangle ,\text{ }\left\langle \Gamma ^{public},\Psi
^{public}\right\rangle \right\} $ and any $I\in \left\{ \mathcal{A}%
,P,F,R\right\} $. Since $C^{\mathcal{A}}\sqsupset ^{\ast \ast }C^{F}$, Lemma %
\ref{lem:extension:off} implies $Z^{\mathcal{E}^{\left\langle \mathcal{A},%
\text{ }\Gamma ,\text{ }\Psi \right\rangle \text{-}\left[ C^{I},C^{\mathcal{A%
}}\right] }}\supset Z^{\mathcal{E}^{\left\langle \mathcal{A},\text{ }\Gamma ,%
\text{ }\Psi \right\rangle \text{-}\left[ C^{I},C^{F}\right] }}$. To show
the converse, i.e., $Z^{\mathcal{E}^{\left\langle \mathcal{A},\text{ }\Gamma
,\text{ }\Psi \right\rangle \text{-}\left[ C^{I},C^{\mathcal{A}}\right]
}}\subset Z^{\mathcal{E}^{\left\langle \mathcal{A},\text{ }\Gamma ,\text{ }%
\Psi \right\rangle \text{-}\left[ C^{I},C^{F}\right] }}$, let us fix any $%
\left( c,s,t\right) \in \mathcal{E}^{\left\langle \mathcal{A},\text{ }\Gamma
,\text{ }\Psi \right\rangle \text{-}\left[ C^{I},C^{\mathcal{A}}\right] }$,
and we aim to show $\left( c,s,t\right) $ is also an $\left[ C^{I},C^{F}%
\right] $-equilibrium. Suppose that principal $j$ unilaterally deviates to
some $c_{j}^{\prime }\in C_{j}^{F}$. Since $c_{j}^{\prime }\in C_{j}^{F}$,
we have a pair of $L_{j}\subset 2^{Y_{j}}$ and $H_{j}=\left\{ \left[
E_{j},y_{j}\right] :y_{j}\in E_{j}\right\} $ satisfying Definition \ref%
{def:menu_of_menu_with_full_recommendation}. Thus, we consider any $\widehat{%
c}_{j}\in C_{j}^{\mathcal{A}}$ such that%
\begin{gather}
\widehat{c}_{j}:M_{j}^{\mathcal{A}}\rightarrow 2^{Y_{j}}\backslash \left\{
\varnothing \right\} \text{, and }\widehat{c}_{j}\left( M_{j}^{\mathcal{A}%
}\right) =L_{j}\cup \left\{ E_{j}\right\} \text{, and}  \label{ttc1} \\
\widehat{c}_{j}^{-1}(D_{j})\equiv \left\{ m_{j}:\widehat{c}%
_{j}(m_{j})=D_{j}\right\} \text{ is a singleton, }\forall D_{j}\in L_{j}%
\text{.}  \notag
\end{gather}%
I.e., we embeds $L_{j}\cup \left\{ E_{j}\right\} $ into $M_{j}^{\mathcal{A}}$%
, and any redundant messages in $M_{j}^{\mathcal{A}}\diagdown \left[
L_{j}\cup \left\{ E_{j}\right\} \right] $ are mapped to $E_{j}$ by $\widehat{%
c}_{j}$. Thus, $M_{j}^{\mathcal{A}}$ is partitioned into two parts: $M_{j}^{%
\mathcal{A}\text{-}L}\equiv \left\{ m_{j}\in M_{j}^{\mathcal{A}}:\widehat{c}%
_{j}(m_{j})\in L_{j}\right\} $ and $M_{j}^{\mathcal{A}}\diagdown M_{j}^{%
\mathcal{A}\text{-}L}$, and furthermore, $\widehat{c}_{j}|_{M_{j}^{\mathcal{A%
}\text{-}L}}$ is a bijection from $M_{j}^{\mathcal{A}\text{-}L}$ to $L$,
i.e.,%
\begin{gather*}
\widehat{c}_{j}\left( M_{j}^{\mathcal{A}\text{-}L}\right) =L_{j}\text{ and }%
\widehat{c}_{j}^{-1}(D_{j})\equiv \left\{ m_{j}:\widehat{c}%
_{j}(m_{j})=D_{j}\right\} \text{ is a singleton, }\forall D_{j}\in L_{j}%
\text{,} \\
\text{and }\widehat{c}_{j}\left( M_{j}^{\mathcal{A}}\diagdown M_{j}^{%
\mathcal{A}\text{-}L}\right) =\left\{ E_{j}\right\} \text{.}
\end{gather*}%
We can replace $M_{j}^{\mathcal{A}}\diagdown M_{j}^{\mathcal{A}\text{-}L}$
with the message set below:%
\begin{equation*}
K_{j}\equiv \left\{ \left( \widehat{c}_{j}\left( m_{j}\right) =E_{j},\text{ }%
t_{j}\left[ \Gamma _{j}\left( \widehat{c}_{j},\widetilde{c}_{-j}\right)
,\Psi _{j}\left( m_{j},m_{-j}\right) \right] \right) :\left[ \widetilde{c}%
_{-j},\text{ }\left( m_{j},m_{-j}\right) \right] \in C_{-j}^{\mathcal{A}%
}\times \left( M_{j}^{\mathcal{A}}\diagdown M_{j}^{\mathcal{A}\text{-}%
L}\right) \times M_{-j}^{\mathcal{A}}\right\} \text{,}
\end{equation*}%
and define a new contract%
\begin{gather*}
\overline{c}_{j}:M_{j}^{\mathcal{A}\text{-}L}\cup K_{j}\rightarrow
2^{Y_{j}}\backslash \{\varnothing \}\text{, and }\overline{c}_{j}\left(
m_{j}\right) =\widehat{c}_{j}\left( m_{j}\right) \text{, }\forall m_{j}\in
M_{j}^{\mathcal{A}\text{-}L}\text{, } \\
\text{and }\overline{c}_{j}\left( m_{j}\right) =E_{j}\text{, }\forall
m_{j}\in K_{j}\text{.}
\end{gather*}%
$\overline{c}_{j}$ can be viewed as a menu-of-menu-with-recommendation
contract, i.e., $\overline{c}_{j}\in C_{j}^{R}$. Applying a similar argument
as in Section \ref{sec:resulst:illustrate:reduction} to off the equilibrium
path, we can replace $\widehat{c}_{j}\in C_{j}^{\mathcal{A}}$ with $%
\overline{c}_{j}\in C_{j}^{R}$, replicating the continuation equilibrium
following $\left( \widehat{c}_{j},c_{-j}\right) $ with a continuation
equilibrium following $\left( \overline{c}_{j},c_{-j}\right) $. Since $%
\left( c,s,t\right) \in \mathcal{E}^{\left\langle \mathcal{A},\text{ }\Gamma
,\text{ }\Psi \right\rangle \text{-}\left[ C^{I},C^{\mathcal{A}}\right] }$,
deviating to $\widehat{c}_{j}\in C_{j}^{\mathcal{A}}$ is not a profitable
deviation for principal $j$, and hence, $\overline{c}_{j}$ is not a
profitable deviation for $j$ either.

Furthermore, we have $c_{j}^{\prime }\geq \overline{c}_{j}$ because $%
K_{j}\subset H_{j}$ and any subset in $L_{j}$ is induced by a single message
in both $c_{j}^{\prime }$ and $\overline{c}_{j}$. Then, applying a similar
argument as in Section \ref{sec:resulst:illustrate:extension} to off the
equilibrium path, we can replace $\overline{c}_{j}\in C_{j}^{R}$ with $%
c_{j}^{\prime }\in C_{j}^{F}$, replicating the continuation equilibrium
following $\left( \overline{c}_{j},c_{-j}\right) $ with a continuation
equilibrium following $\left( c_{j}^{\prime },c_{-j}\right) $, and hence, $%
c_{j}^{\prime }\in C_{j}^{F}$ is not a profitable deviation for $j$ either.
Since this is true for any $c_{j}^{\prime }\in C_{j}^{F}$, we conclude that $%
\left( c,s,t\right) $ is an $\left[ C^{I},C^{F}\right] $-equilibrium.

\subsection{Private announcement and private communication\label%
{sec:private_private}}

Given $\left\langle \mathcal{A}^{non-delegated},\text{ }\Gamma ^{private},%
\text{ }\Psi ^{private}\right\rangle $, the full characterization can be
refined as follows.

\begin{theo}
\label{thm:full-private-private}Suppose $\left\langle \mathcal{A},\text{ }%
\Gamma ,\text{ }\Psi \right\rangle =\left\langle \mathcal{A}^{non-delegated},%
\text{ }\Gamma ^{private},\text{ }\Psi ^{private}\right\rangle $. We have 
\begin{equation}
Z^{\mathcal{E}^{\left\langle \mathcal{A},\text{ }\Gamma ,\text{ }\Psi
\right\rangle \text{-}\left[ C^{\mathcal{A}},C^{\mathcal{A}}\right] }}=Z^{%
\mathcal{E}^{\left\langle \mathcal{A},\text{ }\Gamma ,\text{ }\Psi
\right\rangle \text{-}\left[ C^{P},C^{F}\right] }}\text{.}  \label{tth1}
\end{equation}
\end{theo}

Given $\left\langle \mathcal{A}^{non-delegated},\text{ }\Gamma ^{private},%
\text{ }\Psi ^{private}\right\rangle $, Theorem \ref%
{thm:full-private-private} says that it suffers no loss of generality for
principals to offer menu contracts on the equilibrium path, which sharpens
Theorem \ref{thm:three}, due to $C^{P}\subset C^{R}$ (in (\ref{ktt1})). In
this sense, the menu theorem holds partially (i.e., on the equilibrium
path). However, the refinement of the full characterization in Theorem \ref%
{thm:full-private-private} does not apply to the other three models as shown
by the examples in Sections \ref{sec:examples} and \ref%
{sec:private-public:counterexample}. Furthermore, with $\mathcal{A}=\mathcal{%
A}^{delegated}$, the menu theorem (i.e., Theorem \ref{thm:peters}) says%
\begin{equation*}
Z^{\mathcal{E}^{\left\langle \mathcal{A},\text{ }\Gamma ,\text{ }\Psi
\right\rangle \text{-}\left[ C^{\mathcal{A}},C^{\mathcal{A}}\right] }}=Z^{%
\mathcal{E}^{\left\langle \mathcal{A},\text{ }\Gamma ,\text{ }\Psi
\right\rangle \text{-}\left[ C^{P},C^{P}\right] }}\text{,}
\end{equation*}%
i.e., $Z^{\mathcal{E}^{\left\langle \mathcal{A},\text{ }\Gamma ,\text{ }\Psi
\right\rangle \text{-}\left[ C^{P},C^{P}\right] }}$ is both an upper bound
and a lower bound for $Z^{\mathcal{E}^{\left\langle \mathcal{A},\text{ }%
\Gamma ,\text{ }\Psi \right\rangle \text{-}\left[ C^{\mathcal{A}},C^{%
\mathcal{A}}\right] }}$. Given $\left\langle \mathcal{A}^{non-delegated},%
\text{ }\Gamma ^{private},\text{ }\Psi ^{private}\right\rangle $, Theorem %
\ref{thm:non-delegated_vs_delegated} below shows that $Z^{\mathcal{E}%
^{\left\langle \mathcal{A},\text{ }\Gamma ,\text{ }\Psi \right\rangle \text{-%
}\left[ C^{P},C^{P}\right] }}$ remains an upper bound. This establishes a
second sense that the menu theorem holds partially.

\begin{theo}
\label{thm:non-delegated_vs_delegated}In the model $\left\langle \mathcal{A},%
\text{ }\Gamma ,\text{ }\Psi \right\rangle =\left\langle \mathcal{A}%
^{non-delegated},\text{ }\Gamma ^{private},\text{ }\Psi
^{private}\right\rangle $, we have%
\begin{equation}
Z^{\mathcal{E}^{\left\langle \mathcal{A},\text{ }\Gamma ,\text{ }\Psi
\right\rangle \text{-}\left[ C^{\mathcal{A}},C^{\mathcal{A}}\right]
}}\subset Z^{\mathcal{E}^{\left\langle \mathcal{A},\text{ }\Gamma ,\text{ }%
\Psi \right\rangle \text{-}\left[ C^{P},C^{P}\right] }}\text{.}
\label{hhtt1}
\end{equation}
\end{theo}

The proofs of Theorem \ref{thm:full-private-private} and \ref%
{thm:non-delegated_vs_delegated} are relegated to Appendix \ref%
{sec:thm:full-private-private} and \ref{sec:thm:non-delegated_vs_delegated},
respectively. The converse of (\ref{hhtt1}) in Theorem \ref%
{thm:non-delegated_vs_delegated} fails, which is proved by an example in
Appendix \ref{sec:menu:fail:private-private}.

\section{Full equilibrium characterization II}

\label{sec:private-public}

Throughout this section, we focus on $\left\langle \mathcal{A}%
^{non-delegated}\text{, }\Gamma ^{private},\text{ }\Psi
^{public}\right\rangle $. Theorem \ref{thm:three} shows%
\begin{equation}
Z^{\mathcal{E}^{\left\langle \mathcal{A},\text{ }\Gamma ,\text{ }\Psi
\right\rangle \text{-}\left[ C^{\mathcal{A}},C^{\mathcal{A}}\right] }}=Z^{%
\mathcal{E}^{\left\langle \mathcal{A},\text{ }\Gamma ,\text{ }\Psi
\right\rangle \text{-}\left[ C^{R},C^{F}\right] }}  \label{tft1}
\end{equation}%
for the other three models. Thus, it is natural to conjecture that (\ref%
{tft1}) still holds under $\left\langle \mathcal{A}^{non-delegated}\text{, }%
\Gamma ^{private},\text{ }\Psi ^{public}\right\rangle $. However, we use an
example to disprove this in Section \ref{sec:private-public:counterexample}.
Furthermore, we provide a full characterization of $Z^{\mathcal{E}%
^{\left\langle \mathcal{A},\text{ }\Gamma ,\text{ }\Psi \right\rangle \text{-%
}\left[ C^{\mathcal{A}},C^{\mathcal{A}}\right] }}$ in Section \ref%
{sec:private-public:characterization}.

\subsection{A counterexample}

\label{sec:private-public:counterexample}

Before presenting our counterexample, we describe the intuition for failure
of (\ref{tft1}). Precisely, the problem is that it suffers loss of
generality to focus on $C^{F}$ on off-equilibrium paths. With public
communication, principals observe deviations if the agent sends an
off-equilibrium message at Stage 2. Furthermore, with private announcement,
principals do not know the meaning of such off-equilibrium messages for
general contracts (i.e., in $Z^{\mathcal{E}^{\left\langle \mathcal{A},\text{ 
}\Gamma ,\text{ }\Psi \right\rangle \text{-}\left[ C^{\mathcal{A}},C^{%
\mathcal{A}}\right] }}$). Thus, in order to sustain an equilibrium,
principals have freedom to form beliefs regarding other principals'
contracts, upon observing off-equilibrium messages. Such freedom is not
shared in $Z^{\mathcal{E}^{\left\langle \mathcal{A},\text{ }\Gamma ,\text{ }%
\Psi \right\rangle \text{-}\left[ C^{R},C^{F}\right] }}$. For instance,
given $C^{F}$ on off-equilibrium paths, principal $j$ must believe that
principal $j^{\prime }$ plays $a$ if the agent sends the message $\left(
\left\{ a\right\} ,a\right) $ to principal $j^{\prime }$. As an result of
this problem, equilibria in $\mathcal{E}^{\left\langle \mathcal{A},\text{ }%
\Gamma ,\text{ }\Psi \right\rangle \text{-}\left[ C^{R},C^{F}\right] }$ are
not able to replicate all of the equilibria in $\mathcal{E}^{\left\langle 
\mathcal{A},\text{ }\Gamma ,\text{ }\Psi \right\rangle \text{-}\left[ C^{%
\mathcal{A}},C^{\mathcal{A}}\right] }$. We provide such an example as
follows.

There are two principals, $\mathcal{J}=\left\{ 1,2\right\} $ and each
principal can take one of four possible actions $Y_{1}=Y_{2}=\left\{
1,2,3,4\right\} $. There are two states $\Theta =\left\{ 1,2\right\} $. The
common prior on the state is $p\left( 1\right) =p\left( 2\right) =1/2$. The
preferences for both principals are identical: 
\begin{equation}
v_{j}\left[ \left( y_{1},y_{2}\right) ,\text{ }\theta =1\right] =\left\{ 
\begin{tabular}{ll}
$8$, & if $\left( y_{1},y_{2}\right) =\left( 1,1\right) $, \\ 
$9$, & if $\left( y_{1},y_{2}\right) \in \left\{ \left( 3,1\right) ,\left(
4,1\right) \right\} $, \\ 
$y_{1}-y_{2}+4$, & if $\left\{ y_{1},y_{2}\right\} \subset \left\{
1,2\right\} $ or $\left\{ y_{1},y_{2}\right\} \subset \left\{ 3,4\right\} $,
\\ 
$y_{2}-y_{1}$, & otherwise,%
\end{tabular}%
\right. \forall j\in \mathcal{J}\text{,}  \label{btk1}
\end{equation}%
\begin{equation}
v_{j}\left[ \left( y_{1},y_{2}\right) ,\text{ }\theta =4\right] =\left\{ 
\begin{tabular}{ll}
$8$, & if $\left( y_{1},y_{2}\right) =\left( 4,4\right) $, \\ 
$9$, & if $\left( y_{1},y_{2}\right) \in \left\{ \left( 4,1\right) ,\left(
4,2\right) \right\} $, \\ 
$y_{1}-y_{2}+4$, & if $\left\{ y_{1},y_{2}\right\} \subset \left\{
1,2\right\} $ or $\left\{ y_{1},y_{2}\right\} \subset \left\{ 3,4\right\} $,
\\ 
$y_{2}-y_{1}$, & otherwise,%
\end{tabular}%
\right. \forall j\in \mathcal{J}\text{,}  \label{btk2}
\end{equation}%
The agent's preference is described as follow:%
\begin{equation}
u\left[ \left( y_{1},y_{2}\right) ,\theta \right] =\left\{ 
\begin{tabular}{ll}
$1$, & if $\left( y_{1},y_{2}\right) =\left( 1,4\right) $; \\ 
$0$, & otherwise.%
\end{tabular}%
\right. \forall \theta \in \Theta \text{.}  \label{btk3}
\end{equation}%
Consider the outcome%
\begin{equation}
z^{\ast }\equiv \left[ \left( y_{1},y_{2}\right) =\left( 1,1\right) \text{
at }\theta =1\text{, and }\left( y_{1},y_{2}\right) =\left( 4,4\right) \text{
at }\theta =4\right] \text{.}  \label{hhtt3}
\end{equation}%
The outcome $z^{\ast }$ in (\ref{hhtt3}) cannot be induced by any
equilibrium in the $\left[ C^{R},C^{F}\right] $ game. We prove by
contradiction. Suppose we can achieve the allocation with a $\left[
C^{R},C^{F}\right] $-equilibrium. Let us focus on the equilibrium path. At
State $1$, the agent chooses $E_{1}^{1}$ for principal 1 and $E_{2}^{1}$ for
principal 2, and at State $4$, the agent chooses $E_{1}^{4}$ for principal $%
1 $ and $E_{2}^{4}$ for principal $2$. Since it implements $\left(
1,1\right) $ at $\theta ^{1}$ and $\left( 4,4\right) $ at $\theta ^{4}$, we
have $1\in E_{1}^{1}$ and $4\in E_{2}^{4}$.

At state $1$, on the equilibrium path, principal $1$ expects that principal $%
2$ would choose $1$ at Stage 3. By the preference of principal $1$ (i.e., (%
\ref{btk1})), we have $\left\{ 3,4\right\} \cap E_{1}^{1}=\varnothing $.
Similarly, at state $4$, on the equilibrium path, principal $2$ expects that
principal $1$ would choose $4$ at Stage 3. By the preference of principal 2
(i.e., (\ref{btk2})), we have $\left\{ 1,2\right\} \cap
E_{2}^{4}=\varnothing $. That is, $1\in E_{1}^{1}\subset \left\{ 1,2\right\} 
$ and $4\in E_{2}^{4}\subset \left\{ 3,4\right\} $. Then, at State $1$ (also
at State $4$), the agent finds it profitable to deviate to choose $\left(
E_{1}^{1},E_{2}^{4}\right) $ for the principals. Upon observing this,
principals know that they must choose an action profile in $\left(
y_{1},y_{2}\right) \in E_{1}^{1}\times E_{2}^{4}=\left\{ 1,2\right\} \times
\left\{ 3,4\right\} $. Given this, the dominant strategy for principal $1$
is to choose $1$ and the dominant strategy for principal $2$ is to choose $4$%
. Thus, the principals would chooses $\left( 1,4\right) $, which is strictly
better than the equilibrium allocation $\left( 1,1\right) $ for the agent,
i.e., we reach a contradiction.

However, the outcome $z^{\ast }$ in (\ref{hhtt3}) can be induced by any
equilibrium in the $\left[ C^{\mathcal{A}},C^{\mathcal{A}}\right] $ game. To
see this, we fix three distinct messages, $m_{j}^{1}$, $m_{j}^{2}$ and $%
m_{j}^{4}$ for each $j\in \mathcal{J}$. Define a contract, $c_{j}^{\ast
}:M_{j}^{\mathcal{A}}\rightarrow 2^{Y_{j}}$:%
\begin{equation*}
c_{j}^{\ast }\left( m_{j}\right) =\left\{ 
\begin{tabular}{ll}
$\left\{ 1,2\right\} $, & if $m_{j}=m_{j}^{1}$, \\ 
$\left\{ 3,4\right\} $, & if $m_{j}=m_{j}^{4}$, \\ 
$\left\{ 2\right\} $, & if $m_{j}=m_{j}^{2}$, \\ 
$\left\{ 2\right\} $, & otherwise%
\end{tabular}%
\right. \text{.}
\end{equation*}%
Consider the following equilibrium, which induces $z^{\ast }$ in (\ref{hhtt3}%
).

\begin{enumerate}
\item Each principal $j\in \mathcal{J}$ offers $c_{j}^{\ast }$ on the
equilibrium path.

\item The agent sends $m_{j}^{\theta }$ to each principal $j\in \mathcal{J}$
at each state $\theta \in \Theta $.

\item Principals choose $(\theta ,\theta )$ upon receiving $\left(
m_{1}^{\theta },m_{2}^{\theta }\right) $ for every $\theta \in \Theta $.
\end{enumerate}

To sustain this as an equilibrium, the agent and principals take the
following (behavioral) strategies at Stages 2 and 3. At stage 2, on
off-equilibrium paths, 
\begin{gather*}
\text{the agent takes }s\equiv \lbrack s_{k}:C^{\mathcal{A}}\times \Theta
\rightarrow M_{k}^{\mathcal{A}}]_{k\in \mathcal{J}}\text{ such that} \\
c\left[ s\left( c,\theta \right) \right] \neq \left\{ 1\right\} \times
\left\{ 4\right\} \Rightarrow \left( 
\begin{array}{c}
\nexists m\in M^{\mathcal{A}}\text{ such that }c\left( m\right) =\left\{
1\right\} \times \left\{ 4\right\} \text{,} \\ 
\text{and }s\left( c,\theta \right) =\left( m_{1}^{2},m_{2}^{2}\right)%
\end{array}%
\right) \text{,} \\
\text{ }\forall \left( c,\theta \right) \in \left[ C^{\mathcal{A}}\diagdown
\left\{ \left( c_{1}^{\ast },c_{2}^{\ast }\right) \right\} \right] \times
\Theta \text{.}
\end{gather*}%
That is, we consider two cases: (1) if there exists $m\in M^{\mathcal{A}}$
such that $c\left( m\right) =\left\{ 1\right\} \times \left\{ 4\right\} $,
we let the agent send $m$, which achieves the maximal utility for the agent;
(2) otherwise, the agent always sends $\left( m_{1}^{2},m_{2}^{2}\right) $.
If there exists no $m\in M^{\mathcal{A}}$ such that $c\left( m\right)
=\left\{ 1\right\} \times \left\{ 4\right\} $, as will be clear, principals
will never play $\left( 1,4\right) $ at Stage 3, regardless the agent's
messages at Stage 2. By (\ref{btk3}), the agent's incentive compatibility
holds.

At Stage 3, let each principal $j$ adopt the (behavioral) strategy $%
t_{j}:C_{j}^{\mathcal{A}}\times M_{1}^{\mathcal{A}}\times M_{2}^{\mathcal{A}%
}\rightarrow Y_{j}$, and the corresponding beliefs described as follows.
First, on the equilibrium path,%
\begin{equation*}
\left[ c_{j},\left( m_{1},m_{2}\right) \right] =\left[ c_{j}^{\ast },\left(
m_{1}^{\theta },m_{2}^{\theta }\right) \right] \Rightarrow t_{j}\left[
c_{j},\left( m_{1},m_{2}\right) \right] =\theta \text{, }\forall j\in 
\mathcal{J}\text{, }\forall \theta \in \Theta \text{.}
\end{equation*}%
Second, consider any $j\in \mathcal{J}$, and suppose $j$ unilaterally
deviates to $c_{j}\neq c_{j}^{\ast }$, and the agent sends $\left(
m_{1}^{2},m_{2}^{2}\right) $ at Stage 2. Then, principal $j$ cannot confirm
deviation from the agent or the other principal. Thus, $j$ must believe that
the other principal offers $c_{-j}^{\ast }$ at Stage 1, and there is equal
probability for $\theta =1$ and $\theta =4$. Given $c_{-j}^{\ast }$, the
message $m_{-j}^{2}$ pins down the action of $2$ for principal $-j$ at Stage
3. Thus, let principal $j$ take the following action at Stage 3.%
\begin{equation*}
c_{j}\notin C_{j}^{\mathcal{A}}\diagdown \left\{ c_{j}^{\ast }\right\}
\Rightarrow t_{j}\left[ c_{j},\left( m_{1}^{2},m_{2}^{2}\right) \right] \in
\arg \max_{y_{j}\in c_{j}\left( m_{j}^{2}\right) }\left( \frac{1}{2}v_{j}%
\left[ \left( y_{j},y_{-j}=2\right) ,\theta ^{1}\right] +\frac{1}{2}v_{j}%
\left[ \left( y_{j},y_{-j}=2\right) ,\theta ^{4}\right] \right) \text{.}
\end{equation*}%
Clearly, in this case, principals' incentive compatibility holds at Stage 3.
Furthermore, by (\ref{btk1}) and (\ref{btk2})), such a unilateral deviation
would induce a payoff less than 8 (i.e., the equilibrium payoff) at both
states, i.e., this is not a profitable deviation.

Third, consider all of the other off-equilibrium paths. For each $j\in 
\mathcal{J}$ and each $y_{j}\in Y_{j}$, let $c_{j}^{y_{j}}:M_{j}^{\mathcal{A}%
}\rightarrow 2^{Y_{j}}$ denote the following degenerate contract: $%
c_{j}^{y_{j}}\left( m_{j}\right) =\left\{ y_{j}\right\} $ for all $m_{j}\in
M_{j}^{\mathcal{A}}$. Principal $1$ takes $t_{1}:C_{1}^{\mathcal{A}}\times
M_{1}^{\mathcal{A}}\times M_{2}^{\mathcal{A}}\rightarrow Y_{1}$ such that%
\begin{equation*}
\left( 
\begin{array}{c}
\left( c_{1},m_{1},m_{2}\right) \notin \left\{ \left( c_{1}^{\ast
},m_{1}^{1},m_{2}^{1}\right) ,\left( c_{1}^{\ast
},m_{1}^{4},m_{2}^{4}\right) \right\} \text{,} \\ 
\text{or }\left( c_{1},m_{1},m_{2}\right) \notin \left( C_{1}^{\mathcal{A}%
}\diagdown \left\{ c_{1}^{\ast }\right\} \right) \times \left\{ \left(
m_{1}^{2},m_{2}^{2}\right) \right\}%
\end{array}%
\right) \Rightarrow \left( 
\begin{array}{c}
t_{1}\left( c_{1},m_{1},m_{2}\right) =\max c_{1}\left( m_{1}\right) \equiv 
\widehat{y}\text{,} \\ 
1\text{ believes }\left[ \theta =1\text{ and }c_{2}=c_{2}^{\widehat{y}}%
\right] \\ 
\text{with probability 1 }%
\end{array}%
\right) \text{,}
\end{equation*}%
and principal $2$ takes $t_{2}:C_{2}^{\mathcal{A}}\times M_{1}^{\mathcal{A}%
}\times M_{2}^{\mathcal{A}}\rightarrow Y_{2}$ such that%
\begin{equation*}
\left( 
\begin{array}{c}
\left( c_{2},m_{1},m_{2}\right) \notin \left\{ \left( c_{2}^{\ast
},m_{1}^{1},m_{2}^{1}\right) ,\left( c_{2}^{\ast
},m_{1}^{4},m_{1}^{4}\right) \right\} \text{,} \\ 
\text{or }\left( c_{2},m_{1},m_{2}\right) \notin \left( C_{2}^{\mathcal{A}%
}\diagdown \left\{ c_{2}^{\ast }\right\} \right) \times \left\{ \left(
m_{1}^{2},m_{2}^{2}\right) \right\}%
\end{array}%
\right) \Rightarrow \left( 
\begin{array}{c}
t_{2}\left( c_{2},m_{1},m_{2}\right) =\min c_{2}\left( m_{2}\right) \equiv 
\widetilde{y}\text{,} \\ 
2\text{ believes }\left[ \theta =4\text{ and }c_{j_{1}}=c_{j_{1}}^{%
\widetilde{y}}\right] \\ 
\text{with probability 1 }%
\end{array}%
\right) \text{.}
\end{equation*}%
It is easy to check that principals' incentive compatibility at Stage 3
holds, and that principals will never play $\left( 1,4\right) $ at Stage 3,
if there does not exist $m\in M^{\mathcal{A}}$ such that $c\left( m\right)
=\left\{ 1\right\} \times \left\{ 4\right\} $.

\subsection{A full characterization}

\label{sec:private-public:characterization}

Nevertheless, the generalized menu theorem can be easily adapted in $%
\left\langle \mathcal{A}^{non-delegated}\text{, }\Gamma ^{private},\text{ }%
\Psi ^{public}\right\rangle $. For each $j\in \mathcal{J}$, let $M_{j}^{R-F}$
be the set of messages that are used contracts in $C_{j}^{R}\cup C_{j}^{F}$.
For each $y_{j}\in Y_{j}$, let $c_{j}^{y_{j}}:M_{j}^{R-F}\rightarrow
2^{Y_{j}}\diagdown \left\{ \varnothing \right\} $ denote the following
degenerate contract.%
\begin{equation}
c_{j}^{y_{j}}\left( m_{j}\right) =\left\{ y_{j}\right\} \text{, }\forall
m_{j}\in M_{j}^{R-F}\text{.}  \label{tft3}
\end{equation}%
Define%
\begin{equation}
C^{F^{\ast }}\equiv \times _{j\in \mathcal{J}}C_{j}^{F^{\ast }}\equiv \times
_{j\in \mathcal{J}}\left[ C_{j}^{F}\cup \left\{ c_{j}^{y_{j}}:y_{j}\in
Y_{j}\right\} \right] .  \label{tttf3}
\end{equation}

\begin{theo}
\label{thm:private-public}Given $\left\langle \mathcal{A}\text{, }\Gamma ,%
\text{ }\Psi \right\rangle =\left\langle \mathcal{A}^{non-delegated}\text{, }%
\Gamma ^{private},\text{ }\Psi ^{public}\right\rangle $, we have%
\begin{equation*}
Z^{\mathcal{E}^{\left\langle \mathcal{A},\text{ }\Gamma ,\text{ }\Psi
\right\rangle \text{-}\left[ C^{\mathcal{A}},C^{\mathcal{A}}\right] }}=Z^{%
\mathcal{E}^{\left\langle \mathcal{A},\text{ }\Gamma ,\text{ }\Psi
\right\rangle \text{-}\left[ C^{R},C^{F^{\ast }}\right] }}\text{.}
\end{equation*}
\end{theo}

The proof of Theorem \ref{thm:private-public} is relegated to Appendix \ref%
{sec:theorem}. It still suffers no loss of generality to focus on $C^{R}$ on
the equilibrium path, and the intuition is the same as that of Proposition %
\ref{prop:other:on}. By adding degenerate contracts in $\left\{
c_{j}^{y_{j}}:y_{j}\in Y_{j}\right\} $ to $C^{F^{\ast }}$, principals can no
longer infer other principals' subset of actions by observing the agents'
messages to other principals, which resolves the problem discussed in
Section \ref{sec:private-public:counterexample}. Therefore, it suffers no
loss of generality to focus on $C^{F^{\ast }}$ on off-equilibrium paths.

\section{Common agency with imperfect commitment}

\label{sec:commitment}

In this section, we adapt our model to describe imperfect commitment \emph{%
\`{a} la} \cite{bs2000, bs2001, bs2007}. To achieve this,\textbf{\ }we just
need to make three changes to the model in Section \ref{sec:model}. First,
each principal $j$'s action is decomposed to contractible and
non-contractible parts, i.e., $y_{j}=(y_{j}^{1},y_{j}^{2})\in
Y_{j}=Y_{j}^{1}\times Y_{j}^{2}$, where $y_{j}^{1}$ is contractible and $%
y_{j}^{2}$ is not contractible. Second, for each $j\in \mathcal{J}$, the
contractible part (i.e., $y_{j}^{1}$) of an action determines a subset of
non-contractible parts (i.e., $y_{j}^{2}$) that the principal can choose
later and it is specified by an exogenous function $\phi _{j}:$ $%
Y_{j}^{1}\longrightarrow 2^{Y_{j}^{2}}\diagdown \left\{ \varnothing \right\} 
$. Third, principal $j$'s contract $c_{j}:M_{j}^{\mathcal{A}}\longrightarrow
Y_{j}^{1}$ specifies only the contractible part of an action as a function
of the agent's message. Let $\widetilde{C}_{j}^{\mathcal{A}}\equiv \left(
Y_{j}^{1}\right) ^{M_{j}^{\mathcal{A}}}$ denote $j$'s contract space, and $%
\widetilde{C}^{\mathcal{A}}=\left( \widetilde{C}_{j}^{\mathcal{A}}\right)
_{j\in \mathcal{J}}$.

In this game, we follow the same timeline in Section \ref%
{sec:model:game:timeline}: each principal $j$ simultaneously offers $%
c_{j}\in \widetilde{C}_{j}^{\mathcal{A}}$ at Stage 1; the agent sends $%
m_{j}\in M_{j}^{\mathcal{A}}$ to each principal $j$ at Stage 2, which pins
down the contractible part of action $c_{j}\left( m_{j}\right) \in Y_{j}^{1}$
and the subset $\phi _{j}\left[ c_{j}\left( m_{j}\right) \right] \in
2^{Y_{j}^{2}}\diagdown \left\{ \varnothing \right\} $ for principal $j$; at
Stage 3, each principal $j$ simultaneously takes an action $\left(
c_{j}\left( m_{j}\right) ,\text{ }y_{j}^{2}\in \phi _{j}\left[ c_{j}\left(
m_{j}\right) \right] \right) \in Y_{j}^{1}\times Y_{j}^{2}=Y_{j}$.

In fact, the model in Section \ref{sec:model} can be viewed as a special
case of the model with imperfect commitment. Though $Y_{j}$ in our model in
Section \ref{sec:model} may not be directly decomposed to contractible and
non-contractible components, we can define a new action space for principal $%
j$ as follows.%
\begin{gather*}
\widetilde{Y}_{j}^{1}=2^{Y_{j}}\diagdown \left\{ \varnothing \right\} \text{%
, \ \ }\widetilde{Y}_{j}^{2}=Y_{j}\text{, \ \ }\widetilde{Y}_{j}=\widetilde{Y%
}_{j}^{1}\times \widetilde{Y}_{j}^{2}\text{, \ }\widetilde{\phi }_{j}:%
\widetilde{Y}_{j}^{1}\longrightarrow 2^{\widetilde{Y}_{j}^{2}}\diagdown
\left\{ \varnothing \right\} . \\
\widetilde{\phi }_{j}\left( y_{j}^{1}\right) =y_{j}^{1}\text{, }\forall
y_{j}^{1}\in \widetilde{Y}_{j}^{1}\text{.}
\end{gather*}%
Thus, the common-agency-with-imperfect-commitment model with $\left( 
\widetilde{Y}_{j},\widetilde{\phi }_{j}\right) _{j\in \mathcal{J}}$ is
equivalent to the model in Section \ref{sec:model}.\footnote{%
One superficial difference between the model in Section \ref{sec:model} and
the model with imperfect commitment is that the degree of principals'
commitment is endogenous in the former, but seems exogenous in the latter
(as described by the exogenous $\phi _{j}$). In fact, $\phi _{j}$ can
accommodate endogenous commitment. To see this, take $Y_{j}=Y_{j}^{1}\times
Y_{j}^{2}$\ as principal $j$'s underlying action space.\textbf{\ }%
Conditional on choosing $y_{j}^{1}$, let $\digamma _{j}\left(
y_{j}^{1}\right) \subset 2^{Y_{j}^{2}}\diagdown \left\{ \varnothing \right\} 
$ denote the set of subsets of $Y_{j}^{2}$ to which principal $j$ can commit
for Stage 3. If $\digamma _{j}\left( y_{j}^{1}\right)
=2^{Y_{j}^{2}}\diagdown \left\{ \varnothing \right\} $ for every $%
y_{j}^{1}\in Y_{j}^{1}$, principal $j$'s commitment is fully endogenous, and
if $\left\vert \digamma _{j}\left( y_{j}^{1}\right) \right\vert =1$ for
every $y_{j}^{1}\in Y_{j}^{1}$, principal $j$'s commitment is fully
exogenous. $\phi _{j}$ can describe any $\digamma _{j}$ by changing the
action space as follows.%
\begin{gather*}
\widehat{Y}_{j}^{1}=\left\{ \left( y_{j}^{1},P_{j}^{2}\right) :y_{j}^{1}\in
Y_{j}^{1}\text{ and }P_{j}^{2}\in \digamma _{j}\left( y_{j}^{1}\right)
\right\} \text{, }\widehat{Y}_{j}=\widehat{Y}_{j}^{1}\times Y_{j}^{2}\text{,}
\\
\widehat{\phi }_{j}\left( y_{j}^{1},P_{j}^{2}\right) =P_{j}^{2}\text{, }%
\forall \left( y_{j}^{1},P_{j}^{2}\right) \in \widehat{Y}_{j}^{1}\text{.}
\end{gather*}%
Thus, by picking different $\digamma _{j}$, the
common-agency-with-imperfect-commitment model with $\left( \widehat{Y}_{j},%
\widehat{\phi }_{j}\right) _{j\in \mathcal{J}}$ can describe fully exogenous
commitment, or fully endogenous commitment, or any intermediate ones.}

Theorem \ref{thm:three} can be easily extended to a
common-agency-with-imperfect-commitment model with $\left(
Y_{j}=Y_{j}^{1}\times Y_{j}^{2},\text{ }\phi _{j}\right) _{j\in \mathcal{J}}$%
. Consider $\Xi _{j}\equiv \left\{ \left( y_{j}^{1},y_{j}^{2}\right)
:y_{j}^{1}\in Y_{j}^{1}\text{ and }y_{j}^{2}\in \phi _{j}\left(
y_{j}^{1}\right) \right\} $. In this setup, a
menu-of-menu-with-recommendation contract is represented by a non-empty
subset $E_{j}\in 2^{\Xi _{j}}\diagdown \left\{ \varnothing \right\} $, or
more precisely, the contract, $c_{j}:E_{j}\longrightarrow Y_{j}^{1}$ such
that $c_{j}\left( y_{j}^{1},y_{j}^{2}\right) \equiv y_{j}^{1}$. The
interpretation is that, the message $\left( y_{j}^{1},y_{j}^{2}\right) $
pins down $y_{j}^{1}\in Y_{j}^{1}$ and $\phi _{j}\left( y_{j}^{1}\right)
\subset Y_{j}^{2}$, and the agent recommends $y_{j}^{2}\in \phi _{j}\left(
y_{j}^{1}\right) $ for Stage 3. Let $\widetilde{C}_{j}^{R}$ denote the set
of $j$'s menu-of-menu-with-recommendation contracts, and $\widetilde{C}%
^{R}\equiv \left( \widetilde{C}_{j}^{R}\right) _{j\in \mathcal{J}}$.

Furthermore, consider $\Sigma _{j}\equiv \left\{ \left\{ \left(
y_{j}^{1},y_{j}^{2}\right) :y_{j}^{2}\in \phi _{j}\left( y_{j}^{1}\right)
\right\} :\text{ \ \ }y_{j}^{1}\in Y_{j}^{1}\right\} $. A
menu-of-menu-with-full-recommendation contract is represented by a subset $%
L_{j}\in 2^{Y_{j}^{1}}$ and an element $H_{j}\in \Sigma _{j}$, or more
precisely, the contract, $c_{j}:L_{j}\cup H_{j}\longrightarrow Y_{j}^{1}$
such that%
\begin{equation*}
c_{j}\left( y_{j}^{1}\right) =y_{j}^{1}\text{, }\forall y_{j}^{1}\in L_{j}%
\text{ and }c_{j}\left( y_{j}^{1},y_{j}^{2}\right) =y_{j}^{1}\text{, }%
\forall \left( y_{j}^{1},y_{j}^{2}\right) \in H_{j}\text{.}
\end{equation*}%
Let $\widetilde{C}_{j}^{F}$ denote the set of $j$'s
menu-of-menu-with-full-recommendation contracts, and $\widetilde{C}%
^{F}\equiv \left( \widetilde{C}_{j}^{F}\right) _{j\in \mathcal{J}}$. Then,
following a similar argument, it is straightforward to prove%
\begin{gather*}
Z^{\mathcal{E}^{\left\langle \mathcal{A},\text{ }\Gamma ,\text{ }\Psi
\right\rangle \text{-}\left[ \widetilde{C}^{\mathcal{A}},\widetilde{C}^{%
\mathcal{A}}\right] }}=Z^{\mathcal{E}^{\left\langle \mathcal{A},\text{ }%
\Gamma ,\text{ }\Psi \right\rangle \text{-}\left[ \widetilde{C}^{R},%
\widetilde{C}^{F}\right] }}\text{,} \\
\forall \left\langle \Gamma ,\text{ }\Psi \right\rangle \in \left\{
\left\langle \Gamma ^{private},\Psi ^{private}\right\rangle ,\text{ }%
\left\langle \Gamma ^{public},\Psi ^{private}\right\rangle ,\text{ }%
\left\langle \Gamma ^{public},\Psi ^{public}\right\rangle \right\} \text{.}
\end{gather*}

\section{Conclusion}

\label{sec:conclude}

While the delegation structure of a contract does not affect the set of
equilibrium outcomes in the model with a single principal, it does in
common-agency models. When a principal can offer a non-delegated contract,
the announcement and communication structures also matter because a
principal can choose his action conditional on what he observes regarding
other principals' contracts and the agent's messages to them. Therefore,
there are various common-agency models that differ in delegation,
announcement, and communication structures.

We believe that these structures are determined by regulations and laws in
practice. This leads us to identify canonical contracts for each possible
common-agency model. When the announcement and the communication structures
are both private, the menu theorem partially holds, in the sense that there
is no additional equilibrium allocation even if non-delegated contracts are
allowed. For the other three announcement and communication structures, we
show that there are additional equilibrium allocations if non-delegated
contracts are allowed. For full equilibrium analysis in these three
common-agency models with non-delegated contracts, we generalize the menu
theorem to establish the menu-of-menu-with-recommendation theorem. It
identifies two classes of canonical contracts:
menu-of-menu-with-recommendation contracts on the equilibrium path and
menu-of-menu-with-full-recommendation contracts on off-equilibrium paths.

Non-delegated contracts are related to contracts with imperfect commitment
that specify only the contractible part of an action as a function of the
agent's message (\cite{bs2000, bs2001, bs2007}), while the remaining part of
an action is non-contractible. In fact, common agency with non-delegated
contracts can be thought of as a special case of common agency with
imperfect commitment. Section \ref{sec:commitment} provides a rigorous
relationship between the two models.\ All of our analysis and full
characterization in the former model can be easily extended to the latter.

\appendix

\section{Proofs}

\subsection{Proof of Lemma \protect\ref{lem:on-path:public}}

\label{sec:lem:on-path:public}

Fix\textbf{\ }$\mathcal{A}=\mathcal{A}^{non-delegated}$ and any $%
\left\langle \Gamma ,\text{ }\Psi \right\rangle \in \left\{ \Gamma
^{private},\Gamma ^{public}\right\} \times \left\{ \Psi ^{private},\Psi
^{public}\right\} $. Fix any $\left( c,s,t\right) \in \mathcal{E}%
^{\left\langle \mathcal{A},\text{ }\Gamma ,\text{ }\Psi \right\rangle \text{-%
}\left[ C^{\mathcal{A}},C^{\mathcal{A}}\right] }$. We aim to replicate $%
\left( c,s,t\right) $ with a $\left[ C^{R},C^{\mathcal{A}}\right] $%
-equilibrium that induces $z^{\left( c,s,t\right) }$.

\underline{Replicating $c$ with $c^{\left( c,s,t\right) }\in C^{R}$:}

On the equilibrium path, for each $j\in \mathcal{J}$, offering $c_{j}$ is
equivalent to offering the menu-of-menu-with-recommendation contract, $%
c_{j}^{\left( c,s,t\right) }:M_{j}^{\left( c,s,t\right) }\longrightarrow
2^{Y_{j}}\backslash \{\varnothing \}$ with%
\begin{gather}
M_{j}^{\left( c,s,t\right) }\equiv \left\{ \left[ E_{j}=c_{j}\left[
s_{j}\left( c^{\prime },\theta \right) \right] ,\text{ }y_{j}=t_{j}\left[
\Gamma _{j}\left( c^{\prime }\right) ,\Psi _{j}\left( s\left( c^{\prime
},\theta \right) \right) \right] \right] :%
\begin{tabular}{l}
$c_{j}^{\prime }=c_{j}$, \\ 
$\left( c_{-j}^{\prime },\theta \right) \in C_{-j}^{\mathcal{A}}\times
\Theta $%
\end{tabular}%
\right\} \text{,}  \label{tit1} \\
c_{j}^{\left( c,s,t\right) }\left[ E_{j},\text{ }y_{j}\right] =E_{j}\text{, }%
\forall \left[ E_{j},\text{ }y_{j}\right] \in M_{j}^{\left( c,s,t\right) }%
\text{.}  \notag
\end{gather}%
Given $\left( \left( c_{j},c_{-j}^{\prime }\right) ,\theta \right) \in C^{%
\mathcal{A}}\times \Theta $, if all players follow $\left( s,t\right) $, the
subset $E_{j}=c_{j}\left[ s_{j}\left( \left( c_{j},c_{-j}^{\prime }\right)
,\theta \right) \right] $ is fixed for $j$ at Stage 2, and $j$ takes the
action $y_{j}=t_{j}\left[ \Gamma _{j}\left( \left( c_{j},c_{-j}^{\prime
}\right) \right) ,\Psi _{j}\left( s\left( \left( c_{j},c_{-j}^{\prime
}\right) ,\theta \right) \right) \right] $ at Stage 3. $M_{j}^{\left(
c,s,t\right) }$ is the set of all such profiles. Define $c^{\left(
c,s,t\right) }\equiv \left( c_{k}^{\left( c,s,t\right) }\right) _{k\in 
\mathcal{J}}\in C^{R}$, and let 
\begin{equation*}
\widehat{C}\equiv \times _{k\in \mathcal{J}}\widehat{C}_{k}\equiv \times
_{k\in \mathcal{J}}\left( \left\{ c_{k}^{\left( c,s,t\right) }\right\} \cup
C_{k}^{\mathcal{A}}\right) \text{ and }\widehat{M}\equiv \times _{k\in 
\mathcal{J}}\widehat{M}_{k}\equiv \times _{k\in \mathcal{J}}\left(
M_{k}^{\left( c,s,t\right) }\cup M_{k}^{\mathcal{A}}\right)
\end{equation*}%
denote the relevant contract space and the relevant message space
respectively in the $\left[ C^{R},C^{\mathcal{A}}\right] $-game. In the $%
\left[ C^{\mathcal{A}},C^{\mathcal{A}}\right] $-game, $C^{\mathcal{A}}$ and $%
M^{\mathcal{A}}$ are the relevant contract space and message space,
respectively.

\underline{Replicating $s\equiv \left( s_{k}\right) _{k\in \mathcal{J}}$
with $s^{\left( c,s,t\right) }\equiv \left( s_{k}^{\left( c,s,t\right)
}\right) _{k\in \mathcal{J}}$:}

When the agent observes $c_{k}^{\left( c,s,t\right) }$ in the $\left[
C^{R},C^{\mathcal{A}}\right] $-game, he interprets it as $c_{k}$ in the $%
\left[ C^{\mathcal{A}},C^{\mathcal{A}}\right] $-game, due to the replication
process above. Also, when the agent observes $c_{j}^{\prime }\in C_{j}^{%
\mathcal{A}}$ in the $\left[ C^{R},C^{\mathcal{A}}\right] $-game, he
interprets it as $c_{j}^{\prime }\in C_{j}^{\mathcal{A}}$ in the $\left[ C^{%
\mathcal{A}},C^{\mathcal{A}}\right] $-game. To record this interpretation,
define $\gamma _{j}:\widehat{C}_{j}\longrightarrow C_{j}^{\mathcal{A}}$ for
each $j\in \mathcal{J}$ as%
\begin{equation}
\gamma _{j}\left( \widehat{c}_{j}\right) \equiv \left\{ 
\begin{tabular}{ll}
$c_{j}$, & if $\widehat{c}_{j}=c_{j}^{\left( c,s,t\right) }$; \\ 
$\widehat{c}_{j}$, & if $\widehat{c}_{j}\in C_{j}^{\mathcal{A}}$,%
\end{tabular}%
\right.  \label{tit2}
\end{equation}%
and denote $\gamma \left( \widehat{c}\right) \equiv \left( \left[ \gamma
_{k}\left( \widehat{c}_{k}\right) \right] _{k\in \mathcal{J}}\right) \in C^{%
\mathcal{A}}$ for all $\widehat{c}=\left( \widehat{c}_{k}\right) _{k\in 
\mathcal{J}}\in \widehat{C}$.

For the agent's strategy in the $\left[ C^{R},C^{\mathcal{A}}\right] $-game,
we replicate $s\equiv \left( s_{k}\right) _{k\in \mathcal{J}}$ with $%
s^{\left( c,s,t\right) }\equiv \left( s_{k}^{\left( c,s,t\right) }\right)
_{k\in \mathcal{J}}$ defined as follows. For each $j\in \mathcal{J}$ and all 
$\left( \widehat{c},\theta \right) \in \widehat{C}\times \Theta $,%
\begin{equation*}
s_{j}^{\left( c,s,t\right) }\left( \widehat{c},\theta \right) =\left\{ 
\begin{tabular}{ll}
$s_{j}\left[ \gamma \left( \widehat{c}\right) ,\text{ }\theta \right] $, & 
if $\widehat{c}_{j}\in C_{j}^{\mathcal{A}}$; \\ 
$\left[ E_{j}=c_{j}\left( s_{j}\left[ \gamma \left( \widehat{c}\right) ,%
\text{ }\theta \right] \right) ,\text{ }y_{j}=t_{j}\left( \Gamma _{j}\left(
\gamma \left( \widehat{c}\right) \right) ,\Psi _{j}\left[ s\left( \gamma
\left( \widehat{c}\right) ,\theta \right) \right] \right) \right] $, & if $%
\widehat{c}_{j}=c_{j}^{\left( c,s,t\right) }$.%
\end{tabular}%
\right.
\end{equation*}%
I.e., when principals offer $\widehat{c}$ in the $\left[ C^{R},C^{\mathcal{A}%
}\right] $-game, the agent regards it as $\gamma \left( \widehat{c}\right) $
in the $\left[ C^{\mathcal{A}},C^{\mathcal{A}}\right] $-game. Given $\gamma
\left( \widehat{c}\right) $ in the $\left[ C^{\mathcal{A}},C^{\mathcal{A}}%
\right] $-game, the agent sends $s_{j}\left[ \gamma \left( \widehat{c}%
\right) ,\text{ }\theta \right] $ to $j$ at Stage 2, which pins down the
subset $c_{j}\left( s_{j}\left[ \gamma \left( \widehat{c}\right) ,\text{ }%
\theta \right] \right) $ for $j$, and $j$ takes the action $t_{j}\left(
\Gamma _{j}\left( \gamma \left( \widehat{c}\right) \right) ,\Psi _{j}\left[
s\left( \gamma \left( \widehat{c}\right) ,\theta \right) \right] \right) $
at Stage 3. Then, in the $\left[ C^{R},C^{\mathcal{A}}\right] $-game, $%
s_{j}^{\left( c,s,t\right) }$ replicates $s_{j}$: if $\widehat{c}_{j}\in
C_{j}^{\mathcal{A}}$, the agent sends $s_{j}\left[ \gamma \left( \widehat{c}%
\right) ,\text{ }\theta \right] $ to $j$; if $\widehat{c}_{j}=c_{j}^{\left(
c,s,t\right) }\in C_{j}^{R}$, the agent chooses the subset $c_{j}\left( s_{j}%
\left[ \gamma \left( \widehat{c}\right) ,\text{ }\theta \right] \right) $
with the recommendation $t_{j}\left( \Gamma _{j}\left( \gamma \left( 
\widehat{c}\right) \right) ,\Psi _{j}\left[ s\left( \gamma \left( \widehat{c}%
\right) ,\theta \right) \right] \right) $.

\underline{Replicating $\left( b_{j},t_{j}\right) $ with $\left(
b_{j}^{\left( c,s,t\right) },t_{j}^{\left( c,s,t\right) }\right) $:}

We aim to replicate an equilibrium in the $\left[ C^{\mathcal{A}},C^{%
\mathcal{A}}\right] $-game with an equilibrium in the $\left[ C^{R},C^{%
\mathcal{A}}\right] $-game. $\gamma \equiv \left( \gamma _{k}\right) _{k\in 
\mathcal{J}}$ defined in (\ref{tit2}) describes how players translate
contracts in the the $\left[ C^{R},C^{\mathcal{A}}\right] $-game to
contracts in the $\left[ C^{\mathcal{A}},C^{\mathcal{A}}\right] $-game. We
still need to define how players translate messages in the the $\left[
C^{R},C^{\mathcal{A}}\right] $-game to messages in the $\left[ C^{\mathcal{A}%
},C^{\mathcal{A}}\right] $-game.

By (\ref{tit1}), there exists a surjective function $\zeta _{j}:C_{-j}^{%
\mathcal{A}}\times \Theta \longrightarrow M_{j}^{\left( c,s,t\right) }$ such
that%
\begin{equation*}
\zeta _{j}\left( c_{-j}^{\prime },\theta \right) =\left[ E_{j}=c_{j}\left[
s_{j}\left( \left( c_{j},c_{-j}^{\prime }\right) ,\theta \right) \right] ,%
\text{ }y_{j}=t_{j}\left[ \Gamma _{j}\left( \left( c_{j},c_{-j}^{\prime
}\right) \right) ,\Psi _{j}\left( s\left( \left( c_{j},c_{-j}^{\prime
}\right) ,\theta \right) \right) \right] \right] \text{,}
\end{equation*}%
i.e., upon observing $\left( \left( c_{j},c_{-j}^{\prime }\right) ,\theta
\right) $, by following $s$, the agent's message would pin down the subset $%
E_{j}=c_{j}\left[ s\left( \left( c_{j},c_{-j}^{\prime }\right) ,\theta
\right) \right] \subset Y_{j}$ at Stage 2, and by following $t_{j}$,
principal $j$ would take the action $y_{j}=t_{j}\left[ \Gamma _{j}\left(
\left( c_{j},c_{-j}^{\prime }\right) \right) ,\Psi _{j}\left( s\left( \left(
c_{j},c_{-j}^{\prime }\right) ,\theta \right) \right) \right] $ at Stage 3.
-- $\zeta \left( c_{-j}^{\prime },\theta \right) $ records this profile, $%
\left[ E_{j},\text{ }y_{j}\right] $.

Fix any injective $\zeta _{j}^{-1}:M_{j}^{\left( c,s,t\right)
}\longrightarrow C_{-j}^{\mathcal{A}}\times \Theta $ such that%
\begin{equation*}
\zeta _{j}\left[ \zeta _{j}^{-1}\left( m_{j}\right) \right] =m_{j}\text{, }%
\forall m_{j}\in M_{j}^{\left( c,s,t\right) }\text{,}
\end{equation*}%
i.e., each $m_{j}\in M_{j}^{\left( c,s,t\right) }$ is mapped to some $\left(
c_{-j}^{\prime },\theta \right) $ such that $\zeta _{j}\left( c_{-j}^{\prime
},\theta \right) =m_{j}$. That is, upon observing $m_{j}\in M_{j}^{\left(
c,s,t\right) }$ in the $\left[ C^{R},C^{\mathcal{A}}\right] $-game,
principals interpret it as the message $s_{j}\left[ c_{j},\zeta
_{j}^{-1}\left( m_{j}\right) \right] $ in the $\left[ C^{\mathcal{A}},C^{%
\mathcal{A}}\right] $-game.

Given $\widehat{c}\in \widehat{C}$ offered at Stage 1, consider%
\begin{equation*}
M_{j}^{\widehat{c}}\equiv \left\{ m_{j}\in \widehat{M}_{j}:%
\begin{tabular}{l}
$\widehat{c}_{j}\in C_{j}^{\mathcal{A}}\Longrightarrow m_{j}\in M_{j}^{%
\mathcal{A}}$, \\ 
$\widehat{c}_{j}=c_{j}^{\left( c,s,t\right) }\Longrightarrow m_{j}\in
M_{j}^{\left( c,s,t\right) }$%
\end{tabular}%
\right\} \text{ and }M^{\widehat{c}}\equiv \left( M_{j}^{\widehat{c}}\right)
_{j\in \mathcal{J}}\text{.}
\end{equation*}%
i.e., $M^{\widehat{c}}$ is the set of message profiles that could be sent by
the agent at Stage 2 in the $\left[ C^{R},C^{\mathcal{A}}\right] $-game.
Given $\widehat{c}\in \widehat{C}$ offered at Stage 1, the function $\tau ^{%
\widehat{c}}\equiv \left( \tau _{j}^{\widehat{c}}:M_{j}^{\widehat{c}%
}\longrightarrow M_{j}^{\mathcal{A}}\right) _{j\in \mathcal{J}}$ describes
how the players translate messages in the $\left[ C^{R},C^{\mathcal{A}}%
\right] $-game to the messages in the $\left[ C^{\mathcal{A}},C^{\mathcal{A}}%
\right] $-game.

\begin{equation*}
\tau _{j}^{\widehat{c}}\left( m_{j}\right) \equiv \left\{ 
\begin{tabular}{ll}
$m_{j}$, & if $\widehat{c}_{j}\in C_{j}^{\mathcal{A}}$; \\ 
$s_{j}\left( c_{j},\zeta ^{-1}\left( m_{j}\right) \right) $, & if $\widehat{c%
}_{j}=c_{j}^{\left( c,s,t\right) }$,%
\end{tabular}%
\right. \text{, }\forall j\in \mathcal{J}\text{,}
\end{equation*}%
i.e., the players re-label $m_{j}$ in the $\left[ C^{R},C^{\mathcal{A}}%
\right] $-game if and only if $\widehat{c}_{j}=c_{j}^{\left( c,s,t\right) }$%
, and when $\widehat{c}_{j}=c_{j}^{\left( c,s,t\right) }$, a message $m_{j}$
in the $\left[ C^{R},C^{\mathcal{A}}\right] $-game is interpreted as $%
s_{j}\left( c_{j},\zeta _{j}^{-1}\left( m_{j}\right) \right) $ in the $\left[
C^{\mathcal{A}},C^{\mathcal{A}}\right] $-game.

We are now ready to replicate $t_{j}$ with $t_{j}^{\left( c,s,t\right) }$.%
\begin{equation*}
t_{j}^{\left( c,s,t\right) }\left[ \Gamma _{j}\left( \widehat{c}\right)
,\Psi _{j}\left( \widehat{m}\right) \right] \equiv t_{j}\left[ \Gamma
_{j}\left( \gamma \left( \widehat{c}\right) \right) ,\Psi _{j}\left( \tau ^{%
\widehat{c}}\left( \widehat{m}\right) \right) \right] \text{, }\forall
\left( \widehat{c},\widehat{m}\right) \in \widehat{C}\times \widehat{M}\text{%
,}
\end{equation*}%
i.e., players translate the profile $\left( \widehat{c},\widehat{m}\right)
\in \widehat{C}\times \widehat{M}$ in the $\left[ C^{R},C^{\mathcal{A}}%
\right] $-game to the profile $\left( \gamma \left( \widehat{c}\right) ,\tau
^{\widehat{c}}\left( \widehat{m}\right) \right) \in C^{\mathcal{A}}\times M^{%
\mathcal{A}}$, and $t_{j}^{\left( c,s,t\right) }$ replicates $t_{j}$.

Similarly, we replicate $b_{j}$ with $b_{j}^{\left( c,s,t\right) }$, subject
to the translation of $\left( \gamma ,\tau ^{\widehat{c}}\right) $.
Rigorously, consider%
\begin{equation*}
Q_{j}\equiv \left\{ \left[ \Gamma _{j}\left( \widehat{c}\right) ,\Psi
_{j}\left( \widehat{m}\right) \right] :\widehat{c}\in \widehat{C}\text{ and }%
\widehat{m}\in \widehat{M}\right\} \text{,}
\end{equation*}%
\begin{equation*}
Q_{j}^{\ast }\equiv \left\{ \left[ \Gamma _{j}\left( \widehat{c}\right)
,\Psi _{j}\left( \widehat{m}\right) \right] \in Q_{j}:\widehat{c}%
_{-j}=c_{-j}^{\left( c,s,t\right) }\text{ and }\widehat{m}=s^{\left(
c,s,t\right) }\left[ \widehat{c},\theta \right] \text{ for some }\theta \in
\Theta \right\} \text{.}
\end{equation*}%
i.e., $Q_{j}$ is the set of all possible information that principal $j$ may
observe before $j$ takes an action at Stage 3; $Q_{j}^{\ast }$ is the subset
of information by which $j$ cannot confirm that the other players have
deviated.

When principal $j$ observes $q_{j}=\left[ \Gamma _{j}\left( \widehat{c}%
_{j},c_{-j}^{\left( c,s,t\right) }\right) ,\Psi _{j}\left( s^{\left(
c,s,t\right) }\left[ \left( \widehat{c}_{j},c_{-j}^{\left( c,s,t\right)
}\right) ,\theta \right] \right) \right] \in Q_{j}^{\ast }$ for some $\theta
\in \Theta $, $j$'s belief is induced by Bayes' rule, i.e.,%
\begin{equation*}
b_{j}^{\left( c,s,t\right) }\left( q_{j}\right) \left[ \left\{ \left[ \left( 
\widehat{c}_{j},c_{-j}^{\left( c,s,t\right) }\right) ,\text{ }s^{\left(
c,s,t\right) }\left[ \left( \widehat{c}_{j},c_{-j}^{\left( c,s,t\right)
}\right) ,\theta ^{\prime }\right] ,\text{ }\theta ^{\prime }\right]
\right\} \right] \equiv \left\{ 
\begin{array}{cc}
\frac{p\left( \theta ^{\prime }\right) }{p\left[ \Upsilon \left( \theta
\right) \right] } & \text{if }\theta ^{\prime }\in \Upsilon \left( \theta
\right) ; \\ 
&  \\ 
0 & \text{otherwise}%
\end{array}%
\right. \text{,}
\end{equation*}%
where%
\begin{equation*}
\Upsilon \left( \theta \right) \equiv \left\{ \widetilde{\theta }\in \Theta
:\Psi _{j}\left( s^{\left( c,s,t\right) }\left[ \left( \widehat{c}%
_{j},c_{-j}^{\left( c,s,t\right) }\right) ,\widetilde{\theta }\right]
\right) =\Psi _{j}\left( s^{\left( c,s,t\right) }\left[ \left( \widehat{c}%
_{j},c_{-j}^{\left( c,s,t\right) }\right) ,\theta \right] \right) \right\} 
\text{.}
\end{equation*}%
When principal $j$ observes $q_{j}=\left[ \Gamma _{j}\left( \widehat{c}%
\right) ,\Psi _{j}\left( \widehat{m}\right) \right] \in Q_{j}\diagdown
Q_{j}^{\ast }$, define%
\begin{gather*}
b_{j}^{\left( c,s,t\right) }\left( q_{j}\right) \left[ \left\{ \widehat{c}%
\right\} \times \widehat{M}\times \Theta \right] =1\text{,} \\
b_{j}^{\left( c,s,t\right) }\left[ \Gamma _{j}\left( \widehat{c}\right)
,\Psi _{j}\left( \widehat{m}\right) \right] \left[ \left\{ \widehat{c}%
\right\} \times E\times \left\{ \theta ^{\prime }\right\} \right] \\
\equiv b_{j}\left[ \Gamma _{j}\left( \gamma \left( \widehat{c}\right)
\right) ,\Psi _{j}\left( \tau ^{\widehat{c}}\left( \widehat{m}\right)
\right) \right] \left[ \left\{ \gamma \left( \widehat{c}\right) \right\}
\times \tau ^{\widehat{c}}\left( E\right) \times \left\{ \theta ^{\prime
}\right\} \right] \text{, }\forall E\in 2^{\widehat{M}}\text{.}
\end{gather*}

It is straightforward to see that $\left( c,s,t\right) $ is replicated by $%
\left( c^{\left( c,s,t\right) },s^{\left( c,s,t\right) },t^{\left(
c,s,t\right) }\right) $, and all the players inherit incentive compatibility
from $\left( c,s,t\right) $, i.e., $\left( c^{\left( c,s,t\right)
},s^{\left( c,s,t\right) },t^{\left( c,s,t\right) }\right) $ is a $\left[
C^{R},C^{\mathcal{A}}\right] $-equilibrium and $z^{\left( c,s,t\right)
}=z^{\left( c^{\left( c,s,t\right) },s^{\left( c,s,t\right) },t^{\left(
c,s,t\right) }\right) }$.$\blacksquare $

\subsection{Proof of Lemma \protect\ref{lem:on-path:private-private}}

\label{sec:lem:on-path:private-private}

Fix\textbf{\ }$\left\langle \mathcal{A},\text{ }\Gamma ,\text{ }\Psi
\right\rangle =\left\langle \mathcal{A}^{non-delegated},\text{ }\Gamma
^{private},\text{ }\Psi ^{private}\right\rangle $. Fix any $\left(
c,s,t\right) \in \mathcal{E}^{\left\langle \mathcal{A},\text{ }\Gamma ,\text{
}\Psi \right\rangle \text{-}\left[ C^{\mathcal{A}},C^{\mathcal{A}}\right] }$%
. We aim to replicate $\left( c,s,t\right) $ with a $\left[ C^{P},C^{%
\mathcal{A}}\right] $-equilibrium that induces the allocation $z^{\left(
c,s,t\right) }$.

\underline{Replicating $c$ with $c^{\left( c,s,t\right) }\in C^{P}$:}

On the equilibrium path, for each $j\in \mathcal{J}$, offering $c_{j}$ is
equivalent to offering the menu contract, $c_{j}^{\left( c,s,t\right)
}:M_{j}^{\left( c,s,t\right) }\longrightarrow Y_{j}$ with%
\begin{gather*}
M_{j}^{\left( c,s,t\right) }\equiv \left\{ t_{j}\left[ \Gamma _{j}\left(
\left( c_{j},c_{-j}^{\prime }\right) \right) ,\Psi _{j}\left( s\left( \left(
c_{j},c_{-j}^{\prime }\right) ,\theta \right) \right) \right] :\left(
c_{-j}^{\prime },\theta \right) \in C_{-j}^{\mathcal{A}}\times \Theta
\right\} \text{,} \\
c_{j}^{\left( c,s,t\right) }\left( y_{j}\right) =y_{j}\text{, }\forall
y_{j}\in M_{j}^{\left( c,s,t\right) }\text{.}
\end{gather*}%
Given $\left( \left( c_{j},c_{-j}^{\prime }\right) ,\theta \right) \in C^{%
\mathcal{A}}\times \Theta $, if all players follow $\left( s,t\right) $, the
subset $E_{j}=c_{j}\left[ s_{j}\left( \left( c_{j},c_{-j}^{\prime }\right)
,\theta \right) \right] $ is pinned down for $j$ at Stage 2, and $j$ takes
the action $y_{j}=t_{j}\left[ \Gamma _{j}\left( \left( c_{j},c_{-j}^{\prime
}\right) \right) ,\Psi _{j}\left( s\left( \left( c_{j},c_{-j}^{\prime
}\right) ,\theta \right) \right) \right] $ at Stage 3. $M_{j}^{\left(
c,s,t\right) }$ is the set of all such $y_{j}=t_{j}\left[ \Gamma _{j}\left(
\left( c_{j},c_{-j}^{\prime }\right) \right) ,\Psi _{j}\left( s\left( \left(
c_{j},c_{-j}^{\prime }\right) ,\theta \right) \right) \right] $.

In the $\left[ C^{\mathcal{A}},C^{\mathcal{A}}\right] $-game, $C^{\mathcal{A}%
}$ and $M^{\mathcal{A}}$ are the contract space and the message space,
respectively. Define $c^{\left( c,s,t\right) }\equiv \left( c_{k}^{\left(
c,s,t\right) }\right) _{k\in \mathcal{J}}\in C^{P}$ and let%
\begin{equation*}
\widehat{C}\equiv \times _{k\in \mathcal{J}}\widehat{C}_{k}\equiv \times
_{k\in \mathcal{J}}\left( \left\{ c_{k}^{\left( c,s,t\right) }\right\} \cup
C_{k}^{\mathcal{A}}\right) \text{ and }\widehat{M}\equiv \times _{k\in 
\mathcal{J}}\widehat{M}_{k}\equiv \times _{k\in \mathcal{J}}\left(
M_{k}^{\left( c,s,t\right) }\cup M_{k}^{\mathcal{A}}\right)
\end{equation*}%
denote the relevant contract space and message space in the $\left[ C^{P},C^{%
\mathcal{A}}\right] $-game.

\underline{Replicating $s\equiv \left( s_{k}\right) _{k\in \mathcal{J}}$
with $s^{\left( c,s,t\right) }\equiv \left( s_{k}^{\left( c,s,t\right)
}\right) _{k\in \mathcal{J}}$:}

When the agent observes $c_{k}^{\left( c,s,t\right) }$ in the $\left[
C^{P},C^{\mathcal{A}}\right] $-game, he interprets it as $c_{k}$ in the $%
\left[ C^{\mathcal{A}},C^{\mathcal{A}}\right] $-game, due to the replication
process above. Also, when the agent observes $c_{j}^{\prime }\in C_{j}^{%
\mathcal{A}}$ in the $\left[ C^{P},C^{\mathcal{A}}\right] $-game, he
interprets it as $c_{j}^{\prime }\in C_{j}^{\mathcal{A}}$ in the $\left[ C^{%
\mathcal{A}},C^{\mathcal{A}}\right] $-game. Such interpretations of
contracts in $\widehat{C}$ is denoted by $\gamma _{j}:\widehat{C}%
_{j}\longrightarrow C_{j}^{\mathcal{A}}$ for each $j\in \mathcal{J}$:%
\begin{equation*}
\gamma _{j}\left( \widehat{c}_{j}\right) \equiv \left\{ 
\begin{tabular}{ll}
$c_{j}$, & if $\widehat{c}_{j}=c_{j}^{\left( c,s,t\right) }$; \\ 
$\widehat{c}_{j}$, & if $\widehat{c}_{j}\in C_{j}^{\mathcal{A}}$.%
\end{tabular}%
\right.
\end{equation*}%
Let $\gamma \left( \widehat{c}\right) \equiv \left( \left[ \gamma _{k}\left( 
\widehat{c}_{k}\right) \right] _{k\in \mathcal{J}}\right) \in C^{\mathcal{A}%
} $ for all $\widehat{c}=\left( \widehat{c}_{k}\right) _{k\in \mathcal{J}%
}\in \widehat{C}$.

For the agent's strategy in the $\left[ C^{P},C^{\mathcal{A}}\right] $-game,
we replicate $s\equiv \left( s_{k}\right) _{k\in \mathcal{J}}$ with $%
s^{\left( c,s,t\right) }\equiv \left( s_{k}^{\left( c,s,t\right) }\right)
_{k\in \mathcal{J}}$ defined as follows. For each $j\in \mathcal{J}$ and
each $\left( \widehat{c},\theta \right) \in \widehat{C}\times \Theta $,%
\begin{equation*}
s_{j}^{\left( c,s,t\right) }\left( \widehat{c},\theta \right) =\left\{ 
\begin{tabular}{ll}
$s_{j}\left[ \gamma \left( \widehat{c}\right) ,\text{ }\theta \right] $, & 
if $\widehat{c}_{j}\in C_{j}^{\mathcal{A}}$; \\ 
$t_{j}\left( \Gamma _{j}\left( \gamma \left( \widehat{c}\right) \right)
,\Psi _{j}\left[ s\left( \gamma \left( \widehat{c}\right) ,\theta \right) %
\right] \right) $, & if $\widehat{c}_{j}=c_{j}^{\left( c,s,t\right) }$.%
\end{tabular}%
\right.
\end{equation*}%
When principals offer $\widehat{c}$ at Stage 1 in the $\left[ C^{P},C^{%
\mathcal{A}}\right] $-game, the agent regards it as $\gamma \left( \widehat{c%
}\right) $ in the $\left[ C^{\mathcal{A}},C^{\mathcal{A}}\right] $-game.
Given $\gamma \left( \widehat{c}\right) $ in the $\left[ C^{\mathcal{A}},C^{%
\mathcal{A}}\right] $-game, the agent sends $s_{j}\left[ \gamma \left( 
\widehat{c}\right) ,\text{ }\theta \right] $ to $j$ at Stage 2, and at Stage
3, $j$ would choose the action $t_{j}\left( \Gamma _{j}\left( \gamma \left( 
\widehat{c}\right) \right) ,\Psi _{j}\left[ s\left( \gamma \left( \widehat{c}%
\right) ,\theta \right) \right] \right) $. Then, in the $\left[ C^{P},C^{%
\mathcal{A}}\right] $-game, $s_{j}^{\left( c,s,t\right) }$ replicates $s_{j}$%
: if $\widehat{c}_{j}\in C_{j}^{\mathcal{A}}$, the agent sends $s_{j}\left[
\gamma \left( \widehat{c}\right) ,\text{ }\theta \right] $ to $j$; if $%
\widehat{c}_{j}=c_{j}^{\left( c,s,t\right) }\in C_{j}^{P}$, the agent
chooses the action $t_{j}\left( \Gamma _{j}\left( \gamma \left( \widehat{c}%
\right) \right) ,\Psi _{j}\left[ s\left( \gamma \left( \widehat{c}\right)
,\theta \right) \right] \right) $ from the menu contract $c_{j}^{\left(
c,s,t\right) }$.

\underline{Replicating $\left( b_{j},t_{j}\right) $ with $\left(
b_{j}^{\left( c,s,t\right) },t_{j}^{\left( c,s,t\right) }\right) $:}

For each $j\in \mathcal{J}$, consider%
\begin{eqnarray*}
Q_{j} &\equiv &\left\{ \left[ \Gamma _{j}\left( \widehat{c}\right) ,\Psi
_{j}\left( \widehat{m}\right) \right] :\widehat{c}\in \widehat{C}\text{ and }%
\widehat{m}\in \widehat{M}\right\} \text{, }Q_{j}^{\ast }\equiv \left\{ %
\left[ \Gamma _{j}\left( \widehat{c}\right) ,\Psi _{j}\left( \widehat{m}%
\right) \right] \in Q_{j}:\widehat{c}_{j}=c_{j}^{\left( c,s,t\right)
}\right\} , \\
Q_{j}^{\ast \ast } &\equiv &\left\{ \left[ \Gamma _{j}\left( \widehat{c}%
\right) ,\Psi _{j}\left( s^{\left( c,s,t\right) }\left[ \widehat{c},\theta %
\right] \right) \right] \in Q_{j}:\widehat{c}_{j}\in C_{j}^{\mathcal{A}}%
\text{, }\widehat{c}_{-j}=c_{-j}^{\left( c,s,t\right) }\text{ and }\theta
\in \Theta \right\} ,
\end{eqnarray*}%
i.e., in the $\left[ C^{P},C^{\mathcal{A}}\right] $-game, $Q_{j}$ is the set
of all possible information that principal $j$ may observe before $j$ takes
an action at Stage 3; $Q_{j}^{\ast }$ is the subset of information with
which principal $j$ offers the menu contract $c_{j}^{\left( c,s,t\right) }$; 
$Q_{j}^{\ast \ast }$ is the subset of information with which principal $j$
offers $\widehat{c}_{j}\in C_{j}^{\mathcal{A}}$ and cannot confirm that the
other players have deviated.

First, when principal $j$ observes $\left[ \Gamma _{j}\left( \widehat{c}%
\right) ,\Psi _{j}\left( \widehat{m}\right) \right] \in Q_{j}^{\ast }$, we
have $\widehat{c}_{j}=c_{j}^{\left( c,s,t\right) }$, which is a menu
contract (i.e., a delegated contract). As a result, $j$'s decision at Stage
3 is degenerate and the belief is irrelevant. Second, when principal $j$
observes $\left[ \Gamma _{j}\left( \widehat{c}\right) ,\Psi _{j}\left( 
\widehat{m}\right) \right] \in Q_{j}\diagdown \left( Q_{j}^{\ast }\cup
Q_{j}^{\ast \ast }\right) $, principal $j$ is on an off-equilibrium path and 
$\widehat{c}_{j}\in C_{j}^{\mathcal{A}}$. Define%
\begin{eqnarray*}
t_{j}^{\left( c,s,t\right) }\left[ \Gamma _{j}\left( \widehat{c}\right)
,\Psi _{j}\left( \widehat{m}\right) \right] &\equiv &t_{j}\left[ \Gamma
_{j}\left( \widehat{c}\right) ,\Psi _{j}\left( \widehat{m}\right) \right] 
\text{,} \\
b_{j}^{\left( c,s,t\right) }\left[ \Gamma _{j}\left( \widehat{c}\right)
,\Psi _{j}\left( \widehat{m}\right) \right] &\equiv &b_{j}\left[ \Gamma
_{j}\left( \widehat{c}\right) ,\Psi _{j}\left( \widehat{m}\right) \right] 
\text{,}
\end{eqnarray*}%
i.e., $\left( b_{j}^{\left( c,s,t\right) },t_{j}^{\left( c,s,t\right)
}\right) $ copies $\left( b_{j},t_{j}\right) $.

Finally, when principal $j$ observes $q_{j}=\left[ \Gamma _{j}\left( 
\widehat{c}_{j},c_{-j}^{\left( c,s,t\right) }\right) ,\Psi _{j}\left(
s^{\left( c,s,t\right) }\left[ \left( \widehat{c}_{j},c_{-j}^{\left(
c,s,t\right) }\right) ,\theta \right] \right) \right] \in Q_{j}^{\ast \ast }$%
, i.e., $j$ cannot confirm that the other players have deviated. Define 
\begin{equation*}
t_{j}^{\left( c,s,t\right) }\left( q_{j}\right) \equiv t_{j}\left[ \Gamma
_{j}\left( \widehat{c}_{j},c_{-j}\right) ,\Psi _{j}\left( s\left[ \left( 
\widehat{c}_{j},c_{-j}\right) ,\theta \right] \right) \right] \text{,}
\end{equation*}%
\begin{equation*}
b_{j}^{\left( c,s,t\right) }\left( q_{j}\right) \left[ \left\{ \left( 
\widehat{c}_{j},c_{-j}^{\left( c,s,t\right) }\right) \right\} \times 
\widehat{M}\times \Theta \right] =1\text{ and}
\end{equation*}%
\begin{equation*}
b_{j}^{\left( c,s,t\right) }\left( q_{j}\right) \left[ \left\{ \left[ \left( 
\widehat{c}_{j},c_{-j}^{\left( c,s,t\right) }\right) ,\text{ }s^{\left(
c,s,t\right) }\left[ \left( \widehat{c}_{j},c_{-j}^{\left( c,s,t\right)
}\right) ,\theta ^{\prime }\right] ,\text{ }\theta ^{\prime }\right]
\right\} \right] \equiv \left\{ 
\begin{array}{cc}
\frac{p\left( \theta ^{\prime }\right) }{p\left[ \Upsilon \left( \theta
\right) \right] } & \text{if }\theta ^{\prime }\in \Upsilon \left( \theta
\right) ; \\ 
&  \\ 
0 & \text{otherwise}%
\end{array}%
\right. \text{,}
\end{equation*}%
\begin{equation*}
\text{where }\Upsilon \left( \theta \right) \equiv \left\{ \widetilde{\theta 
}\in \Theta :\Psi _{j}\left( s^{\left( c,s,t\right) }\left[ \left( \widehat{c%
}_{j},c_{-j}^{\left( c,s,t\right) }\right) ,\widetilde{\theta }\right]
\right) =\Psi _{j}\left( s^{\left( c,s,t\right) }\left[ \left( \widehat{c}%
_{j},c_{-j}^{\left( c,s,t\right) }\right) ,\theta \right] \right) \right\} 
\text{,}
\end{equation*}%
i.e., $b_{j}^{\left( c,s,t\right) }\left( q_{j}\right) $ believes in $\left( 
\widehat{c}_{j},c_{-j}^{\left( c,s,t\right) }\right) $ with probability 1
and its belief on $\widehat{M}\times \Theta $ is derived by Bayes' rule.

It is straightforward to see that $\left( c,s,t\right) $ is replicated by $%
\left( c^{\left( c,s,t\right) },s^{\left( c,s,t\right) },t^{\left(
c,s,t\right) }\right) $, and all the players inherit incentive compatibility
from $\left( c,s,t\right) $, i.e., $\left( c^{\left( c,s,t\right)
},s^{\left( c,s,t\right) },t^{\left( c,s,t\right) }\right) $ is a $\left[
C^{P},C^{\mathcal{A}}\right] $-equilibrium and $z^{\left( c,s,t\right)
}=z^{\left( c^{\left( c,s,t\right) },s^{\left( c,s,t\right) },t^{\left(
c,s,t\right) }\right) }$.$\blacksquare $

\subsection{\qquad Proof of Lemma \protect\ref{lem:extension:on}}

\label{sec:lem:extension:on}

Let $\mathcal{A=A}^{non-delegated}$. Fix any $I,II,III\in \left\{ \mathcal{A}%
,P,F,F^{\ast },R\right\} $ such that $C^{I}\sqsupset ^{\ast }C^{III}$. Let $%
M^{I}$, $M^{II}$ and $M^{III}$ denote the message spaces for $C^{I}$, $%
C^{II} $ and $C^{III}$, respectively. Fix any $\left( c,s,t\right) \in 
\mathcal{E}^{\left\langle \mathcal{A},\text{ }\Gamma ,\text{ }\Psi
\right\rangle \text{-}\left[ C^{III},C^{II}\right] }$. We aim to construct a 
$\left[ C^{I},C^{II}\right] $-equilibrium that induces the allocation $%
z^{\left( c,s,t\right) }$.

Since $C^{I}\sqsupset ^{\ast }C^{III}$, there exits $c^{\left( c,s,t\right)
}\in C^{I}$ such that $c_{j}^{\left( c,s,t\right) }\geq c_{j}$ for every $%
j\in \mathcal{J}$, and as a result, there exists a surjective function $%
\iota _{j}:M_{j}^{I}\longrightarrow M_{j}^{III}$ such that%
\begin{equation*}
c_{j}^{\left( c,s,t\right) }\left( m_{j}\right) =c_{j}\left( \iota
_{j}\left( m_{j}\right) \right) \text{, }\forall m_{j}\in M_{j}^{I}\text{.}
\end{equation*}%
We will replicate $\left( c,s,t\right) $ with $\left( c^{\left( c,s,t\right)
},s^{\left( c,s,t\right) },t^{\left( c,s,t\right) }\right) \in \mathcal{E}%
^{\left\langle \mathcal{A},\text{ }\Gamma ,\text{ }\Psi \right\rangle \text{-%
}\left[ C^{I},C^{II}\right] }$ in two steps. First, with slight abuse of
notation, let $\iota _{j}^{-1}:M_{j}^{III}\longrightarrow M_{j}^{I}$ denote
any injective function such that%
\begin{equation}
c_{j}^{\left( c,s,t\right) }\left( \iota _{j}^{-1}\left( m_{j}\right)
\right) =c_{j}\left( m_{j}\right) \text{, }\forall m_{j}\in M_{j}^{III}\text{%
.}  \label{ttf1}
\end{equation}%
Consider the contract $\widehat{c_{j}^{\left( c,s,t\right) }}:\iota
_{j}^{-1}\left( M_{j}^{III}\right) \longrightarrow 2^{Y_{j}}\diagdown
\left\{ \varnothing \right\} $ with the restricted domain $\iota
_{j}^{-1}\left( M_{j}^{III}\right) $: 
\begin{equation*}
\widehat{c_{j}^{\left( c,s,t\right) }}\left( m_{j}\right) =c_{j}^{\left(
c,s,t\right) }\left( m_{j}\right) \text{, }\forall m_{j}\in \iota
_{j}^{-1}\left( M_{j}^{III}\right) \text{,}
\end{equation*}%
which, together with (\ref{ttf1}), implies%
\begin{gather*}
\widehat{c_{j}^{\left( c,s,t\right) }}\left( m_{j}\right) =c_{j}\left( \iota
_{j}\left( m_{j}\right) \right) \text{, }\forall m_{j}\in \iota
_{j}^{-1}\left( M_{j}^{III}\right) \text{,} \\
\text{or equivalently, }\widehat{c_{j}^{\left( c,s,t\right) }}\left( \iota
_{j}^{-1}\left( m_{j}\right) \right) =c_{j}\left( m_{j}\right) \text{, }%
\forall m_{j}\in M_{j}^{III}\text{.}
\end{gather*}%
That is, each $\widehat{c_{j}^{\left( c,s,t\right) }}$ is the same as $c_{j}$%
, where each message $m_{j}\in M_{j}^{III}$ for the latter is translated to $%
\iota _{j}^{-1}\left( m_{j}\right) \in \iota _{j}^{-1}\left(
M_{j}^{III}\right) $ for the former. Denote $\widehat{c^{\left( c,s,t\right)
}}\equiv \left( \widehat{c_{j}^{\left( c,s,t\right) }}\right) _{j\in 
\mathcal{J}}$. We will construct an equilibrium $\left( \widehat{c^{\left(
c,s,t\right) }},s^{\left( c,s,t\right) },\widehat{t^{\left( c,s,t\right) }}%
\right) $. However, $\widehat{c^{\left( c,s,t\right) }}\notin C^{I}$, and in
the second step below, we extend $\left( \widehat{c^{\left( c,s,t\right) }}%
,s^{\left( c,s,t\right) },\widehat{t^{\left( c,s,t\right) }}\right) $ to $%
\left( c^{\left( c,s,t\right) },s^{\left( c,s,t\right) },t^{\left(
c,s,t\right) }\right) \in \mathcal{E}^{\left\langle \mathcal{A},\text{ }%
\Gamma ,\text{ }\Psi \right\rangle \text{-}\left[ C^{I},C^{II}\right] }$.

\underline{Replicating $s$ with $s^{\left( c,s,t\right) }$:}

For each $j\in \mathcal{J}$, define the bijection, $\eta _{j}:\left\{ 
\widehat{c_{j}^{\left( c,s,t\right) }}\right\} \cup
C_{j}^{II}\longrightarrow \left\{ c_{j}\right\} \cup C_{j}^{II}$ as follows.%
\begin{equation*}
\eta _{j}\left( c_{j}^{\prime }\right) \equiv \left\{ 
\begin{tabular}{ll}
$c_{j}$, & if $c_{j}^{\prime }=\widehat{c_{j}^{\left( c,s,t\right) }}$; \\ 
$c_{j}^{\prime }$, & if $c_{j}^{\prime }\in C_{j}^{II}\diagdown \left\{ 
\widehat{c_{j}^{\left( c,s,t\right) }}\right\} $,%
\end{tabular}%
\right.
\end{equation*}%
i.e., we identify $\widehat{c_{j}^{\left( c,s,t\right) }}$ with $c_{j}$.
Define%
\begin{equation*}
s_{j}^{\left( c,s,t\right) }\left( \left( c_{k}^{\prime }\right) _{k\in 
\mathcal{J}},\theta \right) \equiv \left\{ 
\begin{tabular}{ll}
$\iota _{j}^{-1}\left( s_{j}\left[ \left( \eta _{k}\left( c_{k}^{\prime
}\right) \right) _{k\in \mathcal{J}},\text{ }\theta \right] \right) $, & if $%
c_{j}^{\prime }=\widehat{c_{j}^{\left( c,s,t\right) }}$; \\ 
$s_{j}\left[ \left( \eta _{k}\left( c_{k}^{\prime }\right) \right) _{k\in 
\mathcal{J}},\text{ }\theta \right] $, & if $c_{j}^{\prime }\in
C_{j}^{II}\diagdown \left\{ \widehat{c_{j}^{\left( c,s,t\right) }}\right\} $.%
\end{tabular}%
\right.
\end{equation*}%
That is, $s_{j}^{\left( c,s,t\right) }$ replicates $s_{j}$: the agent
identifies $\left( c_{k}^{\prime }\right) _{k\in \mathcal{J}}$ with $\left(
\eta _{k}\left( c_{k}^{\prime }\right) \right) _{k\in \mathcal{J}}$, and
follow $s_{j}$ to send the message $s_{j}\left[ \left( \eta _{k}\left(
c_{k}^{\prime }\right) \right) _{k\in \mathcal{J}},\text{ }\theta \right] $
when $c_{j}^{\prime }\in C_{j}^{II}\diagdown \left\{ \widehat{c_{j}^{\left(
c,s,t\right) }}\right\} $; the message is re-labeled to $\iota
_{j}^{-1}\left( s_{j}\left[ \left( \eta _{k}\left( c_{k}^{\prime }\right)
\right) _{k\in \mathcal{J}},\text{ }\theta \right] \right) $ when $%
c_{j}^{\prime }=\widehat{c_{j}^{\left( c,s,t\right) }}$.

\underline{Replicating $t$ with $\widehat{t^{\left( c,s,t\right) }}$:}

For each $\left( c_{k}^{\prime }\right) _{k\in \mathcal{J}}$ \ and each $%
j\in \mathcal{J}$, define $\xi _{j}^{\left( c_{k}^{\prime }\right) _{k\in 
\mathcal{J}}}:M_{j}^{I}\cup M_{j}^{II}\longrightarrow M_{j}^{III}\cup
M_{j}^{II}$: 
\begin{equation*}
\xi _{j}^{\left( c_{k}^{\prime }\right) _{k\in \mathcal{J}}}\left(
m_{j}\right) =\left\{ 
\begin{array}{cc}
\iota _{j}\left( m_{j}\right) & \text{if }c_{j}^{\prime }=\widehat{%
c_{j}^{\left( c,s,t\right) }}\text{;} \\ 
m_{j} & \text{if }c_{j}^{\prime }\in C_{j}^{II}\diagdown \left\{ \widehat{%
c_{j}^{\left( c,s,t\right) }}\right\} \text{.}%
\end{array}%
\right.
\end{equation*}%
That is, $\xi _{j}^{\left( c_{k}^{\prime }\right) _{k\in \mathcal{J}}}\left(
m_{j}\right) $ re-label $m_{j}$ if and only if $c_{j}^{\prime }=\widehat{%
c_{j}^{\left( c,s,t\right) }}$. Denote $\xi ^{\left( c_{k}^{\prime }\right)
_{k\in \mathcal{J}}}\equiv \left( \xi _{j}^{\left( c_{k}^{\prime }\right)
_{k\in \mathcal{J}}}\right) _{j\in \mathcal{J}}$.

For each $j\in \mathcal{J}$, define%
\begin{equation*}
\widehat{t_{j}^{\left( c,s,t\right) }}\left( \Gamma _{j}\left( \left(
c_{k}^{\prime }\right) _{k\in \mathcal{J}}\right) ,\Psi _{j}\left( m\right)
\right) \equiv t_{j}\left( \Gamma _{j}\left( \left( \eta _{k}\left(
c_{k}^{\prime }\right) \right) _{k\in \mathcal{J}}\right) ,\Psi _{j}\left(
\xi ^{\left( c_{k}^{\prime }\right) _{k\in \mathcal{J}}}\left( m\right)
\right) \right) \text{,}
\end{equation*}%
i.e., $\widehat{t_{j}^{\left( c,s,t\right) }}$ replicates $t_{j}$ subject to
re-labeling of the messages (by $\xi _{j}^{\left( c_{k}^{\prime }\right)
_{k\in \mathcal{J}}}$).

Clearly, $\left( \widehat{c^{\left( c,s,t\right) }},s^{\left( c,s,t\right) },%
\widehat{t^{\left( c,s,t\right) }}\right) $ is the "same" as $\left(
c,s,t\right) $ subject to re-labeling of the messages. Similarly, we can
transform the principals' beliefs at Stage 3 for $\left( c,s,t\right) $ to
their beliefs, denoted by $\widehat{b^{\left( c,s,t\right) }}$, for $\left( 
\widehat{c^{\left( c,s,t\right) }},s^{\left( c,s,t\right) },\widehat{%
t^{\left( c,s,t\right) }}\right) $ subject to re-labeling of the messages.
Therefore, $\left( \widehat{c^{\left( c,s,t\right) }},s^{\left( c,s,t\right)
},\widehat{t^{\left( c,s,t\right) }}\right) $ is an equilibrium. The only
problem is $\widehat{c^{\left( c,s,t\right) }}\notin C^{I}$. We now extend $%
\left( \widehat{c^{\left( c,s,t\right) }},s^{\left( c,s,t\right) },\widehat{%
t^{\left( c,s,t\right) }}\right) $ to $\left( c^{\left( c,s,t\right)
},s^{\left( c,s,t\right) },t^{\left( c,s,t\right) }\right) \in \mathcal{E}%
^{\left\langle \mathcal{A},\text{ }\Gamma ,\text{ }\Psi \right\rangle \text{-%
}\left[ C^{I},C^{II}\right] }$.

\underline{Extending $\left( \widehat{t^{\left( c,s,t\right) }}\text{, }%
\widehat{b^{\left( c,s,t\right) }}\right) $ to $\left( t^{\left(
c,s,t\right) }\text{, }b^{\left( c,s,t\right) }\right) $:}

For each $j\in \mathcal{J}$, define $\varphi _{j}:\left( \left\{
c_{j}^{\left( c,s,t\right) }\right\} \cup C_{j}^{II}\right) \longrightarrow
\left( \left\{ \widehat{c_{j}^{\left( c,s,t\right) }}\right\} \cup
C_{j}^{II}\right) $ as%
\begin{equation*}
\varphi _{j}\left( c_{j}^{\prime }\right) =\left\{ 
\begin{tabular}{ll}
$\widehat{c_{j}^{\left( c,s,t\right) }}$ & $\text{if }c_{j}^{\prime
}=c_{j}^{\left( c,s,t\right) }\text{;}$ \\ 
$c_{j}^{\prime }$ & $\text{if }c_{j}^{\prime }\in C_{j}^{II}\diagdown
\left\{ c_{j}^{\left( c,s,t\right) }\right\} $.%
\end{tabular}%
\right.
\end{equation*}%
I.e., $\varphi _{j}$ re-label $c_{j}^{\prime }$ to $\widehat{c_{j}^{\left(
c,s,t\right) }}$ if and only if $c_{j}^{\prime }=c_{j}^{\left( c,s,t\right)
} $.

For each $j\in \mathcal{J}$ and each $\left( c_{k}^{\prime }\right) _{k\in 
\mathcal{J}}$, define $\pi _{j}^{\left( c_{k}^{\prime }\right) _{k\in 
\mathcal{J}}}:M_{j}^{I}\cup M_{j}^{II}\longrightarrow M_{j}^{I}\cup
M_{j}^{II}$ as%
\begin{equation*}
\pi _{j}^{\left( c_{k}^{\prime }\right) _{k\in \mathcal{J}}}\left(
m_{j}\right) =\left\{ 
\begin{tabular}{ll}
$m_{j}$ & $\text{if }c_{j}^{\prime }\in C_{j}^{II}\diagdown \left\{
c_{j}^{\left( c,s,t\right) }\right\} \text{;}$ \\ 
$\iota _{j}^{-1}\left( \iota _{j}\left( m_{j}\right) \right) $ & $\text{if }%
c_{j}^{\prime }=c_{j}^{\left( c,s,t\right) }$.%
\end{tabular}%
\right.
\end{equation*}%
That is, $\pi _{j}^{\left( c_{k}^{\prime }\right) _{k\in \mathcal{J}}}\left(
m_{j}\right) $ re-labels $m_{j}$ to $\iota _{j}^{-1}\left( \iota _{j}\left(
m_{j}\right) \right) \in \iota _{j}^{-1}\left( M_{j}^{III}\right) $ if and
only if $c_{j}^{\prime }=c_{j}^{\left( c,s,t\right) }$.\footnote{$%
c_{j}^{\left( c,s,t\right) }\left( m_{j}\right) =c_{j}^{\left( c,s,t\right)
}\left( \iota _{j}^{-1}\left( \iota _{j}\left( m_{j}\right) \right) \right) $
for every $m_{j}\in M_{j}^{I}$, i.e., $m_{j}$ and $\iota _{j}^{-1}\left(
\iota _{j}\left( m_{j}\right) \right) $ pin down the same subset of actions
under the contract $c_{j}^{\left( c,s,t\right) }$.} Denote $\pi ^{\left(
c_{k}^{\prime }\right) _{k\in \mathcal{J}}}\equiv \left( \pi _{j}^{\left(
c_{k}^{\prime }\right) _{k\in \mathcal{J}}}\right) _{j\in \mathcal{J}}$. For
each $j\in \mathcal{J}$, define%
\begin{equation*}
t_{j}^{\left( c,s,t\right) }\left( \Gamma _{j}\left( \left( c_{k}^{\prime
}\right) _{k\in \mathcal{J}}\right) ,\Psi _{j}\left( m\right) \right) \equiv 
\widehat{t_{j}^{\left( c,s,t\right) }}\left( \Gamma _{j}\left( \varphi
_{k}\left( c_{k}^{\prime }\right) _{k\in \mathcal{J}}\right) ,\Psi
_{j}\left( \pi ^{\left( c_{k}^{\prime }\right) _{k\in \mathcal{J}}}\left(
m\right) \right) \right) \text{,}
\end{equation*}%
i.e., $t_{j}^{\left( c,s,t\right) }$ replicates $\widehat{t_{j}^{\left(
c,s,t\right) }}$ subject to identifying each $m_{j}\in M_{j}^{I}\diagdown
\iota _{j}^{-1}\left( M_{j}^{III}\right) $ with $\iota _{j}^{-1}\left( \iota
_{j}\left( m_{j}\right) \right) $. Define beliefs as follows.%
\begin{equation*}
b_{j}^{\left( c,s,t\right) }\left( \Gamma _{j}\left( \left( c_{k}^{\prime
}\right) _{k\in \mathcal{J}}\right) ,\Psi _{j}\left( m\right) \right) \equiv 
\widehat{b_{j}^{\left( c,s,t\right) }}\left( \Gamma _{j}\left( \varphi
_{k}\left( c_{k}^{\prime }\right) _{k\in \mathcal{J}}\right) ,\Psi
_{j}\left( \pi ^{\left( c_{k}^{\prime }\right) _{k\in \mathcal{J}}}\left(
m\right) \right) \right) \text{.}
\end{equation*}%
Clearly, $\left( c^{\left( c,s,t\right) },s^{\left( c,s,t\right) },t^{\left(
c,s,t\right) }\right) $ replicates $\left( \widehat{c^{\left( c,s,t\right) }}%
,s^{\left( c,s,t\right) },\widehat{t^{\left( c,s,t\right) }}\right) $, and
principals and the agent inherit incentive compatibility. Note that, upon
receiving $c_{j}^{\left( c,s,t\right) }$ from principal $j$, the agent does
not find it profitable to deviate to sending messages in $M_{j}^{I}\diagdown
\iota _{j}^{-1}\left( M_{j}^{III}\right) $, because sending $m_{j}\in
M_{j}^{I}\diagdown \iota _{j}^{-1}\left( M_{j}^{III}\right) $ is equivalent
to sending $\iota _{j}^{-1}\left( \iota _{j}\left( m_{j}\right) \right) \in
\iota _{j}^{-1}\left( M_{j}^{III}\right) $, which is not a profitable
deviation in the equilibrium $\left( \widehat{c^{\left( c,s,t\right) }}%
,s^{\left( c,s,t\right) },\widehat{t^{\left( c,s,t\right) }}\right) $.
Finally, all of the argument above works for any $\left\langle \Gamma ,\text{
}\Psi \right\rangle \in \left\{ \Gamma ^{public},\text{ }\Gamma
^{private}\right\} \times \left\{ \Psi ^{public},\Psi ^{private}\right\} $.$%
\blacksquare $

\subsection{Proof of Lemma \protect\ref{lem:extension:off}}

\label{sec:lem:extension:off}

Let $\mathcal{A=A}^{non-delegated}$. Fix any $I,II,IV\in \left\{ \mathcal{A}%
,P,F,R\right\} $ such that $C^{II}\sqsupset ^{\ast \ast }C^{IV}$. Let $M^{I}$%
, $M^{II}$ and $M^{IV}$ denote the message spaces for $C^{I}$, $C^{II}$ and $%
C^{IV}$, respectively. Fix any $\left( c,s,t\right) \in \mathcal{E}%
^{\left\langle \mathcal{A},\text{ }\Gamma ,\text{ }\Psi \right\rangle \text{-%
}\left[ C^{I},C^{IV}\right] }$. We aim to construct a $\left[ C^{I},C^{II}%
\right] $-equilibrium that induces $z^{\left( c,s,t\right) }$.

Since $C^{II}\sqsupset ^{\ast \ast }C^{IV}$, there exists a function $\psi
_{j}:$ $C_{j}^{II}\longrightarrow C_{j}^{IV}$ for each $j\in \mathcal{J}$
such that $c_{j}^{\prime }\geq \psi _{j}\left( c_{j}^{\prime }\right) $ for
all $c_{j}^{\prime }\in C_{j}^{II}$. Thus, for each $c_{j}^{\prime }\in
C_{j}^{II}$, there exists a surjective $\iota _{j}^{c_{j}^{\prime
}}:M_{j}^{II}\longrightarrow M_{j}^{IV}$ such that%
\begin{equation*}
c_{j}^{\prime }\left( m_{j}\right) =\psi _{j}\left( c_{j}^{\prime }\right)
\left( \iota _{j}^{c_{j}^{\prime }}\left( m_{j}\right) \right) \text{, }%
\forall m_{j}\in M_{j}^{II}\text{.}
\end{equation*}%
As in Appendix \ref{sec:lem:extension:on}, let $\left( \iota
_{j}^{c_{j}^{\prime }}\right) ^{-1}:M_{j}^{IV}\longrightarrow M_{j}^{II}$
denote any injective function such that%
\begin{equation*}
c_{j}^{\prime }\left( \left( \iota _{j}^{c_{j}^{\prime }}\right) ^{-1}\left(
m_{j}\right) \right) =\psi _{j}\left( c_{j}^{\prime }\right) \left(
m_{j}\right) \text{, }\forall m_{j}\in M_{j}^{IV}\text{.}
\end{equation*}%
Consider the contract $\widehat{c_{j}^{\prime }}:\left( \iota
_{j}^{c_{j}^{\prime }}\right) ^{-1}\left( M_{j}^{IV}\right) \longrightarrow
2^{Y_{j}}\diagdown \left\{ \varnothing \right\} $ with the restricted
domain: 
\begin{equation*}
\widehat{c_{j}^{\prime }}\left( m_{j}\right) =c_{j}^{\prime }\left(
m_{j}\right) \text{, }\forall m_{j}\in \left( \iota _{j}^{c_{j}^{\prime
}}\right) ^{-1}\left( M_{j}^{IV}\right) \text{,}
\end{equation*}%
i.e., each $\widehat{c_{j}^{\prime }}$ is the same as $\psi _{j}\left(
c_{j}^{\prime }\right) $, where each message $m_{j}\in M_{j}^{IV}$ for the
latter is translated to $\left( \iota _{j}^{c_{j}^{\prime }}\right)
^{-1}\left( m_{j}\right) \in \left( \iota _{j}^{c_{j}^{\prime }}\right)
^{-1}\left( M_{j}^{IV}\right) $ for the former. Denote $\widehat{c^{\prime }}%
\equiv \left( \widehat{c_{j}^{\prime }}\right) _{j\in \mathcal{J}}$. Define%
\begin{equation*}
\widehat{C}_{j}^{II}\equiv \left\{ \widehat{c_{j}^{\prime }}:c_{j}^{\prime
}\in C_{j}^{II}\right\} \text{, }\forall j\in \mathcal{J}\text{ and }%
\widehat{C}^{II}\equiv \times _{j\in \mathcal{J}}\widehat{C}_{j}^{II}\text{.}
\end{equation*}%
Clearly, we have $Z^{\mathcal{E}^{\left\langle \mathcal{A},\text{ }\Gamma ,%
\text{ }\Psi \right\rangle \text{-}\left[ C^{I},\widehat{C}^{II}\right]
}}\supset Z^{\mathcal{E}^{\left\langle \mathcal{A},\text{ }\Gamma ,\text{ }%
\Psi \right\rangle \text{-}\left[ C^{I},C^{IV}\right] }}$, i.e., we can
replicate any $\left( c,s,t\right) \in \mathcal{E}^{\left\langle \mathcal{A},%
\text{ }\Gamma ,\text{ }\Psi \right\rangle \text{-}\left[ C^{I},C^{IV}\right]
}$ with some $\left( c,\widehat{s},\widehat{t}\right) \in \mathcal{E}%
^{\left\langle \mathcal{A},\text{ }\Gamma ,\text{ }\Psi \right\rangle \text{-%
}\left[ C^{I},\widehat{C}^{II}\right] }$, which summarizes the re-labeling
of messages (i.e., $\left( \iota _{j}^{c_{j}^{\prime }}\right) ^{-1}$) as in
Appendix \ref{sec:lem:extension:on}.

As in Appendix \ref{sec:lem:extension:on}, we can extend each $\widehat{%
c_{j}^{\prime }}\in \widehat{C}_{j}^{II}$ to $c_{j}^{\prime }\in C_{j}^{II}$%
. The difference between $\widehat{c_{j}^{\prime }}$ and $c_{j}^{\prime }$
is that the agent cannot send messages in $M_{j}^{II}\diagdown \left( \iota
_{j}^{c_{j}^{\prime }}\right) ^{-1}\left( M_{j}^{IV}\right) $ under $%
\widehat{c_{j}^{\prime }}$, while he can under $c_{j}^{\prime }$. Suppose
principal $j$ identify each $m_{j}\in M_{j}^{II}\diagdown \left( \iota
_{j}^{c_{j}^{\prime }}\right) ^{-1}\left( M_{j}^{IV}\right) $ with 
\begin{equation*}
\left( \iota _{j}^{c_{j}^{\prime }}\right) ^{-1}\left[ \iota
_{j}^{c_{j}^{\prime }}\left( m_{j}\right) \right] \in \left( \iota
_{j}^{c_{j}^{\prime }}\right) ^{-1}\left( M_{j}^{IV}\right) \text{,}
\end{equation*}%
and then replicate $\widehat{t}$. Given this, the agent does not have
incentive to send any message $m_{j}\in M_{j}^{II}\diagdown \left( \iota
_{j}^{c_{j}^{\prime }}\right) ^{-1}\left( M_{j}^{IV}\right) $, because
sending $m_{j}$ is equivalent to sending $\left( \iota _{j}^{c_{j}^{\prime
}}\right) ^{-1}\left[ \iota _{j}^{c_{j}^{\prime }}\left( m_{j}\right) \right]
\in \left( \iota _{j}^{c_{j}^{\prime }}\right) ^{-1}\left( M_{j}^{IV}\right) 
$. Rigorously,

\underline{Extending $\left( \widehat{t}\text{, }\widehat{b}\right) $ to $%
\left( t^{\ast }\text{, }b^{\ast }\right) $:}

For each $j\in \mathcal{J}$, define $\varphi _{j}:\left( \left\{
c_{j}\right\} \cup C_{j}^{II}\right) \longrightarrow \left( \left\{
c_{j}\right\} \cup \widehat{C}_{j}^{II}\right) $,%
\begin{equation*}
\varphi _{j}\left( c_{j}^{\prime }\right) =\left\{ 
\begin{tabular}{ll}
$\widehat{c_{j}^{\prime }}$ & if $c_{j}^{\prime }\in C_{j}^{II}\diagdown
\left\{ c_{j}\right\} $; \\ 
$c_{j}$ & if $c_{j}^{\prime }=c_{j}$.%
\end{tabular}%
\right.
\end{equation*}%
For each $j\in \mathcal{J}$ and each $\left( c_{k}^{\prime }\right) _{k\in 
\mathcal{J}}$, define $\pi _{j}^{\left( c_{k}^{\prime }\right) _{k\in 
\mathcal{J}}}:M_{j}^{I}\cup M_{j}^{II}\longrightarrow M_{j}^{I}\cup
M_{j}^{II}$: 
\begin{equation*}
\pi _{j}^{\left( c_{k}^{\prime }\right) _{k\in \mathcal{J}}}\left(
m_{j}\right) =\left\{ 
\begin{tabular}{ll}
$\left( \iota _{j}^{c_{j}^{\prime }}\right) ^{-1}\left[ \iota
_{j}^{c_{j}^{\prime }}\left( m_{j}\right) \right] $ & if $c_{j}^{\prime }\in
C_{j}^{II}\diagdown \left\{ c_{j}\right\} $; \\ 
$m_{j}$ & $\text{if }c_{j}^{\prime }=c_{j}$.%
\end{tabular}%
\right.
\end{equation*}%
For each $j\in \mathcal{J}$, define%
\begin{eqnarray*}
t_{j}^{\ast }\left( \Gamma _{j}\left( \left( c_{k}^{\prime }\right) _{k\in 
\mathcal{J}}\right) ,\Psi _{j}\left( m\right) \right) &\equiv &\widehat{t_{j}%
}\left( \Gamma _{j}\left( \varphi _{k}\left( c_{k}^{\prime }\right) _{k\in 
\mathcal{J}}\right) ,\Psi _{j}\left( \pi ^{\left( c_{k}^{\prime }\right)
_{k\in \mathcal{J}}}\left( m\right) \right) \right) \text{,} \\
b_{j}^{\ast }\left( \Gamma _{j}\left( \left( c_{k}^{\prime }\right) _{k\in 
\mathcal{J}}\right) ,\Psi _{j}\left( m\right) \right) &\equiv &\widehat{b_{j}%
}\left( \Gamma _{j}\left( \varphi _{k}\left( c_{k}^{\prime }\right) _{k\in 
\mathcal{J}}\right) ,\Psi _{j}\left( \pi ^{\left( c_{k}^{\prime }\right)
_{k\in \mathcal{J}}}\left( m\right) \right) \right) \text{.}
\end{eqnarray*}%
Therefore, $\left( c,\widehat{s},t^{\ast }\right) \in \mathcal{E}%
^{\left\langle \mathcal{A},\text{ }\Gamma ,\text{ }\Psi \right\rangle \text{-%
}\left[ C^{I},C^{II}\right] }$ replicates $\left( c,s,t\right) \in \mathcal{E%
}^{\left\langle \mathcal{A},\text{ }\Gamma ,\text{ }\Psi \right\rangle \text{%
-}\left[ C^{I},C^{IV}\right] }$.$\blacksquare $

\subsection{Proof of Proposition \protect\ref{prop:deviation}}

\label{sec:prop:deviation}

Fix $\mathcal{A}=\mathcal{A}^{non-delegated}$. Fix any $\left\langle \Gamma ,%
\text{ }\Psi \right\rangle \in \left\{ \left\langle \Gamma ^{private},\Psi
^{private}\right\rangle ,\text{ }\left\langle \Gamma ^{public},\Psi
^{private}\right\rangle ,\text{ }\left\langle \Gamma ^{public},\Psi
^{public}\right\rangle \right\} $ and any $I\in \left\{ \mathcal{A}%
,P,F,R\right\} $. Since $C^{\mathcal{A}}\sqsupset ^{\ast \ast }C^{F}$, Lemma %
\ref{lem:extension:off} implies $Z^{\mathcal{E}^{\left\langle \mathcal{A},%
\text{ }\Gamma ,\text{ }\Psi \right\rangle \text{-}\left[ C^{I},C^{\mathcal{A%
}}\right] }}\supset Z^{\mathcal{E}^{\left\langle \mathcal{A},\text{ }\Gamma ,%
\text{ }\Psi \right\rangle \text{-}\left[ C^{I},C^{F}\right] }}$. We now
prove%
\begin{equation*}
Z^{\mathcal{E}^{\left\langle \mathcal{A},\text{ }\Gamma ,\text{ }\Psi
\right\rangle \text{-}\left[ C^{I},C^{\mathcal{A}}\right] }}\subset Z^{%
\mathcal{E}^{\left\langle \mathcal{A},\text{ }\Gamma ,\text{ }\Psi
\right\rangle \text{-}\left[ C^{I},C^{F}\right] }}\text{.}
\end{equation*}%
Fix any $\left( c,s,t\right) \in \mathcal{E}^{\left\langle \mathcal{A},\text{
}\Gamma ,\text{ }\Psi \right\rangle \text{-}\left[ C^{I},C^{\mathcal{A}}%
\right] }$. We will replicate $\left( c,s,t\right) $ with some $\left( c,%
\overline{s},\overline{t}\right) \in \mathcal{E}^{\left\langle \mathcal{A},%
\text{ }\Gamma ,\text{ }\Psi \right\rangle \text{-}\left[ C^{I},C^{F}\right]
}$ such that $z^{\left( c,s,t\right) }=z^{\left( c,\overline{s},\overline{t}%
\right) }$.

\underline{Replicating $s\equiv \left( s_{k}\right) _{k\in \mathcal{J}}$
with $\overline{s}\equiv \left( \overline{s}_{k}\right) _{k\in \mathcal{J}}$:%
}

We replicate $s$ in the $\left[ C^{I},C^{\mathcal{A}}\right] $-game with $%
\overline{s}$ in the $\left[ C^{I},C^{F}\right] $-game. Since $s$ is defined
on $\left( \left\{ c_{k}\right\} \cup C_{k}^{\mathcal{A}}\right) _{k\in 
\mathcal{J}}$ and $\overline{s}$ is defined on $\left( \left\{ c_{k}\right\}
\cup C_{k}^{F}\right) _{k\in \mathcal{J}}$, we need to define a function $%
\gamma :\left( \left\{ c_{k}\right\} \cup C_{k}^{F}\right) _{k\in \mathcal{J}%
}\longrightarrow \left( \left\{ c_{k}\right\} \cup C_{k}^{\mathcal{A}%
}\right) _{k\in \mathcal{J}}$, such that, upon observing $c^{\prime }\in
\left( \left\{ c_{k}\right\} \cup C_{k}^{F}\right) _{k\in \mathcal{J}}$ in
the $\left[ C^{I},C^{F}\right] $-game, the agent regards it as $\gamma
\left( c^{\prime }\right) $ in the $\left[ C^{I},C^{\mathcal{A}}\right] $%
-game, and then, $\overline{s}\left( c^{\prime }\right) $ mimics $s\left[
\gamma \left( c^{\prime }\right) \right] $. Thus, for each $j\in \mathcal{J}$%
, consider $\gamma _{j}:\left\{ c_{j}\right\} \cup C_{j}^{F}\longrightarrow
\left\{ c_{j}\right\} \cup C_{j}^{\mathcal{A}}$ such that 
\begin{equation*}
\gamma _{j}\left( c_{j}\right) =c_{j},
\end{equation*}
and for each $c_{j}^{\prime }\in C_{j}^{F}\diagdown \left\{ c_{j}\right\} $,
we define $\gamma _{j}\left( c_{j}^{\prime }\right) $ as follows. Since $%
c_{j}^{\prime }\in C_{j}^{F}$, we have a pair of $\left[ L_{j}\subset
2^{Y_{j}}\text{ and }H_{j}=\left\{ \left[ E_{j},y_{j}\right] :y_{j}\in
E_{j}\right\} \right] $ satisfying Definition \ref%
{def:menu_of_menu_with_full_recommendation}. Fix any injective function $%
\phi _{j}^{c_{j}^{\prime }}:L_{j}\cup H_{j}\longrightarrow M_{j}^{\mathcal{A}%
}$, i.e., given $c_{j}^{\prime }\in C_{j}^{F}$, we identify a message $%
m_{j}^{\prime }\in L_{j}\cup H_{j}$ in the $\left[ C^{I},C^{F}\right] $-game
to the message $\phi _{j}^{c_{j}^{\prime }}\left( m_{j}^{\prime }\right) \in
M_{j}^{\mathcal{A}}$ in the $\left[ C^{I},C^{\mathcal{A}}\right] $-game.
With slight abuse of notation, let $\left( \phi _{j}^{c_{j}^{\prime
}}\right) ^{-1}:\phi _{j}^{c_{j}^{\prime }}\left[ L_{j}\cup H_{j}\right]
\longrightarrow L_{j}\cup H_{j}$ denote the inverse function of $\phi
_{j}^{c_{j}^{\prime }}$, i.e.,%
\begin{equation*}
\phi _{j}^{c_{j}^{\prime }}\left[ \left( \phi _{j}^{c_{j}^{\prime }}\right)
^{-1}\left( m_{j}\right) \right] =m_{j}\text{, }\forall m_{j}\in \phi
_{j}^{c_{j}^{\prime }}\left[ L_{j}\cup H_{j}\right] \text{.}
\end{equation*}%
We thus define $\gamma _{j}\left( c_{j}^{\prime }\right) $ for each $%
c_{j}^{\prime }\in C_{j}^{F}\diagdown \left\{ c_{j}\right\} $ as follows.%
\begin{equation*}
\gamma _{j}\left( c_{j}^{\prime }\right) \left[ m_{j}\right] =\left\{ 
\begin{array}{cc}
c_{j}^{\prime }\left( \left( \phi _{j}^{c_{j}^{\prime }}\right) ^{-1}\left[
m_{j}\right] \right) & \text{if }m_{j}\in \phi _{j}^{c_{j}^{\prime }}\left[
L_{j}\cup H_{j}\right] ; \\ 
E_{j} & \text{otherwise,}%
\end{array}%
\right. \text{ }\forall m_{j}\in M_{j}^{\mathcal{A}}\text{,}
\end{equation*}%
i.e., we first embed the message space $L_{j}\cup H_{j}$ into $M_{j}^{%
\mathcal{A}}$ by $\phi _{j}^{c_{j}^{\prime }}$; second, we copy $%
c_{j}^{\prime }$ with $\gamma _{j}\left( c_{j}^{\prime }\right) $ on the
embedded message space; third, all of the other messages in $M_{j}^{\mathcal{%
A}}$ are mapped to $E_{j}$. Furthermore, denote $\gamma \left( c^{\prime
}\right) \equiv \left( \gamma _{k}\left( c_{k}^{\prime }\right) \right)
_{k\in \mathcal{J}}$ and $\phi ^{c^{\prime }}:\equiv \left[ \phi
_{j}^{c_{j}^{\prime }}:L_{j}\cup H_{j}\longrightarrow M_{j}^{\mathcal{A}}%
\right] _{k\in \mathcal{J}}$.

We are now ready to define $\overline{s}\equiv \left( \overline{s}%
_{k}\right) _{k\in \mathcal{J}}$. For each $j\in \mathcal{J}$, define 
\begin{equation*}
\overline{s}_{j}\left( c^{\prime },\theta \right) \equiv \left\{ 
\begin{tabular}{ll}
$s_{j}\left( \gamma \left( c^{\prime }\right) ,\theta \right) $ & if $%
c_{j}^{\prime }=c_{j}$; \\ 
&  \\ 
$\left( \phi _{j}^{c_{j}^{\prime }}\right) ^{-1}\left[ s_{j}\left( \gamma
\left( c^{\prime }\right) ,\theta \right) \right] $ & if $c_{j}^{\prime
}\neq c_{j}$ and $s_{j}\left( \gamma \left( c^{\prime }\right) ,\theta
\right) \in \phi _{j}^{c_{j}^{\prime }}\left[ L_{j}\right] ;$ \\ 
&  \\ 
$\left[ E_{j},\text{ }y_{j}=t_{j}\left( \Gamma _{j}\left( \gamma \left(
c^{\prime }\right) \right) ,\Psi _{j}\left( s\left( \gamma \left( c^{\prime
}\right) ,\theta \right) \right) \right) \right] $ & otherwise.%
\end{tabular}%
\right.
\end{equation*}%
When the agent observes $c^{\prime }\in \left( \left\{ c_{k}\right\} \cup
C_{k}^{F}\right) _{k\in \mathcal{J}}$ in the $\left[ C^{I},C^{F}\right] $%
-game, the agent translates it to $\gamma \left( c^{\prime }\right) \in
\left( \left\{ c_{k}\right\} \cup C_{k}^{\mathcal{A}}\right) _{k\in \mathcal{%
J}}$ being offered in the $\left[ C^{I},C^{\mathcal{A}}\right] $-game. Then, 
$\overline{s}_{j}\left( c^{\prime },\theta \right) $ replicates $s_{j}\left(
\gamma \left( c^{\prime }\right) ,\theta \right) $: if $c_{j}^{\prime
}=c_{j} $, we have $\overline{s}_{j}\left( c^{\prime },\theta \right)
=s_{j}\left( \gamma \left( c^{\prime }\right) ,\theta \right) $; if $%
c_{j}^{\prime }\neq c_{j}$ and $s_{j}\left( \gamma \left( c^{\prime }\right)
,\theta \right) \in \phi _{j}^{c_{j}^{\prime }}\left[ L_{j}\right] $, $%
\overline{s}_{j}\left( c^{\prime },\theta \right) $ mimics $s_{j}\left(
\gamma \left( c^{\prime }\right) ,\theta \right) $ subject to re-labeling of
message by $\left( \phi _{j}^{c_{j}^{\prime }}\right) ^{-1}$; otherwise, the
message $s_{j}\left( \gamma \left( c^{\prime }\right) ,\theta \right) $ pins
down the subset $E_{j}$ at Stage 2 in the $\left[ C^{I},C^{\mathcal{A}}%
\right] $-game, and at stage 3, $j$ would take the action $t_{j}\left(
\Gamma _{j}\left( \gamma \left( c^{\prime }\right) \right) ,\Psi _{j}\left(
s\left( \gamma \left( c^{\prime }\right) ,\theta \right) \right) \right) $,
and hence, $\overline{s}_{j}\left( c^{\prime },\theta \right) $ mimics this
by choosing 
\begin{equation*}
\left[ E_{j},\text{ }y_{j}=t_{j}\left( \Gamma _{j}\left( \gamma \left(
c^{\prime }\right) \right) ,\Psi _{j}\left( s\left( \gamma \left( c^{\prime
}\right) ,\theta \right) \right) \right) \right]
\end{equation*}%
in the menu-of-menu-with-full-recommendation contract $c_{j}^{\prime }\in
C_{j}^{F}$.

\underline{Replicating $\left( b_{j},t_{j}\right) $ with $\left( \overline{b}%
_{j},\overline{t}_{j}\right) $:}

Given $c_{j}^{\prime }\in \left\{ c_{j}\right\} \cup C_{j}^{F}$, let $%
M_{j}^{c_{j}^{\prime }}$ denote the domain of $c_{j}^{\prime }$. For each $%
j\in \mathcal{J}$, define%
\begin{equation*}
Q_{j}^{F}\equiv \left\{ \left( \Gamma _{j}\left( c^{\prime }\right) ,\Psi
_{j}\left( m\right) \right) :%
\begin{array}{c}
c^{\prime }=\left( c_{k}^{\prime }\right) _{k\in \mathcal{J}}\in \times
_{k\in \mathcal{J}}\left( \left\{ c_{k}\right\} \cup C_{k}^{F}\right) \\ 
m=\left( m_{k}\right) _{k\in \mathcal{J}}\in \times _{k\in \mathcal{J}%
}\left( M_{k}^{c_{k}^{\prime }}\right)%
\end{array}%
\right\} \text{,}
\end{equation*}%
\begin{equation*}
Q_{j}^{F\text{-}\Gamma ^{public}}\equiv \left\{ \left( \Gamma _{j}\left(
c\right) ,\Psi _{j}\left( \overline{s}\left( c,\theta \right) \right)
\right) :\theta \in \Theta \right\} \text{,}
\end{equation*}%
\begin{equation*}
Q_{j}^{F\text{-}\Gamma ^{private}}\equiv \left\{ \left( \Gamma _{j}\left(
c_{j}^{\prime },c_{-j}\right) ,\Psi _{j}\left( \overline{s}\left( \left(
c_{j}^{\prime },c_{-j}\right) ,\theta \right) \right) \right) :%
\begin{array}{c}
c_{j}^{\prime }\in \left\{ c_{j}\right\} \cup C_{j}^{F} \\ 
\theta \in \Theta%
\end{array}%
\right\} \text{,}
\end{equation*}%
i.e., in the $\left[ C^{I},C^{F}\right] $-game, $Q_{j}^{F}$ is the set of
all possible information that principal $j$ may observe before $j$ takes an
action at Stage 3. Given public announcement, $Q_{j}^{F\text{-}\Gamma
^{public}}$ is the set of information where principal $j$ cannot confirm
that the other principal or the agent have deviated from the equilibrium.
Similarly, given private announcement, $Q_{j}^{F\text{-}\Gamma ^{private}}$
is the set of information where principal $j$ cannot confirm that the other
principal or the agent have deviated from the equilibrium. Define 
\begin{equation*}
Q_{j}^{\mathcal{A}}\equiv \left\{ \left( \Gamma _{j}\left( c^{\prime
}\right) ,\Psi _{j}\left( m\right) \right) :%
\begin{array}{c}
c^{\prime }=\left( c_{k}^{\prime }\right) _{k\in \mathcal{J}}\in \times
_{k\in \mathcal{J}}\left( \left\{ c_{k}\right\} \cup C_{k}^{\mathcal{A}%
}\right) \\ 
m=\left( m_{k}\right) _{k\in \mathcal{J}}\in \times _{k\in \mathcal{J}%
}\left( M_{k}^{c_{k}^{\prime }}\right)%
\end{array}%
\right\} \text{,}
\end{equation*}%
i.e., the $\left[ C^{I},C^{\mathcal{A}}\right] $-game, $Q_{j}^{\mathcal{A}}$
is the set of all possible information that principal $j$ may observe before 
$j$ takes an action at Stage 3. Since $\left( b_{j},t_{j}\right) $ is
defined on $Q_{j}^{\mathcal{A}}$ and $\left( \overline{b}_{j},\overline{t}%
_{j}\right) $ is defined on $Q_{j}^{F}$, we need to translate information in 
$Q_{j}^{F}$ to information in $Q_{j}^{\mathcal{A}}$. We proceed by consider
three cases.

\paragraph{Case 1:}

Fix $\left\langle \Gamma ,\Psi \right\rangle =\left\langle \Gamma
^{public},\Psi ^{public}\right\rangle $. For each $q_{j}\in Q_{j}^{F\text{-}%
\Gamma ^{public}}$, fix any $\theta ^{q_{j}}\in \Theta $ such that $q_{j}=%
\left[ c,\overline{s}\left( c,\theta ^{q_{j}}\right) \right] $. Define $%
\Sigma _{j}^{\left\langle \Gamma ^{public},\Psi ^{public}\right\rangle
}:Q_{j}^{F}\longrightarrow Q_{j}^{\mathcal{A}}$ as%
\begin{equation*}
\Sigma _{j}^{\left\langle \Gamma ^{public},\Psi ^{public}\right\rangle
}\left( q_{j}=\left( \overline{c},\overline{m}\right) \right) \equiv \left\{ 
\begin{array}{cc}
\left[ c,s\left( c,\theta ^{q_{j}}\right) \right] & \text{if }q_{j}\in
Q_{j}^{F\text{-}\Gamma ^{public}}; \\ 
&  \\ 
\left[ \gamma \left( \overline{c}\right) ,\text{ }\phi ^{\overline{c}}\left( 
\overline{m}\right) \right] & \text{if }q_{j}\in Q_{j}^{F}\diagdown Q_{j}^{F%
\text{-}\Gamma ^{public}}%
\end{array}%
\right. \text{, }\forall q_{j}=\left( \overline{c},\overline{m}\right) \in
Q_{j}^{F}\text{.}
\end{equation*}%
Thus, we define $\overline{t}_{j}\left( q_{j}\right) \equiv t_{j}\left(
\Sigma _{j}^{\left\langle \Gamma ^{public},\Psi ^{public}\right\rangle
}\left( q_{j}\right) \right) $. \ Furthermore, given $\left\langle \Gamma
^{public},\Psi ^{public}\right\rangle $, all principals observe all
contracts and all messages. Thus, each principal has a degenerate belief on $%
C\times M$, and hence, we only describe the marginal belief on $\Theta $.
Define $\overline{b}_{j}\left( q_{j}\right) \left[ \left\{ \left( \theta
\right) \right\} \right] \equiv b_{j}\left( \Sigma _{j}^{\left\langle \Gamma
^{public},\Psi ^{public}\right\rangle }\left( q_{j}\right) \right) \left[
\left\{ \left( \theta \right) \right\} \right] $, i.e., $\overline{b}%
_{j}\left( q_{j}\right) $ mimics $b_{j}\left( \Sigma _{j}^{\left\langle
\Gamma ^{public},\Psi ^{public}\right\rangle }\left( q_{j}\right) \right) $
on the belief on $\Theta $.

\paragraph{Case 2:}

Fix $\left\langle \Gamma ,\Psi \right\rangle =\left\langle \Gamma
^{private},\Psi ^{private}\right\rangle $. For each $q_{j}\in Q_{j}^{F\text{-%
}\Gamma ^{private}}$, fix any $\left( c_{j}^{q_{j}},\theta ^{q_{j}}\right)
\in C_{j}^{F}\times \Theta $ such that $q_{j}=\left( \Gamma _{j}\left(
c_{j}^{q_{j}},c_{-j}\right) ,\Psi _{j}\left( \overline{s}\left(
c_{j}^{q_{j}},c_{-j},\theta ^{q_{j}}\right) \right) \right) $. For each $%
q_{j}\in Q_{j}^{F}\diagdown Q_{j}^{F\text{-private}}$, fix any $\left( 
\overline{c^{q_{j}}},\overline{m}\right) $ such that $q_{j}=\left( \Gamma
_{j}\left( \overline{c^{q_{j}}}\right) ,\Psi _{j}\left( \overline{m}\right)
\right) $. Define $\Sigma _{j}^{\left\langle \Gamma ^{private},\Psi
^{private}\right\rangle }:Q_{j}^{F}\longrightarrow Q_{j}^{\mathcal{A}}$ as%
\begin{equation*}
\Sigma _{j}^{\left\langle \Gamma ^{private},\Psi ^{private}\right\rangle
}\left( q_{j}\right) \equiv \left\{ 
\begin{array}{cc}
\left[ \Gamma _{j}\left[ \gamma \left( c_{j}^{q_{j}},c_{-j}\right) \right]
,\Psi _{j}\left( s\left( \gamma \left( c_{j}^{q_{j}},c_{-j}\right) ,\theta
^{q_{j}}\right) \right) \right] & \text{if }q_{j}\in Q_{j}^{F\text{-private}%
}; \\ 
&  \\ 
\left[ \Gamma _{j}\left[ \gamma \left( \overline{c^{q_{j}}}\right) \right]
,\Psi _{j}\left[ \phi ^{\overline{c^{q_{j}}}}\left( \overline{m}\right) %
\right] \right] & \text{if }q_{j}\in Q_{j}^{F}\diagdown Q_{j}^{F\text{%
-private}}%
\end{array}%
\right. \text{.}
\end{equation*}%
Thus, we define $\overline{t}_{j}\left( q_{j}\right) \equiv t_{j}\left(
\Sigma _{j}^{\left\langle \Gamma ^{private},\Psi ^{private}\right\rangle
}\left( q_{j}\right) \right) $.

For each $q_{j}=\left[ \Gamma _{j}\left[ \gamma \left(
c_{j}^{q_{j}},c_{-j}\right) \right] ,\Psi _{j}\left( s\left( \gamma \left(
c_{j}^{q_{j}},c_{-j}\right) ,\theta ^{q_{j}}\right) \right) \right] \in
Q_{j}^{F\text{-private}}$, define%
\begin{multline*}
\overline{b}_{j}\left( q_{j}\right) \left[ \left\{ \left(
c_{j}^{q_{j}},c_{-j}\right) \right\} \times \left\{ \overline{s}\left(
\left( c_{j}^{q_{j}},c_{-j}\right) ,\theta \right) \right\} \times \left\{
\theta \right\} \right] \\
=b_{j}\left( \Sigma _{j}^{\left\langle \Gamma ^{private},\Psi
^{private}\right\rangle }\left( q_{j}\right) \right) \left[ \left\{ \gamma
\left( c_{j}^{q_{j}},c_{-j}\right) \right\} \times \left\{ s\left( \gamma
\left( c_{j}^{q_{j}},c_{-j}\right) ,\theta \right) \right\} \times \left\{
\theta \right\} \right] \text{, }\forall \theta \in \Theta \text{.}
\end{multline*}%
i.e., $\overline{b}_{j}\left( q_{j}\right) $ mimics $b_{j}\left( \Sigma
_{j}^{\left\langle \Gamma ^{private},\Psi ^{private}\right\rangle }\left(
q_{j}\right) \right) $ regarding the belief on contracts, messages and $%
\Theta $, subject to re-labeling of contracts and messages.

For each $j\in \mathcal{J}$ and each $y_{j}\in Y_{j}$, let $c_{j}^{y_{j}}$ \
denote the degenerate menu contract $\left\{ y_{j}\right\} $, i.e., it is a
menu containing only $y_{j}$. Let $m_{j}^{y_{j}}=y_{j}$ be the unique
message in this contract. Clearly, $c_{j}^{y_{j}}$ $\in C_{j}^{P}\subset
C_{j}^{F}$.

Given $\left\langle \Gamma ^{private},\Psi ^{private}\right\rangle $, each
principal $j$ observes only $\left( c_{j}^{\prime },m_{j}^{\prime }\right) $%
, and for notational simplicity, we write $t_{j}\left[ c_{j}^{\prime
},m_{j}^{\prime }\right] $ for $t_{j}\left( \Gamma _{j}\left( c^{\prime
}\right) ,\Psi _{j}\left( m^{\prime }\right) \right) $. Furthermore, $j$ has
a degenerate belief on $C_{j}\times M_{j}$, and hence, we describe only the
marginal belief on $C_{-j}\times M_{-j}\times \Theta $. For each $q_{j}\in
Q_{j}^{F}\diagdown Q_{j}^{F\text{-private}}$, define%
\begin{gather*}
\overline{b}_{j}\left( q_{j}\right) \left[ \left\{ \left( \left(
c_{k}^{y_{k}}\right) _{k\in \mathcal{J\diagdown }\left\{ j\right\} },\left(
m_{k}^{y_{k}}\right) _{k\in \mathcal{J\diagdown }\left\{ j\right\} },\theta
\right) \right\} \right] \\
\equiv b_{j}\left( \Sigma _{j}\left( q_{j}\right) \right) \left[ \left\{
\left( \left( c_{k}^{\prime }\right) _{k\in \mathcal{J\diagdown }\left\{
j\right\} },\left( m_{k}^{\prime }\right) _{k\in \mathcal{J\diagdown }%
\left\{ j\right\} },\theta \right) :\left( y_{k}\right) _{k\in \mathcal{%
J\diagdown }\left\{ j\right\} }=\left( t_{k}\left[ c_{k}^{\prime
},m_{k}^{\prime }\right] \right) _{k\in \mathcal{J\diagdown }\left\{
j\right\} }\right\} \right] \text{, } \\
\forall \left[ \left( y_{k}\right) _{k\in \mathcal{J\diagdown }\left\{
j\right\} },\theta \right] \in \left[ \times _{k\in \mathcal{J\diagdown }%
\left\{ j\right\} }Y_{k}\right] \times \Theta \text{,}
\end{gather*}%
i.e., $\overline{b}_{j}\left( q_{j}\right) $ mimics $b_{j}\left( \Sigma
_{j}\left( q_{j}\right) \right) $ regarding induced belief on $\left[ \times
_{k\in \mathcal{J\diagdown }\left\{ j\right\} }Y_{k}\right] \times \Theta $.

\paragraph{Case 3:}

Fix $\left\langle \Gamma ,\Psi \right\rangle =\left\langle \Gamma
^{public},\Psi ^{private}\right\rangle $. For each $q_{j}\in Q_{j}^{F\text{-}%
\Gamma ^{public}}$, fix any $\theta ^{q_{j}}\in \Theta $ such that $q_{j}=%
\left[ c,\overline{s}\left( c,\theta ^{q_{j}}\right) \right] $. For each $%
q_{j}\in Q_{j}^{F}\diagdown Q_{j}^{F\text{-}\Gamma ^{public}}$, fix any $%
\left( \overline{c^{q_{j}}},\overline{m}\right) $ such that $q_{j}=\left(
\Gamma _{j}\left( \overline{c^{q_{j}}}\right) ,\Psi _{j}\left( \overline{m}%
\right) \right) $. Define $\Sigma _{j}^{\left\langle \Gamma ^{public},\Psi
^{private}\right\rangle }:Q_{j}^{F}\longrightarrow Q_{j}^{\mathcal{A}}$ as%
\begin{equation*}
\Sigma _{j}^{\left\langle \Gamma ^{public},\Psi ^{private}\right\rangle
}\left( q_{j}\right) \equiv \left\{ 
\begin{array}{cc}
\left[ c,\Psi _{j}\left( s\left( c,\theta ^{q_{j}}\right) \right) \right] & 
\text{if }q_{j}\in Q_{j}^{F\text{-}\Gamma ^{public}}; \\ 
&  \\ 
\left[ \gamma \left( \overline{c^{q_{j}}}\right) ,\Psi _{j}\left[ \phi ^{%
\overline{c^{q_{j}}}}\left( \overline{m}\right) \right] \right] & \text{if }%
q_{j}=\left( \overline{c^{q_{j}}},\overline{m}\right) \in Q_{j}^{F}\diagdown
Q_{j}^{F\text{-}\Gamma ^{public}}%
\end{array}%
\right. \text{.}
\end{equation*}%
Thus, we define $\overline{t}_{j}\left( q_{j}\right) \equiv t_{j}\left(
\Sigma _{j}^{\left\langle \Gamma ^{public},\Psi ^{private}\right\rangle
}\left( q_{j}\right) \right) $.

For each $q_{j}=\left[ \Gamma _{j}\left( c\right) ,\Psi _{j}\left( s\left(
c,\theta ^{q_{j}}\right) \right) \right] \in Q_{j}^{F\text{-private}}$,
define%
\begin{multline*}
\overline{b}_{j}\left( q_{j}\right) \left[ \left\{ c\right\} \times \left\{ 
\overline{s}\left( c,\theta ^{q_{j}}\right) \right\} \times \left\{ \theta
\right\} \right] \\
=b_{j}\left( \Sigma _{j}^{\left\langle \Gamma ^{public},\Psi
^{private}\right\rangle }\left( q_{j}\right) \right) \left[ \left\{
c\right\} \times \left\{ s\left( c,\theta ^{q_{j}}\right) \right\} \times
\left\{ \theta \right\} \right] \text{, }\forall \theta \in \Theta \text{.}
\end{multline*}%
i.e., $\overline{b}_{j}\left( q_{j}\right) $ mimics $b_{j}\left( \Sigma
_{j}^{\left\langle \Gamma ^{private},\Psi ^{private}\right\rangle }\left(
q_{j}\right) \right) $ regarding the belief on contracts, messages and $%
\Theta $, subject to re-labeling of contracts and messages.

Given $\left\langle \Gamma ^{public},\Psi ^{private}\right\rangle $, each
principal $j$ observes only $\left( c^{\prime },m_{j}^{\prime }\right) $,
and for notational simplicity, we write $t_{j}\left[ c^{\prime
},m_{j}^{\prime }\right] $ for $t_{j}\left( \Gamma _{j}\left( c^{\prime
}\right) ,\Psi _{j}\left( m^{\prime }\right) \right) $. Furthermore, $j$ has
a degenerate belief on $C\times M_{j}$, and hence, we describe only the
marginal belief on $M_{-j}\times \Theta $. For each $c^{\prime }\in \left(
\left\{ c_{k}\right\} \cup C_{k}^{F}\right) _{k\in \mathcal{J}}$ and each $%
j\in \mathcal{J}$, we use $\eta _{j}^{c^{\prime }}:M_{j}^{\gamma _{j}\left(
c_{j}^{\prime }\right) }\longrightarrow M_{j}^{c_{j}^{\prime }}$ defined
below to translate messages in $M_{j}^{\gamma _{j}\left( c_{j}^{\prime
}\right) }$ back to messages in $M_{j}^{c_{j}^{\prime }}$.%
\begin{equation*}
\eta _{j}^{c^{\prime }}\left( m_{j}\right) =\left\{ 
\begin{array}{cc}
m_{j} & \text{if }c_{j}^{\prime }=c_{j}; \\ 
\phi _{j}^{c_{j}^{\prime }}\left( m_{j}\right) & \text{if }\left( 
\begin{array}{c}
c_{j}^{\prime }\in C_{j}^{F}\diagdown \left\{ c_{j}\right\} \text{,} \\ 
\exists \left[ L_{j}\subset 2^{Y_{j}}\text{ and }H_{j}=\left\{ \left[
E_{j},y_{j}\right] :y_{j}\in E_{j}\right\} \right] \text{ satisfying
Definition }\ref{def:menu_of_menu_with_full_recommendation}\text{,} \\ 
\gamma _{j}\left( c_{j}^{\prime }\right) \left[ m_{j}\right] \neq E_{j}%
\end{array}%
\right) \\ 
\left[ E_{j},t_{j}\left[ c^{\prime },m_{j}\right] \right] & \text{if }\left( 
\begin{array}{c}
c_{j}^{\prime }\in C_{j}^{F}\diagdown \left\{ c_{j}\right\} \text{,} \\ 
\exists \left[ L_{j}\subset 2^{Y_{j}}\text{ and }H_{j}=\left\{ \left[
E_{j},y_{j}\right] :y_{j}\in E_{j}\right\} \right] \text{ satisfying
Definition }\ref{def:menu_of_menu_with_full_recommendation}\text{,} \\ 
\gamma _{j}\left( c_{j}^{\prime }\right) \left[ m_{j}\right] =E_{j}%
\end{array}%
\right)%
\end{array}%
\right.
\end{equation*}%
For each $q_{j}\in Q_{j}^{F}\diagdown Q_{j}^{F\text{-private}}$, define%
\begin{gather*}
\overline{b}_{j}\left( q_{j}\right) \left[ N\times \left\{ \theta \right\} %
\right] \equiv b_{j}\left( \Sigma _{j}^{\left\langle \Gamma ^{public},\Psi
^{private}\right\rangle }\left( q_{j}\right) \right) \left[ \left\{ \left(
\left( m_{k}\right) _{k\in \mathcal{J\diagdown }\left\{ j\right\} },\theta
\right) :\left[ \eta _{k}^{c^{q_{j}}}\left( m_{k}\right) \right] _{k\in 
\mathcal{J\diagdown }\left\{ j\right\} }\in N\right\} \right] \text{, } \\
\forall N\subset \times _{k\in \mathcal{J\diagdown }\left\{ j\right\}
}\left( M_{k}^{c^{q_{j}}}\right) \text{, }\forall \theta \in \Theta \text{,}
\end{gather*}%
i.e., $\overline{b}_{j}\left( q_{j}\right) $ mimics $b_{j}\left( \Sigma
_{j}\left( q_{j}\right) \right) $ subject to re-labeling messages via $\eta
_{k}^{c^{q_{j}}}$.

It is straightforward to see that $\left( c,s,t\right) $ is replicated by $%
\left( c,\overline{s},\overline{t}\right) $, and all the players inherit
incentive compatibility from $\left( c,s,t\right) $, i.e., $\left( c,%
\overline{s},\overline{t}\right) $ is a $\left[ C^{I},C^{F}\right] $%
-equilibrium and $z^{\left( c,s,t\right) }=z^{\left( c,\overline{s},%
\overline{t}\right) }$.$\blacksquare $

\subsection{Proof of Theorem \protect\ref{thm:full-private-private}}

\label{sec:thm:full-private-private}

\noindent \textbf{Proof. }Fix\textbf{\ }$\left\langle \mathcal{A},\text{ }%
\Gamma ,\text{ }\Psi \right\rangle =\left\langle \mathcal{A}^{non-delegated},%
\text{ }\Gamma ^{private},\text{ }\Psi ^{private}\right\rangle $. We have%
\begin{eqnarray}
Z^{\mathcal{E}^{\left\langle \mathcal{A},\text{ }\Gamma ,\text{ }\Psi
\right\rangle \text{-}\left[ C^{\mathcal{A}},C^{\mathcal{A}}\right] }}
&\supset &Z^{\mathcal{E}^{\left\langle \mathcal{A},\text{ }\Gamma ,\text{ }%
\Psi \right\rangle \text{-}\left[ C^{P},C^{\mathcal{A}}\right] }}\text{,}
\label{kka1} \\
Z^{\mathcal{E}^{\left\langle \mathcal{A},\text{ }\Gamma ,\text{ }\Psi
\right\rangle \text{-}\left[ C^{\mathcal{A}},C^{\mathcal{A}}\right] }}
&\subset &Z^{\mathcal{E}^{\left\langle \mathcal{A},\text{ }\Gamma ,\text{ }%
\Psi \right\rangle \text{-}\left[ C^{P},C^{\mathcal{A}}\right] }}\text{,}
\label{kka1a} \\
Z^{\mathcal{E}^{\left\langle \mathcal{A},\text{ }\Gamma ,\text{ }\Psi
\right\rangle \text{-}\left[ C^{P},C^{\mathcal{A}}\right] }} &=&Z^{\mathcal{E%
}^{\left\langle \mathcal{A},\text{ }\Gamma ,\text{ }\Psi \right\rangle \text{%
-}\left[ C^{P},C^{F}\right] }}\text{,}  \label{kka1b}
\end{eqnarray}%
where (\ref{kka1}) follows from Lemma \ref{lem:extension:on} and $C^{%
\mathcal{A}}\sqsupset ^{\ast }C^{P}$, (\ref{kka1a}) from Lemma \ref%
{lem:on-path:private-private}, and (\ref{kka1b}) from Proposition \ref%
{prop:deviation}. (\ref{kka1}), (\ref{kka1a}) and (\ref{kka1b}) imply
Theorem \ref{thm:full-private-private}.$\blacksquare $

\subsection{Proof of Theorem \protect\ref{thm:non-delegated_vs_delegated}}

\label{sec:thm:non-delegated_vs_delegated}

\noindent \textbf{Proof. }Fix\textbf{\ }$\left\langle \mathcal{A},\text{ }%
\Gamma ,\text{ }\Psi \right\rangle =\left\langle \mathcal{A}^{non-delegated},%
\text{ }\Gamma ^{private},\text{ }\Psi ^{private}\right\rangle $. We have%
\begin{equation*}
Z^{\mathcal{E}^{\left\langle \mathcal{A},\text{ }\Gamma ,\text{ }\Psi
\right\rangle \text{-}\left[ C^{P},C^{P}\right] }}\supset Z^{\mathcal{E}%
^{\left\langle \mathcal{A},\text{ }\Gamma ,\text{ }\Psi \right\rangle \text{-%
}\left[ C^{P},C^{F}\right] }}=Z^{\mathcal{E}^{\left\langle \mathcal{A},\text{
}\Gamma ,\text{ }\Psi \right\rangle \text{-}\left[ C^{\mathcal{A}},C^{%
\mathcal{A}}\right] }}\text{,}
\end{equation*}%
where "$=$"\ follows from Theorem \ref{thm:full-private-private} and "$%
\supset $" from Lemma \ref{lem:extension:off} and $C^{P}\sqsupset ^{\ast
\ast }C^{F}$.$\blacksquare $

\section{Example: private announcement and private communication}

\label{sec:menu:fail:private-private}

In this section, we provide an example with private announcement and private
communication that shows the failure of the converse of (18) in Theorem \ref%
{thm:non-delegated_vs_delegated}. Consider 
\begin{equation*}
\mathcal{J}=\left\{ 1,2\right\} \text{, }Y_{1}=Y_{2}=\left\{ H,T\right\} 
\text{, }\Theta =\left\{ \theta \right\} \text{.}
\end{equation*}%
Principals' preferences are described as follows.%
\begin{equation}
\begin{tabular}{|c|c|c|}
\hline
$\left( v_{1}\left[ \left( y_{1},y_{2}\right) ,\theta \right] ,\text{ }v_{2}%
\left[ \left( y_{1},y_{2}\right) ,\theta \right] \right) :$ & $y_{2}=H$ & $%
y_{2}=T$ \\ \hline
$y_{1}=H$ & $1,-1$ & $-1,1$ \\ \hline
$y_{1}=T$ & $-1,1$ & $1,-1$ \\ \hline
\end{tabular}
\label{tthh2}
\end{equation}%
The agent is indifferent among all action profiles. The outcome $%
(y_{1}=H,y_{2}=H)$ is sustained by the following equilibrium, if we focus on
delegated contract only.

\begin{enumerate}
\item On the equilibrium path, each principal $j\in \mathcal{J}$ offers $%
c_{j}^{\ast }:Y_{j}\rightarrow Y_{j}$ such that $c_{j}^{\ast }\left(
y_{j}\right) =y_{j}$ for any $y_{j}\in Y_{j}$,

\item On the equilibrium path, the agent sends $m_{j}^{\ast }=H$ to each
principal $j\in \mathcal{J}$,

\item If principal $1$ unilaterally deviates to offer $c_{1}:M_{1}%
\rightarrow Y_{1}$, fix any $m_{1}\in M_{1}$, and the agent sends $m_{1}$ to
principal $1$, and sends $y_{2}\in \left\{ H,T\right\} \diagdown \left\{
c_{1}\left( m_{1}\right) \right\} $ to principal $2$,

\item If principal $2$ unilaterally deviates to offer $c_{2}:M_{2}%
\rightarrow Y_{2}$, fix any $m_{2}\in M_{2}$, and the agent sends $m_{2}$ to
principal $2$, and sends $c_{2}\left( m_{2}\right) \in \left\{ H,T\right\} $
to principal $1$.
\end{enumerate}

In this equilibrium of the common-agency game with delegated contracts,
principal 2 achieves the minimal utility of $-1$. However, $%
(y_{1}=H,y_{2}=H) $ cannot be induced by any equilibrium if we allow for
non-delegated contracts. Specifically, in any equilibrium with non-delegated
contracts which induces $(y_{1}=H,y_{2}=H)$, principal 2 can always
unilaterally deviates to offer the degenerate contract $c_{2}^{\prime
}:M_{2}^{\prime }\rightarrow 2^{Y_{2}}\backslash \{\varnothing \}$ such that 
$c_{2}^{\prime }\left( m_{2}^{\prime }\right) =\left\{ H,T\right\} $ for any 
$m_{2}^{\prime }\in M_{2}^{\prime }$. In the continuation equilibrium
following this unilateral deviation, principal 2 must play a best reply to
principal 1's strategy, and hence, principal 2 would achieve utility of at
least the minmax value of the stage game in (\ref{tthh2}), which is $0$ if
we allow for mixed strategies, and is $1$ if we focus on pure strategies
only. The minmax value is greater than principal 2's utility under $%
(y_{1}=H,y_{2}=H)$, and hence, $(y_{1}=H,y_{2}=H)$ cannot be induced by any
equilibrium with non-delegated contracts.

\section{Proof of Theorem \protect\ref{thm:private-public}}

\label{sec:theorem}

We need the following three results to prove Theorem \ref{thm:private-public}%
.

\begin{prop}
\label{prop:other:on1}Given $\left\langle \mathcal{A},\text{ }\Gamma ,\text{ 
}\Psi \right\rangle =\left\langle \mathcal{A}^{non-delegated},\Gamma
^{private},\Psi ^{public}\right\rangle $, we have%
\begin{equation*}
Z^{\mathcal{E}^{\left\langle \mathcal{A},\text{ }\Gamma ,\text{ }\Psi
\right\rangle \text{-}\left[ C^{\mathcal{A}},C^{\mathcal{A}}\right]
}}\subset Z^{\mathcal{E}^{\left\langle \mathcal{A},\text{ }\Gamma ,\text{ }%
\Psi \right\rangle \text{-}\left[ C^{R},C^{\mathcal{A}}\right] }}\text{.}
\end{equation*}
\end{prop}

\begin{prop}
\label{prop:other:on2}Given $\left\langle \mathcal{A},\text{ }\Gamma ,\text{ 
}\Psi \right\rangle =\left\langle \mathcal{A}^{non-delegated},\Gamma
^{private},\Psi ^{public}\right\rangle $, we have%
\begin{equation*}
Z^{\mathcal{E}^{\left\langle \mathcal{A},\text{ }\Gamma ,\text{ }\Psi
\right\rangle \text{-}\left[ C^{R},C^{\mathcal{A}}\right] }}\subset Z^{%
\mathcal{E}^{\left\langle \mathcal{A},\text{ }\Gamma ,\text{ }\Psi
\right\rangle \text{-}\left[ C^{R},C^{F^{\ast }}\right] }}\text{.}
\end{equation*}
\end{prop}

\begin{prop}
\label{prop:deviation1}Given $\left\langle \mathcal{A},\text{ }\Gamma ,\text{
}\Psi \right\rangle =\left\langle \mathcal{A}^{non-delegated},\Gamma
^{private},\Psi ^{public}\right\rangle $, we have%
\begin{equation*}
Z^{\mathcal{E}^{\left\langle \mathcal{A},\text{ }\Gamma ,\text{ }\Psi
\right\rangle \text{-}\left[ C^{\mathcal{A}},C^{F^{\ast }}\right] }}\subset
Z^{\mathcal{E}^{\left\langle \mathcal{A},\text{ }\Gamma ,\text{ }\Psi
\right\rangle \text{-}\left[ C^{\mathcal{A}},C^{\mathcal{A}}\right] }}\text{.%
}
\end{equation*}
\end{prop}

\noindent \textbf{Proof of Theorem \ref{thm:private-public}.} Given $%
\left\langle \mathcal{A},\text{ }\Gamma ,\text{ }\Psi \right\rangle
=\left\langle \mathcal{A}^{non-delegated},\Gamma ^{private},\Psi
^{public}\right\rangle $, Propositions \ref{prop:other:on1} and \ref%
{prop:other:on2} imply 
\begin{equation}
Z^{\mathcal{E}^{\left\langle \mathcal{A},\text{ }\Gamma ,\text{ }\Psi
\right\rangle \text{-}\left[ C^{\mathcal{A}},C^{\mathcal{A}}\right]
}}\subset Z^{\mathcal{E}^{\left\langle \mathcal{A},\text{ }\Gamma ,\text{ }%
\Psi \right\rangle \text{-}\left[ C^{R},C^{\mathcal{A}}\right] }}\subset Z^{%
\mathcal{E}^{\left\langle \mathcal{A},\text{ }\Gamma ,\text{ }\Psi
\right\rangle \text{-}\left[ C^{R},C^{F^{\ast }}\right] }}\text{.}
\label{tat1}
\end{equation}%
Furthermore, we have%
\begin{equation}
Z^{\mathcal{E}^{\left\langle \mathcal{A},\text{ }\Gamma ,\text{ }\Psi
\right\rangle \text{-}\left[ C^{\mathcal{A}},C^{\mathcal{A}}\right]
}}\supset Z^{\mathcal{E}^{\left\langle \mathcal{A},\text{ }\Gamma ,\text{ }%
\Psi \right\rangle \text{-}\left[ C^{\mathcal{A}},C^{F^{\ast }}\right]
}}\supset Z^{\mathcal{E}^{\left\langle \mathcal{A},\text{ }\Gamma ,\text{ }%
\Psi \right\rangle \text{-}\left[ C^{R},C^{F^{\ast }}\right] }}\text{.}
\label{tat2}
\end{equation}%
where the first "$\supset $" follows from Proposition \ref{prop:deviation1},
the second "$\supset $" from $C^{\mathcal{A}}\sqsupset ^{\ast }C^{R}$ and
Lemma \ref{lem:on-path:public}. Finally, (\ref{tat1}) and (\ref{tat2}) imply
Theorem \ref{thm:private-public}.$\blacksquare $

Proposition \ref{prop:other:on1} follows Lemma \ref{lem:on-path:public}, so
we present the proofs for Propositions \ref{prop:other:on2} and \ref%
{prop:deviation1} in Sections \ref{sec:prop:other:on2} and \ref%
{sec:prop:deviation1} below.

\subsection{Proof of Proposition \protect\ref{prop:other:on2}}

\label{sec:prop:other:on2}

Fix $\left\langle \mathcal{A},\text{ }\Gamma ,\text{ }\Psi \right\rangle
=\left\langle \mathcal{A}^{non-delegated},\Gamma ^{private},\Psi
^{public}\right\rangle $, and we aim to prove%
\begin{equation*}
Z^{\mathcal{E}^{\left\langle \mathcal{A},\text{ }\Gamma ,\text{ }\Psi
\right\rangle \text{-}\left[ C^{R},C^{\mathcal{A}}\right] }}\subset Z^{%
\mathcal{E}^{\left\langle \mathcal{A},\text{ }\Gamma ,\text{ }\Psi
\right\rangle \text{-}\left[ C^{R},C^{F^{\ast }}\right] }}\text{.}
\end{equation*}%
Fix any $\left( c,s,t\right) \in \mathcal{E}^{\left\langle \mathcal{A},\text{
}\Gamma ,\text{ }\Psi \right\rangle \text{-}\left[ C^{R},C^{\mathcal{A}}%
\right] }$. We aim to replicate $\left( c,s,t\right) $ with some $\left( c,%
\overline{s},\overline{t}\right) \in \mathcal{E}^{\left\langle \mathcal{A},%
\text{ }\Gamma ,\text{ }\Psi \right\rangle \text{-}\left[ C^{R},C^{F^{\ast }}%
\right] }$ such that $z^{\left( c,s,t\right) }=z^{\left( c,\overline{s},%
\overline{t}\right) }$.

\underline{Replicating $s\equiv \left( s_{k}\right) _{k\in \mathcal{J}}$
with $\overline{s}\equiv \left( \overline{s}_{k}\right) _{k\in \mathcal{J}}$:%
}

We replicate $s$ in the $\left[ C^{R},C^{\mathcal{A}}\right] $-game with $%
\overline{s}$ in the $\left[ C^{R},C^{F^{\ast }}\right] $-game. Since $s$ is
defined on $\left( \left\{ c_{k}\right\} \cup C_{k}^{\mathcal{A}}\right)
_{k\in \mathcal{J}}$ and $\overline{s}$ is defined on $\left( \left\{
c_{k}\right\} \cup C_{k}^{F^{\ast }}\right) _{k\in \mathcal{J}}$, we need to
define a function $\gamma :\left( \left\{ c_{k}\right\} \cup C_{k}^{F^{\ast
}}\right) _{k\in \mathcal{J}}\longrightarrow \left( \left\{ c_{k}\right\}
\cup C_{k}^{\mathcal{A}}\right) _{k\in \mathcal{J}}$, such that, upon
observing $c^{\prime }\in \left( \left\{ c_{k}\right\} \cup C_{k}^{F^{\ast
}}\right) _{k\in \mathcal{J}}$ in the $\left[ C^{R},C^{F^{\ast }}\right] $%
-game, the agent regards it as $\gamma \left( c^{\prime }\right) $ in the $%
\left[ C^{R},C^{\mathcal{A}}\right] $-game, and then, $\overline{s}\left(
c^{\prime }\right) $ mimics $s\left[ \gamma \left( c^{\prime }\right) \right]
$. Thus, for each $j\in \mathcal{J}$, we define $\gamma _{j}:\left\{
c_{j}\right\} \cup C_{j}^{F^{\ast }}\longrightarrow \left\{ c_{j}\right\}
\cup C_{j}^{\mathcal{A}}$ in three steps. First,%
\begin{equation*}
\gamma _{j}\left( c_{j}\right) =c_{j}\text{.}
\end{equation*}

For each $y_{j}\in Y_{j}$, let $c_{j}^{y_{j}}$ denote the degenerate
contract such that $c_{k}^{y_{k}}\left( m_{k}\right) \equiv \left\{
y_{k}\right\} $. Second for any $c_{j}^{y_{j}}\in C_{j}^{F^{\ast }}\diagdown
\left\{ c_{j}\right\} $, we define $\gamma _{j}\left( c_{j}^{y_{j}}\right) $
as $c_{j}^{y_{j}}$ in $C_{j}^{\mathcal{A}}$. Third, for any $c_{j}^{\prime
}\in C_{j}^{F}\diagdown \left\{ c_{j}\right\} $, we have a pair of $\left[
L_{j}\subset 2^{Y_{j}}\text{ and }H_{j}=\left\{ \left[ E_{j},y_{j}\right]
:y_{j}\in E_{j}\right\} \right] $ satisfying Definition \ref%
{def:menu_of_menu_with_full_recommendation}. Fix any injective function $%
\phi _{j}^{c_{j}^{\prime }}:L_{j}\cup H_{j}\longrightarrow M_{j}^{\mathcal{A}%
}$, i.e., given $c_{j}^{\prime }\in C_{j}^{F}$, we identify a message $%
m_{j}^{\prime }\in L_{j}\cup H_{j}$ in the $\left[ C^{R},C^{F^{\ast }}\right]
$-game to the message $\phi _{j}^{c_{j}^{\prime }}\left( m_{j}^{\prime
}\right) \in M_{j}^{\mathcal{A}}$ in the $\left[ C^{R},C^{\mathcal{A}}\right]
$-game. With slight abuse of notation, let $\left( \phi _{j}^{c_{j}^{\prime
}}\right) ^{-1}:\phi _{j}^{c_{j}^{\prime }}\left[ L_{j}\cup H_{j}\right]
\longrightarrow L_{j}\cup H_{j}$ denote the inverse function of $\phi
_{j}^{c_{j}^{\prime }}$, i.e.,%
\begin{equation*}
\phi _{j}^{c_{j}^{\prime }}\left[ \left( \phi _{j}^{c_{j}^{\prime }}\right)
^{-1}\left( m_{j}\right) \right] =m_{j}\text{, }\forall m_{j}\in \phi
_{j}^{c_{j}^{\prime }}\left[ L_{j}\cup H_{j}\right] \text{.}
\end{equation*}%
We thus define $\gamma _{j}\left( c_{j}^{\prime }\right) $ for each $%
c_{j}^{\prime }\in C_{j}^{F}\diagdown \left\{ c_{j}\right\} $ as follows.%
\begin{equation*}
\gamma _{j}\left( c_{j}^{\prime }\right) \left[ m_{j}\right] =\left\{ 
\begin{array}{cc}
c_{j}^{\prime }\left( \left( \phi _{j}^{c_{j}^{\prime }}\right) ^{-1}\left[
m_{j}\right] \right)  & \text{if }m_{j}\in \phi _{j}^{c_{j}^{\prime }}\left[
L_{j}\cup H_{j}\right] ; \\ 
E_{j} & \text{otherwise,}%
\end{array}%
\right. \text{ }\forall m_{j}\in M_{j}^{\mathcal{A}}\text{,}
\end{equation*}%
i.e., we first embed the message space $L_{j}\cup H_{j}$ into $M_{j}^{%
\mathcal{A}}$ by $\phi _{j}^{c_{j}^{\prime }}$; second, we copy $%
c_{j}^{\prime }$ with $\gamma _{j}\left( c_{j}^{\prime }\right) $ on the
embedded message space; third, all of the other messages in $M_{j}^{\mathcal{%
A}}$ are mapped to $E_{j}$. Furthermore, denote $\gamma \left( c^{\prime
}\right) \equiv \left( \gamma _{k}\left( c_{k}^{\prime }\right) \right)
_{k\in \mathcal{J}}$ and $\phi ^{c^{\prime }}:\equiv \left[ \phi
_{j}^{c_{j}^{\prime }}:L_{j}\cup H_{j}\longrightarrow M_{j}^{\mathcal{A}}%
\right] _{k\in \mathcal{J}}$.

We are now ready to define $\overline{s}\equiv \left( \overline{s}%
_{k}\right) _{k\in \mathcal{J}}$. For each $j\in \mathcal{J}$, define 
\begin{equation*}
\overline{s}_{j}\left( c^{\prime },\theta \right) \equiv \left\{ 
\begin{tabular}{ll}
$s_{j}\left( \gamma \left( c^{\prime }\right) ,\theta \right) $ & if $%
c_{j}^{\prime }\in \left\{ c_{j}\right\} \cup \left\{ c_{j}^{y_{j}}:y_{j}\in
Y_{j}\right\} $; \\ 
&  \\ 
$\left( \phi _{j}^{c_{j}^{\prime }}\right) ^{-1}\left[ s_{j}\left( \gamma
\left( c^{\prime }\right) ,\theta \right) \right] $ & 
\begin{tabular}{l}
if $c_{j}^{\prime }\notin \left\{ c_{j}\right\} \cup \left\{
c_{j}^{y_{j}}:y_{j}\in Y_{j}\right\} $ \\ 
and $s_{j}\left( \gamma \left( c^{\prime }\right) ,\theta \right) \in \phi
_{j}^{c_{j}^{\prime }}\left[ L_{j}\right] ;$%
\end{tabular}
\\ 
&  \\ 
$\left[ E_{j},\text{ }y_{j}=t_{j}\left( \Gamma _{j}\left( \gamma \left(
c^{\prime }\right) \right) ,\Psi _{j}\left( s\left( \gamma \left( c^{\prime
}\right) ,\theta \right) \right) \right) \right] $ & otherwise.%
\end{tabular}%
\right.
\end{equation*}%
When the agent observes $c^{\prime }\in \left( \left\{ c_{k}\right\} \cup
C_{k}^{F^{\ast }}\right) _{k\in \mathcal{J}}$ in the $\left[
C^{R},C^{F^{\ast }}\right] $-game, the agent translates it to $\gamma \left(
c^{\prime }\right) \in \left( \left\{ c_{k}\right\} \cup C_{k}^{\mathcal{A}%
}\right) _{k\in \mathcal{J}}$ being offered in the $\left[ C^{R},C^{\mathcal{%
A}}\right] $-game. Then, $\overline{s}_{j}\left( c^{\prime },\theta \right) $
replicates $s_{j}\left( \gamma \left( c^{\prime }\right) ,\theta \right) $:
if $c_{j}^{\prime }\in \left\{ c_{j}\right\} \cup \left\{
c_{j}^{y_{j}}:y_{j}\in Y_{j}\right\} $, we have $\overline{s}_{j}\left(
c^{\prime },\theta \right) =s_{j}\left( \gamma \left( c^{\prime }\right)
,\theta \right) $; if $c_{j}^{\prime }\notin \left\{ c_{j}\right\} \cup
\left\{ c_{j}^{y_{j}}:y_{j}\in Y_{j}\right\} $ and $s_{j}\left( \gamma
\left( c^{\prime }\right) ,\theta \right) \in \phi _{j}^{c_{j}^{\prime }}%
\left[ L_{j}\right] $, $\overline{s}_{j}\left( c^{\prime },\theta \right) $
mimics $s_{j}\left( \gamma \left( c^{\prime }\right) ,\theta \right) $
subject to re-labeling of message by $\left( \phi _{j}^{c_{j}^{\prime
}}\right) ^{-1}$; otherwise, the message $s_{j}\left( \gamma \left(
c^{\prime }\right) ,\theta \right) $ pins down the subset $E_{j}$ at Stage 2
in the $\left[ C^{R},C^{\mathcal{A}}\right] $-game, and at stage 3, $j$
would take the action $t_{j}\left( \Gamma _{j}\left( \gamma \left( c^{\prime
}\right) \right) ,\Psi _{j}\left( s\left( \gamma \left( c^{\prime }\right)
,\theta \right) \right) \right) $, and hence, $\overline{s}_{j}\left(
c^{\prime },\theta \right) $ mimics this by choosing 
\begin{equation*}
\left[ E_{j},\text{ }y_{j}=t_{j}\left( \Gamma _{j}\left( \gamma \left(
c^{\prime }\right) \right) ,\Psi _{j}\left( s\left( \gamma \left( c^{\prime
}\right) ,\theta \right) \right) \right) \right]
\end{equation*}%
in the menu-of-menu-with-full-recommendation contract $c_{j}^{\prime }\in
C_{j}^{F}$.

\underline{Replicating $\left( b_{j},t_{j}\right) $ with $\left( \overline{b}%
_{j},\overline{t}_{j}\right) $:}

Given $\left\langle \Gamma ^{private},\Psi ^{public}\right\rangle $, each
principal $j$ observes only $\left( c_{j}^{\prime },m^{\prime }\right) $.
Thus, for notational simplicity, we write $t_{j}\left( c_{j}^{\prime
},m^{\prime }\right) $ and $b_{j}\left( c_{j}^{\prime },m^{\prime }\right) $
for $t_{j}\left( \Gamma _{j}\left( c^{\prime }\right) ,\Psi _{j}\left(
m^{\prime }\right) \right) $ and $b_{j}\left( \Gamma _{j}\left( c^{\prime
}\right) ,\Psi _{j}\left( m^{\prime }\right) \right) $, respectively.
Furthermore, $j$ has a degenerate belief on $C_{j}\times M$, and hence, we
describe only the marginal belief on $C_{-j}\times \Theta $.

Given $c_{j}^{\prime }\in \left\{ c_{j}\right\} \cup C_{j}^{F^{\ast }}$, let 
$M_{j}^{c_{j}^{\prime }}$ denote the domain of $c_{j}^{\prime }$. Consider%
\begin{equation*}
Q_{j}\equiv \left\{ \left( \widehat{c}_{j},\widehat{m}\right) :%
\begin{array}{c}
\exists \widehat{c}=\left( \left\{ c_{k}\right\} \cup C_{k}^{F^{\ast
}}\right) _{k\in \mathcal{J}} \\ 
m=\left( m_{k}\right) _{k\in \mathcal{J}}\in \times _{k\in \mathcal{J}%
}\left( M_{k}^{\widehat{c}_{k}}\right)%
\end{array}%
\right\} \text{,}
\end{equation*}%
\begin{equation*}
Q_{j}^{\ast }\equiv \left\{ \left( \widehat{c}_{j},s^{\ast }\left[ \left( 
\widehat{c}_{j},c_{-j}\right) ,\theta \right] \right) \in Q_{j}:\widehat{c}%
_{j}\in \left( \left\{ c_{j}\right\} \cup C_{j}^{F^{\ast }}\right) \text{
and }\theta \in \Theta \right\} \text{,}
\end{equation*}%
i.e., in the $\left[ C^{R},C^{F^{\ast }}\right] $-game, $Q_{j}$ is the set
of all possible information that principal $j$ may observe before $j$ takes
an action at Stage 3; $Q_{j}^{\ast }$ is the subset of information by which $%
j$ cannot confirm that the other players have deviated.

For each $q_{j}\in Q_{j}^{\ast }$, fix any $\left( c_{j}^{q_{j}},\theta
^{q_{j}}\right) $ such that%
\begin{equation*}
q_{j}=\left( c_{j}^{q_{j}},\overline{s}\left[ \left(
c_{j}^{q_{j}},c_{-j}\right) ,\theta ^{q_{j}}\right] \right) \text{.}
\end{equation*}%
For each $q_{j}=\left( \widehat{c}_{j},\widehat{m}\right) \in Q_{j}\diagdown
Q_{j}^{\ast }$, fix any $\left( \widetilde{c}_{j}^{q_{j}},\widetilde{m}%
^{q_{j}}\right) \in \left( \left\{ c_{j}\right\} \cup C_{j}^{\mathcal{A}%
}\right) \times M^{\mathcal{A}}$ such that%
\begin{equation*}
\widehat{c}_{j}\left( \widehat{m}_{j}\right) =\widetilde{c}%
_{j}^{q_{j}}\left( \widetilde{m}_{j}^{q_{j}}\right) \text{.}
\end{equation*}%
We record this as $\Sigma _{j}:Q_{j}\longrightarrow Q_{j}^{\mathcal{A}}$
such that%
\begin{equation*}
\Sigma _{j}\left( q_{j}\right) \equiv \left\{ 
\begin{array}{cc}
\left[ \gamma _{j}\left( c^{q_{j}}\right) ,\text{ }\overline{s}\left[ \left(
c_{j}^{q_{j}},c_{-j}\right) ,\theta ^{q_{j}}\right] \right] & \text{if }%
q_{j}\in Q_{j}^{\ast }; \\ 
&  \\ 
\left( \widetilde{c}_{j}^{q_{j}},\widetilde{m}^{q_{j}}\right) & \text{if }%
q_{j}\in Q_{j}\diagdown Q_{j}^{\ast }%
\end{array}%
\right. ,
\end{equation*}%
\begin{equation*}
\text{where }Q_{j}^{\mathcal{A}}\equiv \left\{ \left( \Gamma _{j}\left(
c^{\prime }\right) ,\Psi _{j}\left( m\right) \right) :%
\begin{array}{c}
c^{\prime }=\left( c_{k}^{\prime }\right) _{k\in \mathcal{J}}\in \times
_{k\in \mathcal{J}}\left( \left\{ c_{j}\right\} \cup C_{k}^{\mathcal{A}%
}\right) \\ 
m=\left( m_{k}\right) _{k\in \mathcal{J}}\in \times _{k\in \mathcal{J}%
}\left( M_{k}^{c_{k}^{\prime }}\right)%
\end{array}%
\right\} \text{.}
\end{equation*}%
i.e., we translate $j$'s information (at the beginning of Stage 3) in the $%
\left[ C^{R},C^{F^{\ast }}\right] $-game to that in the $\left[ C^{R},C^{%
\mathcal{A}}\right] $-game via the function $\Sigma _{j}$.

We are now ready to replicate $t_{j}$ with $\overline{t}_{j}$.%
\begin{equation*}
\overline{t}_{j}\left[ q_{j}\right] \equiv t_{j}\left[ \Sigma _{j}\left(
q_{j}\right) \right] \text{, }\forall q_{j}\in Q_{j}\text{.}
\end{equation*}%
Similarly, we replicate $b_{j}$ with $\overline{b}_{j}$. When principal $j$
observes $q_{j}=\left( c_{j}^{q_{j}},\overline{s}\left[ \left(
c_{j}^{q_{j}},c_{-j}\right) ,\theta ^{q_{j}}\right] \right) \in Q_{j}^{\ast
} $, principal $j$'s belief is induced by Bayes' rule, i.e.,%
\begin{equation*}
\overline{b}_{j}\left( q_{j}\right) \left[ \left\{ \left[ c_{-j},\text{ }%
\theta ^{\prime }\right] \right\} \right] \equiv \left\{ 
\begin{array}{cc}
\frac{p\left( \theta ^{\prime }\right) }{p\left[ \Upsilon \left( \theta
\right) \right] } & \text{if }\theta ^{\prime }\in \Upsilon \left( \theta
^{q_{j}}\right) ; \\ 
&  \\ 
0 & \text{otherwise}%
\end{array}%
\right. \text{, }\forall \theta ^{\prime }\in \Theta \text{,}
\end{equation*}%
where%
\begin{equation*}
\Upsilon \left( \theta ^{q_{j}}\right) \equiv \left\{ \widetilde{\theta }\in
\Theta :q_{j}=\left( c_{j}^{q_{j}},\overline{s}\left[ \left(
c_{j}^{q_{j}},c_{-j}\right) ,\widetilde{\theta }\right] \right) \right\} 
\text{.}
\end{equation*}

For each $k\in \mathcal{J}$ and each $y_{k}\in Y_{k}$, let $c_{k}^{y_{k}}\in
C_{k}^{F^{\ast }}$ denote the degenerate contract such that $%
c_{k}^{y_{k}}\left( m_{k}\right) \equiv \left\{ y_{k}\right\} $. When
principal $j$ observes $q_{j}\in Q_{j}\diagdown Q_{j}^{\ast }$, we have $%
\Sigma _{j}\left( q_{j}\right) =\left( \widetilde{c}_{j}^{q_{j}},\widetilde{m%
}^{q_{j}}\right) $, and define%
\begin{gather*}
\overline{b}_{j}\left[ q_{j}\right] \left[ \left\{ \left( \left(
c_{k}^{y_{k}}\right) _{k\in \mathcal{J\diagdown }\left\{ j\right\} },\theta
\right) \right\} \right] \\
\equiv b_{j}\left[ \Sigma _{j}\left( q_{j}\right) \right] \left[ \left\{
\left( \left( c_{k}^{\prime }\right) _{k\in \mathcal{J\diagdown }\left\{
j\right\} },\theta \right) :%
\begin{tabular}{l}
$\left( c_{k}^{\prime }\right) _{k\in \mathcal{J\diagdown }\left\{ j\right\}
}\in C_{-j}^{\mathcal{A}}$ and \\ 
$\left( y_{k}\right) _{k\in \mathcal{J\diagdown }\left\{ j\right\} }=\left(
t_{k}\left[ c_{k}^{\prime },\widetilde{m}^{q_{j}}\right] \right) _{k\in 
\mathcal{J\diagdown }\left\{ j\right\} }$%
\end{tabular}%
\right\} \right] \text{, } \\
\forall \left[ \left( y_{k}\right) _{k\in \mathcal{J\diagdown }\left\{
j\right\} },\theta \right] \in \left[ \times _{k\in \mathcal{J\diagdown }%
\left\{ j\right\} }Y_{k}\right] \times \Theta \text{.}
\end{gather*}%
i.e., $\overline{b}_{j}\left[ q_{j}\right] $ mimics $b_{j}\left[ \Sigma
_{j}\left( q_{j}\right) \right] $ regarding induced belief on $\left[ \times
_{k\in \mathcal{J\diagdown }\left\{ j\right\} }Y_{k}\right] \times \Theta $.

It is straightforward to see that $\left( c,s,t\right) $ is replicated by $%
\left( c,\overline{s},\overline{t}\right) $, and all the players inherit
incentive compatibility from $\left( c,s,t\right) $, i.e., $\left( c,%
\overline{s},\overline{t}\right) $ is a $\left[ C^{R},C^{F^{\ast }}\right] $%
-equilibrium and $z^{\left( c,s,t\right) }=z^{\left( c,\overline{s},%
\overline{t}\right) }$.$\blacksquare $

\subsection{Proof of Proposition \protect\ref{prop:deviation1}}

\label{sec:prop:deviation1}

Fix $\left\langle \mathcal{A},\text{ }\Gamma ,\text{ }\Psi \right\rangle
=\left\langle \mathcal{A}^{non-delegated},\Gamma ^{private},\Psi
^{public}\right\rangle $, and we aim to prove%
\begin{equation*}
Z^{\mathcal{E}^{\left\langle \mathcal{A},\text{ }\Gamma ,\text{ }\Psi
\right\rangle \text{-}\left[ C^{\mathcal{A}},C^{F^{\ast }}\right] }}\subset
Z^{\mathcal{E}^{\left\langle \mathcal{A},\text{ }\Gamma ,\text{ }\Psi
\right\rangle \text{-}\left[ C^{\mathcal{A}},C^{\mathcal{A}}\right] }}\text{.%
}
\end{equation*}%
Fix any $\left( c,s,t\right) \in \mathcal{E}^{\left\langle \mathcal{A},\text{
}\Gamma ,\text{ }\Psi \right\rangle \text{-}\left[ C^{\mathcal{A}%
},C^{F^{\ast }}\right] }$. We aim to replicate $\left( c,s,t\right) $ with
some $\left( c,s^{\ast },t^{\ast }\right) \in \mathcal{E}^{\left\langle 
\mathcal{A},\text{ }\Gamma ,\text{ }\Psi \right\rangle \text{-}\left[ C^{%
\mathcal{A}},C^{\mathcal{A}}\right] }$ such that $z^{\left( c,s,t\right)
}=z^{\left( c,s^{\ast },t^{\ast }\right) }$.

\underline{Replicating $s\equiv \left( s_{k}\right) _{k\in \mathcal{J}}$
with $s^{\ast }\equiv \left( s_{k}^{\ast }\right) _{k\in \mathcal{J}}$:}

Since $C^{\mathcal{A}}\sqsupset ^{\ast \ast }C^{F^{\ast }}$, there exists a
function $\psi _{j}:$ $C_{j}^{\mathcal{A}}\longrightarrow C_{j}^{F^{\ast }}$
for each $j\in \mathcal{J}$ such that%
\begin{equation*}
c_{j}^{\prime }\geq \psi _{j}\left( c_{j}^{\prime }\right) \text{, }\forall
c_{j}^{\prime }\in C_{j}^{\mathcal{A}}\text{.}
\end{equation*}%
Thus, for each $c_{j}^{\prime }\in C_{j}^{\mathcal{A}}$, there exists a
surjective $\iota _{j}^{c_{j}^{\prime }}:M_{j}^{\mathcal{A}}\longrightarrow
M_{j}^{F^{\ast }}$ such that%
\begin{equation*}
c_{j}^{\prime }\left( m_{j}\right) =\psi _{j}\left( c_{j}^{\prime }\right)
\left( \iota _{j}^{c_{j}^{\prime }}\left( m_{j}\right) \right) \text{, }%
\forall m_{j}\in M_{j}^{\mathcal{A}}\text{.}
\end{equation*}%
Let $\left( \iota _{j}^{c_{j}^{\prime }}\right) ^{-1}:M_{j}^{F^{\ast
}}\longrightarrow M_{j}^{\mathcal{A}}$ denote any injective function such
that%
\begin{equation*}
c_{j}^{\prime }\left( \left( \iota _{j}^{c_{j}^{\prime }}\right) ^{-1}\left(
m_{j}\right) \right) =\psi _{j}\left( c_{j}^{\prime }\right) \left(
m_{j}\right) \text{, }\forall m_{j}\in M_{j}^{F^{\ast }}\text{.}
\end{equation*}

When the agent observes $c_{j}$ in the $\left[ C^{\mathcal{A}},C^{\mathcal{A}%
}\right] $-game, he interprets it as $c_{j}$ in the $\left[ C^{\mathcal{A}%
},C^{F^{\ast }}\right] $-game. Also, when the agent observes $c_{j}^{\prime
}\in C_{j}^{\mathcal{A}}\diagdown \left\{ c_{j}\right\} $ in the $\left[ C^{%
\mathcal{A}},C^{\mathcal{A}}\right] $-game, he interprets it as $\psi
_{j}\left( c_{j}^{\prime }\right) \in C_{j}^{F^{\ast }}$ in the $\left[ C^{%
\mathcal{A}},C^{F^{\ast }}\right] $-game. To record this interpretation,
define $\gamma _{j}:C_{j}^{\mathcal{A}}\longrightarrow \left( \left\{
c_{j}\right\} \cup C_{j}^{F^{\ast }}\right) $ for each $j\in \mathcal{J}$ as%
\begin{equation*}
\gamma _{j}\left( c_{j}^{\prime }\right) \equiv \left\{ 
\begin{tabular}{ll}
$c_{j}$, & if $c_{j}^{\prime }=c_{j}$; \\ 
$\psi _{j}\left( c_{j}^{\prime }\right) $, & if $c_{j}^{\prime }\in C_{j}^{%
\mathcal{A}}\diagdown \left\{ c_{j}\right\} $,%
\end{tabular}%
\right.
\end{equation*}%
and denote $\gamma \left( c^{\prime }\right) \equiv \left( \left[ \gamma
_{k}\left( c_{k}^{\prime }\right) \right] _{k\in \mathcal{J}}\right) \in
\left( \left\{ c_{k}\right\} \cup C_{k}^{F^{\ast }}\right) _{k\in \mathcal{J}%
}$ for all $c^{\prime }\in C^{\mathcal{A}}$.

For the agent's strategy in the $\left[ C^{\mathcal{A}},C^{\mathcal{A}}%
\right] $-game, we replicate $s\equiv \left( s_{k}\right) _{k\in \mathcal{J}%
} $ with $s^{\ast }\equiv \left( s_{k}^{\ast }\right) _{k\in \mathcal{J}}$
defined as follows. For each $j\in \mathcal{J}$ and all $\left( c^{\prime
},\theta \right) \in C^{\mathcal{A}}\times \Theta $,%
\begin{equation*}
s_{j}^{\ast }\left( c^{\prime },\theta \right) =\left\{ 
\begin{tabular}{ll}
$s_{j}\left[ \gamma \left( c^{\prime }\right) ,\text{ }\theta \right] $, & 
if $c_{j}^{\prime }=c_{j}$; \\ 
$\left( \iota _{j}^{c_{j}^{\prime }}\right) ^{-1}\left( s_{j}\left[ \gamma
\left( c^{\prime }\right) ,\text{ }\theta \right] \right) $, & if $%
c_{j}^{\prime }\in C_{j}^{\mathcal{A}}\diagdown \left\{ c_{j}\right\} $.%
\end{tabular}%
\right.
\end{equation*}%
I.e., when principals offer $c^{\prime }$ in the $\left[ C^{\mathcal{A}},C^{%
\mathcal{A}}\right] $-game, the agent regards it as $\gamma \left( c^{\prime
}\right) $ in the $\left[ C^{\mathcal{A}},C^{F^{\ast }}\right] $-game. Given 
$\gamma \left( c^{\prime }\right) $ in the $\left[ C^{\mathcal{A}%
},C^{F^{\ast }}\right] $-game, the agent sends $s_{j}\left[ \gamma \left(
c^{\prime }\right) ,\text{ }\theta \right] $ to $j$ at Stage 2. If $%
c_{j}^{\prime }=c_{j}$, the agent indeed sends $s_{j}\left[ \gamma \left(
c^{\prime }\right) ,\text{ }\theta \right] $. If $c_{j}^{\prime }\in C_{j}^{%
\mathcal{A}}\diagdown \left\{ c_{j}\right\} $, $c_{j}^{\prime }$ corresponds
to $\gamma _{j}\left( c_{j}^{\prime }\right) $ in the $\left[ C^{\mathcal{A}%
},C^{F^{\ast }}\right] $-game, and the equilibrium message $s_{j}\left[
\gamma \left( c^{\prime }\right) ,\text{ }\theta \right] $ corresponds to
the message $\left( \iota _{j}^{c_{j}^{\prime }}\right) ^{-1}\left( s_{j}%
\left[ \gamma \left( c^{\prime }\right) ,\text{ }\theta \right] \right) $ in
the $\left[ C^{\mathcal{A}},C^{\mathcal{A}}\right] $-game.

\underline{Extending $\left( \widehat{t}\text{, }\widehat{b}\right) $ to $%
\left( t^{\ast }\text{, }b^{\ast }\right) $:}

Given $\left\langle \Gamma ^{private},\Psi ^{public}\right\rangle $, each
principal $j$ observes only $\left( c_{j}^{\prime },m^{\prime }\right) $.
Thus, for notational simplicity, we write $t_{j}\left( c_{j}^{\prime
},m^{\prime }\right) $ and $b_{j}\left( c_{j}^{\prime },m^{\prime }\right) $
for $t_{j}\left( \Gamma _{j}\left( c^{\prime }\right) ,\Psi _{j}\left(
m^{\prime }\right) \right) $ and $b_{j}\left( \Gamma _{j}\left( c^{\prime
}\right) ,\Psi _{j}\left( m^{\prime }\right) \right) $, respectively.
Furthermore, $j$ has a degenerate belief on $C_{j}\times M$, and hence, we
describe only the marginal belief on $C_{-j}\times \Theta $. Consider%
\begin{equation*}
Q_{j}\equiv \left\{ \left( \widehat{c}_{j},\widehat{m}\right) :\widehat{c}%
_{j}\in C_{j}^{\mathcal{A}}\text{ and }\widehat{m}\in M^{\mathcal{A}%
}\right\} \text{,}
\end{equation*}%
\begin{equation*}
Q_{j}^{\ast }\equiv \left\{ \left( \widehat{c}_{j},s^{\ast }\left[ \left( 
\widehat{c}_{j},c_{-j}\right) ,\theta \right] \right) \in Q_{j}:\widehat{c}%
_{j}\in C_{j}^{\mathcal{A}}\text{ and }\theta \in \Theta \right\} \text{,}
\end{equation*}%
i.e., in the $\left[ C^{\mathcal{A}},C^{\mathcal{A}}\right] $-game, $Q_{j}$
is the set of all possible information that principal $j$ may observe before 
$j$ takes an action at Stage 3; $Q_{j}^{\ast }$ is the subset of information
by which $j$ cannot confirm that the other players have deviated.

For each $q_{j}\in Q_{j}^{\ast }$, fix any $\left( c_{j}^{q_{j}},\theta
^{q_{j}}\right) $ such that%
\begin{equation*}
q_{j}=\left( c_{j}^{q_{j}},s^{\ast }\left[ \left(
c_{j}^{q_{j}},c_{-j}\right) ,\theta ^{q_{j}}\right] \right) \text{.}
\end{equation*}%
For each $q_{j}=\left( \widehat{c}_{j},\widehat{m}\right) \in Q_{j}\diagdown
Q_{j}^{\ast }$, fix any $\left( \widetilde{c}_{j}^{q_{j}},\widetilde{m}%
^{q_{j}}\right) \in \left( \left\{ c_{j}\right\} \cup C_{j}^{F^{\ast
}}\right) \times \left( M_{j}^{c_{j}}\cup M_{j}^{F^{\ast }}\right) $ such
that%
\begin{equation*}
\widehat{c}_{j}\left( \widehat{m}_{j}\right) =\widetilde{c}%
_{j}^{q_{j}}\left( \widetilde{m}_{j}^{q_{j}}\right) \text{.}
\end{equation*}%
We record this as $\Sigma _{j}:Q_{j}\longrightarrow Q_{j}^{F^{\ast }}$ such
that%
\begin{equation*}
\Sigma _{j}\left( q_{j}\right) \equiv \left\{ 
\begin{array}{cc}
\left[ \gamma _{j}\left( c^{q_{j}}\right) ,\text{ }s^{\ast }\left[ \left(
c_{j}^{q_{j}},c_{-j}\right) ,\theta ^{q_{j}}\right] \right] & \text{if }%
q_{j}\in Q_{j}^{\ast }; \\ 
&  \\ 
\left( \widetilde{c}_{j}^{q_{j}},\widetilde{m}^{q_{j}}\right) & \text{if }%
q_{j}\in Q_{j}\diagdown Q_{j}^{\ast }%
\end{array}%
\right. ,
\end{equation*}%
\begin{equation*}
\text{where }Q_{j}^{F^{\ast }}\equiv \left\{ \left( \Gamma _{j}\left(
c^{\prime }\right) ,\Psi _{j}\left( m\right) \right) :%
\begin{array}{c}
c^{\prime }=\left( c_{k}^{\prime }\right) _{k\in \mathcal{J}}\in \times
_{k\in \mathcal{J}}\left( \left\{ c_{j}\right\} \cup C_{k}^{F^{\ast }}\right)
\\ 
m=\left( m_{k}\right) _{k\in \mathcal{J}}\in \times _{k\in \mathcal{J}%
}\left( M_{k}^{c_{k}^{\prime }}\right)%
\end{array}%
\right\} \text{.}
\end{equation*}%
i.e., we translate $j$'s information (at the beginning of Stage 3) in the $%
\left[ C^{\mathcal{A}},C^{\mathcal{A}}\right] $-game to that in the $\left[
C^{\mathcal{A}},C^{F^{\ast }}\right] $-game via the function $\Sigma _{j}$.

We are now ready to replicate $t_{j}$ with $t_{j}^{\ast }$.%
\begin{equation*}
t_{j}^{\ast }\left[ q_{j}\right] \equiv t_{j}\left[ \Sigma _{j}\left(
q_{j}\right) \right] \text{, }\forall q_{j}\in Q_{j}\text{.}
\end{equation*}%
Similarly, we replicate $b_{j}$ with $b_{j}^{\ast }$. When principal $j$
observes $q_{j}=\left( c_{j}^{q_{j}},s^{\ast }\left[ \left(
c_{j}^{q_{j}},c_{-j}\right) ,\theta ^{q_{j}}\right] \right) \in Q_{j}^{\ast
} $, principal $j$'s belief is induced by Bayes' rule, i.e.,%
\begin{equation*}
b_{j}^{\ast }\left( q_{j}\right) \left[ \left\{ \left[ c_{-j},\text{ }\theta
^{\prime }\right] \right\} \right] \equiv \left\{ 
\begin{array}{cc}
\frac{p\left( \theta ^{\prime }\right) }{p\left[ \Upsilon \left( \theta
\right) \right] } & \text{if }\theta ^{\prime }\in \Upsilon \left( \theta
^{q_{j}}\right) ; \\ 
&  \\ 
0 & \text{otherwise}%
\end{array}%
\right. \text{, }\forall \theta ^{\prime }\in \Theta \text{,}
\end{equation*}%
where%
\begin{equation*}
\Upsilon \left( \theta ^{q_{j}}\right) \equiv \left\{ \widetilde{\theta }\in
\Theta :q_{j}=\left( c_{j}^{q_{j}},s^{\ast }\left[ \left(
c_{j}^{q_{j}},c_{-j}\right) ,\widetilde{\theta }\right] \right) \right\} 
\text{.}
\end{equation*}

For each $k\in \mathcal{J}$ and each $y_{k}\in Y_{k}$, let $c_{k}^{y_{k}}\in
C_{k}^{\mathcal{A}}$ denote the degenerate contract such that $%
c_{k}^{y_{k}}\left( m_{k}\right) =\left\{ y_{k}\right\} $ for every $%
m_{k}\in M_{k}^{\mathcal{A}}$. When principal $j$ observes $q_{j}\in
Q_{j}\diagdown Q_{j}^{\ast }$, we have $\Sigma _{j}\left( q_{j}\right)
=\left( \widetilde{c}_{j}^{q_{j}},\widetilde{m}^{q_{j}}\right) $, and define%
\begin{gather*}
b_{j}^{\ast }\left[ q_{j}\right] \left[ \left\{ \left( \left(
c_{k}^{y_{k}}\right) _{k\in \mathcal{J\diagdown }\left\{ j\right\} },\theta
\right) \right\} \right] \\
\equiv b_{j}\left[ \Sigma _{j}\left( q_{j}\right) \right] \left[ \left\{
\left( \left( c_{k}^{\prime }\right) _{k\in \mathcal{J\diagdown }\left\{
j\right\} },\theta \right) :%
\begin{tabular}{l}
$\left( c_{k}^{\prime }\right) _{k\in \mathcal{J\diagdown }\left\{ j\right\}
}\in C_{-j}^{\mathcal{A}}$ and \\ 
$\left( y_{k}\right) _{k\in \mathcal{J\diagdown }\left\{ j\right\} }=\left(
t_{k}\left[ c_{k}^{\prime },\widetilde{m}^{q_{j}}\right] \right) _{k\in 
\mathcal{J\diagdown }\left\{ j\right\} }$%
\end{tabular}%
\right\} \right] \text{, } \\
\forall \left[ \left( y_{k}\right) _{k\in \mathcal{J\diagdown }\left\{
j\right\} },\theta \right] \in \left[ \times _{k\in \mathcal{J\diagdown }%
\left\{ j\right\} }Y_{k}\right] \times \Theta \text{.}
\end{gather*}%
i.e., $b_{j}^{\ast }\left[ q_{j}\right] $ mimics $b_{j}\left[ \Sigma
_{j}\left( q_{j}\right) \right] $ regarding induced belief on $\left[ \times
_{k\in \mathcal{J\diagdown }\left\{ j\right\} }Y_{k}\right] \times \Theta $.

It is straightforward to see that $\left( c,s,t\right) $ is replicated by $%
\left( c,s^{\ast },t^{\ast }\right) $, and all the players inherit incentive
compatibility from $\left( c,s,t\right) $, i.e., $\left( c,s^{\ast },t^{\ast
}\right) $ is a $\left[ C^{\mathcal{A}},C^{\mathcal{A}}\right] $-equilibrium
and $z^{\left( c,s,t\right) }=z^{\left( c,s^{\ast },t^{\ast }\right) }$.$%
\blacksquare $

\section{Agent's effort}

\label{sec:agent_effort}

Suppose that there is a Stage 4 in the game, in which the agent chooses his
effort after observing principals' actions at Stage 3. Let $X$ be the set of
all efforts that the agent can choose, and principals' and the agent's
utility depends on the effort. Specifically, each principal $j$'s utility
function is $v_{j}^{e}:X\times Y\times \Theta \rightarrow 
\mathbb{R}
$. The agent's utility function is $u^{e}:X\times Y\times \Theta \rightarrow 
\mathbb{R}
$.

At Stage 4, the agent chooses his effort in $X$ conditional on his type,
principals' contracts, his messages and principals' action choices. Let $%
e:C\times M\times Y\times \Theta \rightarrow X$ denote the agent's effort
strategy. Define 
\begin{equation*}
\Pi \equiv \left\{ e:C\times M\times Y\times \Theta \rightarrow X:%
\begin{array}{c}
e\left( c,m,x,y\right) \in \arg \max_{x\in X}u^{e}\left( x,y,\theta \right) 
\text{, } \\ 
\forall \left( c,m,y,\theta \right) \in C\times M\times Y\times \Theta%
\end{array}%
\right\} \text{.}
\end{equation*}%
Thus, in any equilibrium, the agent's behavior strategy at Stage 4 must be a
function $e^{\ast }:C\times M\times Y\times \Theta \rightarrow X$ in $\Pi $.
Fix any $e^{\ast }\in \Pi $, consider any equilibrium in which the agent
adopt $e^{\ast }$ at Stage 4.---Since this common knowledge among all
players, by using an argument as in backward induction, we can reduce the
4-stage game to a 3-stage game with $e^{\ast }$ embedded to players' utility
functions. Given a strategy and type profile $(c,s,t,\theta ),$ the agent's
utility function $U\left( c,s,t,\theta \right) $ can be redefined as%
\begin{equation*}
U\left( c,s,t,\theta \right) \equiv u^{e}\left( e^{\ast }\left[ c,s\left(
c,\theta \right) ,t\left( \Gamma \left( c\right) ,\Psi \left( s\left(
c,\theta \right) \right) \right) ,\theta \right] ,t\left( \Gamma \left(
c\right) ,\Psi \left( s\left( c,\theta \right) \right) \right) ,\theta
\right) ,
\end{equation*}%
where $s\left( c,\theta \right) =\left[ s_{k}\left( c,\theta \right) \right]
_{k\in \mathcal{J}}$ and $t\left( \Gamma \left( c\right) ,\Psi \left(
s\left( c,\theta \right) \right) \right) =\left[ t_{k}\left( \Gamma
_{k}\left( c\right) ,\Psi _{k}\left( s\left( c,\theta \right) \right)
\right) \right] _{k\in \mathcal{J}}$. Similarly, give a strategy profile $%
(c,s,t),$ principal $j$'s expected utility $V_{j}(c,s,t)$ can be redefined
as 
\begin{equation*}
V_{j}(c,s,t)\equiv \sum_{\theta \in \Theta }p\left( \theta \right)
v_{j}^{e}\left( e^{\ast }\left[ c,s\left( c,\theta \right) ,t\left( \Gamma
\left( c\right) ,\Psi \left( s\left( c,\theta \right) \right) \right)
,\theta \right] ,t\left( \Gamma \left( c\right) ,\Psi \left( s\left(
c,\theta \right) \right) \right) ,\theta \right)
\end{equation*}%
Further, given belief $b_{j}$, principal $j$'s expected utility conditional
on $\left( \alpha _{j},\beta _{j}\right) \in \Gamma _{j}\left( C\right)
\times \Psi _{j}\left( M\right) $ is%
\begin{equation*}
V_{j}\left( t_{j},t_{-j}|\alpha _{j},\beta _{j},b_{j}\right) \equiv
\dint\limits_{b_{j}\left( \alpha _{j},\beta _{j}\right) }v_{j}^{e}\left(
e^{\ast }\left[ c,m,t\left( \Gamma \left( c\right) ,\Psi \left( m\right)
\right) ,\theta \right] ,t\left( \Gamma \left( c\right) ,\Psi \left(
m\right) \right) ,\theta \right) d\left( c,m,\theta \right) \text{,}
\end{equation*}%
where $t\left( \Gamma \left( c\right) ,\Psi \left( m\right) \right) =\left[
t_{k}\left( \Gamma _{k}\left( c\right) ,\Psi _{k}\left( m\right) \right) %
\right] _{k\in \mathcal{J}}$.

As a result, the 3-stage game is the model defined in Section \ref%
{sec:model:game}, and our results imply%
\begin{gather*}
\left\langle \mathcal{A},\text{ }\Gamma ,\text{ }\Psi \right\rangle
=\left\langle \mathcal{A}^{non-delegated},\text{ }\Gamma ^{private},\text{ }%
\Psi ^{private}\right\rangle \Rightarrow Z^{\mathcal{E}^{\left\langle 
\mathcal{A},\text{ }\Gamma ,\text{ }\Psi \right\rangle \text{-}\left[ C^{%
\mathcal{A}},C^{\mathcal{A}}\right] }}=Z^{\mathcal{E}^{\left\langle \mathcal{%
A},\text{ }\Gamma ,\text{ }\Psi \right\rangle \text{-}\left[ C^{P},C^{F}%
\right] }}\text{,} \\
\left( 
\begin{array}{c}
\mathcal{A}=\mathcal{A}^{non-delegated}\text{ and} \\ 
\left\langle \Gamma ,\text{ }\Psi \right\rangle \in \left\{ \left\langle
\Gamma ^{public},\text{ }\Psi ^{private}\right\rangle ,\left\langle \Gamma
^{public},\text{ }\Psi ^{public}\right\rangle \right\} 
\end{array}%
\right) \Rightarrow Z^{\mathcal{E}^{\left\langle \mathcal{A},\text{ }\Gamma ,%
\text{ }\Psi \right\rangle \text{-}\left[ C^{\mathcal{A}},C^{\mathcal{A}}%
\right] }}=Z^{\mathcal{E}^{\left\langle \mathcal{A},\text{ }\Gamma ,\text{ }%
\Psi \right\rangle \text{-}\left[ C^{R},C^{F}\right] }}\text{,} \\
\left\langle \mathcal{A},\text{ }\Gamma ,\text{ }\Psi \right\rangle
=\left\langle \mathcal{A}^{non-delegated},\text{ }\Gamma ^{private},\text{ }%
\Psi ^{public}\right\rangle \Rightarrow Z^{\mathcal{E}^{\left\langle 
\mathcal{A},\text{ }\Gamma ,\text{ }\Psi \right\rangle \text{-}\left[ C^{%
\mathcal{A}},C^{\mathcal{A}}\right] }}=Z^{\mathcal{E}^{\left\langle \mathcal{%
A},\text{ }\Gamma ,\text{ }\Psi \right\rangle \text{-}\left[
C^{R},C^{F^{\ast }}\right] }}\text{.}
\end{gather*}%
Since this is true for any $e^{\ast }\in \Pi $, we conclude that all of the
results still hold for the 4-stage game with the agent's effort.

\section{Mixed-strategy equilibria\label{sec:mixed_strategy_eq}}

Though we focus on pure-strategy equilibria in our analysis, our results can
be easily extended for mixed-strategy equilibria. To accommodate the change,
we need to modify the definitions of menus, menu-of-menu-with-recommendation
contracts and menu-of-menu-with-full-recommendation contracts.

Our presumption on $C^{\mathcal{A}^{non-delegated}}$ is that $M_{j}^{%
\mathcal{A}^{non-delegated}}$ is so general that any equilibrium allocation
in $Z^{\mathcal{E}^{\left\langle \mathcal{A},\text{ }\Gamma ,\text{ }\Psi
\right\rangle \text{-}\left[ C^{\mathcal{A}},C^{\mathcal{A}}\right] }}$ is
robust in the sense that it survives even when a principal deviates to a
more complex contract, which is not in $C^{\mathcal{A}^{non-delegated}}$.
Given that our interest is mixed-strategy equilibrium allocations, that
might not be the case if $\left\vert M_{j}^{\mathcal{A}^{non-delegated}}%
\right\vert <\left\vert \triangle \left( Y_{j}\right) \right\vert $.
Therefore, we assume that $\left\vert M_{j}^{\mathcal{A}^{non-delegated}}%
\right\vert \geq \left\vert \triangle \left( Y_{j}\right) \right\vert $.

\textbf{New definition of menus:} A menu is a contract, $c_{j}^{\triangle
}:E_{j}\rightarrow Y_{j}$, such that%
\begin{equation*}
E_{j}\in 2^{\triangle \left( Y_{j}\right) }\diagdown \left\{ \varnothing
\right\} \text{ and }c_{j}^{\triangle }\left( y_{j}\right) =y_{j}\text{, }%
\forall y_{j}\in E_{j}\text{.}
\end{equation*}%
While principal $j$ delegates the randomization of her action choice to the
agent in a menu contract, she randomizes her action choice in
menu-of-menu-with-recommendation contracts and
menu-of-menu-with-full-recommendation contracts. Let $C^{P\text{-}%
\bigtriangleup }$ denote the set of such menu profiles.

\textbf{New definition of menu-of-menu-with-recommendation contracts: }For
each $j\in \mathcal{J}$, define%
\begin{equation*}
M_{j}^{R\text{-}\triangle }\equiv \left\{ \left[ E_{j},y_{j}\right]
:E_{j}\in 2^{Y_{j}}\diagdown \left\{ \varnothing \right\} \text{ and }%
y_{j}\in \triangle \left( E_{j}\right) \right\} \text{.}
\end{equation*}%
Then, a menu-of-menu-with-recommendation contract for principal $j$ is a
pair of (i) $K_{j}\subset 2^{M_{j}^{R\text{-}\triangle }}\diagdown \left\{
\varnothing \right\} $ and (ii) a function, $c_{j}^{\triangle
}:K_{j}\rightarrow 2^{Y_{j}}\diagdown \left\{ \varnothing \right\} $ such
that%
\begin{equation*}
c_{j}^{\triangle }\left( \left[ E_{j},y_{j}\right] \right) =E_{j}\text{, }%
\forall \left[ E_{j},y_{j}\right] \in K_{j}\text{.}
\end{equation*}%
Note that the agent's recommendation to principal $j$ is a mixed action
(i.e., we replace $M_{j}^{R}$ with $M_{j}^{R\text{-}\triangle }$). Let $C^{R%
\text{-}\bigtriangleup }$ denote the set of such
menu-of-menu-with-recommendation contract profiles.

\textbf{New definition of menu-of-menu-with-full-recommendation contracts: }%
A menu-of-menu-with-full-recommendation contract for principal $j$ is a
function $c_{j}^{\triangle }:L_{j}\cup H_{j}\rightarrow 2^{Y_{j}}\diagdown
\left\{ \varnothing \right\} $ with 
\begin{gather*}
H_{j}=\left\{ \left[ E_{j},y_{j}\right] :y_{j}\in \triangle \left(
E_{j}\right) \right\} \text{ for some }E_{j}\in 2^{Y_{j}}\diagdown \left\{
\varnothing \right\} , \\
L_{j}\subset 2^{Y_{j}}\diagdown \left\{ E_{j}\right\} ,
\end{gather*}%
such that 
\begin{eqnarray*}
c_{j}^{\triangle }\left( E_{j}^{\prime }\right) &=&E_{j}^{\prime }\text{, }%
\forall E_{j}^{\prime }\in L_{j}\text{,} \\
c_{j}^{\triangle }\left( \left[ E_{j},y_{j}\right] \right) &=&E_{j}\text{, }%
\forall \left[ E_{j},y_{j}\right] \in H_{j}\text{.}
\end{eqnarray*}%
Let $C^{F\text{-}\bigtriangleup }\equiv \times _{k\in \mathcal{J}}C_{k}^{F%
\text{-}\bigtriangleup }$ denote the set of such
menu-of-menu-with-full-recommendation contract profiles.

Furthermore, define%
\begin{equation*}
C^{F^{\ast }\text{-}\bigtriangleup }\equiv \times _{k\in \mathcal{J}}\left[
C_{k}^{F\text{-}\bigtriangleup }\cup \left\{ c_{k}^{y_{k}}:y_{k}\in
Y_{k}\right\} \right] \text{,}
\end{equation*}%
where $c_{k}^{y_{k}}$ $:M_{k}^{R-F}\rightarrow 2^{Y_{k}}\diagdown \left\{
\varnothing \right\} $ denote the following degenerate contract:%
\begin{equation*}
c_{k}^{y_{k}}\left( m_{k}\right) =\left\{ y_{k}\right\} ,\forall m_{k}\in
M_{k}^{R-F}
\end{equation*}%
as defined in Section \ref{sec:private-public:characterization}.

With the changes above, it is easy to show 
\begin{gather*}
\left\langle \mathcal{A},\text{ }\Gamma ,\text{ }\Psi \right\rangle
=\left\langle \mathcal{A}^{non-delegated},\text{ }\Gamma ^{private},\text{ }%
\Psi ^{private}\right\rangle \Rightarrow Z^{\mathcal{E}^{\left\langle 
\mathcal{A},\text{ }\Gamma ,\text{ }\Psi \right\rangle \text{-}\left[ C^{%
\mathcal{A}},C^{\mathcal{A}}\right] }}=Z^{\mathcal{E}^{\left\langle \mathcal{%
A},\text{ }\Gamma ,\text{ }\Psi \right\rangle \text{-}\left[ C^{P\text{-}%
\bigtriangleup },C^{F\text{-}\bigtriangleup }\right] }}\text{,} \\
\left( 
\begin{array}{c}
\mathcal{A}=\mathcal{A}^{non-delegated}\text{ and} \\ 
\left\langle \Gamma ,\text{ }\Psi \right\rangle \in \left\{ \left\langle
\Gamma ^{public},\text{ }\Psi ^{private}\right\rangle ,\left\langle \Gamma
^{public},\text{ }\Psi ^{public}\right\rangle \right\} 
\end{array}%
\right) \Rightarrow Z^{\mathcal{E}^{\left\langle \mathcal{A},\text{ }\Gamma ,%
\text{ }\Psi \right\rangle \text{-}\left[ C^{\mathcal{A}},C^{\mathcal{A}}%
\right] }}=Z^{\mathcal{E}^{\left\langle \mathcal{A},\text{ }\Gamma ,\text{ }%
\Psi \right\rangle \text{-}\left[ C^{R\text{-}\bigtriangleup },C^{F\text{-}%
\bigtriangleup }\right] }}\text{,} \\
\left\langle \mathcal{A},\text{ }\Gamma ,\text{ }\Psi \right\rangle
=\left\langle \mathcal{A}^{non-delegated},\text{ }\Gamma ^{private},\text{ }%
\Psi ^{public}\right\rangle \Rightarrow Z^{\mathcal{E}^{\left\langle 
\mathcal{A},\text{ }\Gamma ,\text{ }\Psi \right\rangle \text{-}\left[ C^{%
\mathcal{Ak}},C^{\mathcal{A}}\right] }}=Z^{\mathcal{E}^{\left\langle 
\mathcal{A},\text{ }\Gamma ,\text{ }\Psi \right\rangle \text{-}\left[ C^{R%
\text{-}\bigtriangleup },C^{F^{\ast }\text{-}\bigtriangleup }\right] }}\text{%
.}
\end{gather*}

\bibliographystyle{econometrica}
\bibliography{common-agency}

\end{document}